\documentclass[11pt,a4paper]{article}
\pdfoutput=1
\usepackage{jheppub}
\usepackage{rotating}

\usepackage{wrapfig}
\usepackage{graphicx,graphbox}
\usepackage{amsfonts}
\usepackage{amsmath}
\usepackage{slashed}
\usepackage{amssymb}
\usepackage{latexsym}
\usepackage{multirow}
\usepackage{hyperref}
\usepackage{enumitem}
\hypersetup{colorlinks=true,citecolor=red,linkcolor=magenta,urlcolor=blue}
\usepackage{relsize}
\usepackage{booktabs}
\usepackage{subfigure}
\usepackage{cleveref}
\usepackage{fancyhdr}
\usepackage{float}
\usepackage{graphicx}
\usepackage{microtype}
\usepackage{tabularx}
\usepackage{xcolor}
\usepackage{siunitx,adjustbox,bm}
\usepackage[bottom]{footmisc}
\usepackage{makecell}
\usepackage[normalem]{ulem}

\usepackage{mathtools}
\usepackage[thinc]{esdiff}
\usepackage[euler]{textgreek}
\usepackage{braket}
\usepackage{physics}
\usepackage{xfrac}
\usepackage{soul}
\usepackage{stmaryrd}

\newcommand{\nn}{\nonumber}

\newcommand{\eff}{\text{eff}}
\renewcommand{\L}{\mathcal{L}}

\begin{document}
\allowdisplaybreaks
\title{One-loop Effective Action up to Dimension Eight: Integrating out Heavy Scalar(s) }
\abstract{We present the complete one-loop effective action up to dimension eight after integrating out degenerate scalars using the Heat-Kernel method. The result is provided without assuming any specific form of either UV or low energy  theories, i.e., universal. In this paper, we consider the effects of only heavy scalar propagators in the loops. We also verify  part of the results using the covariant diagram technique.}	
\author{Upalaparna Banerjee, Joydeep Chakrabortty, Shakeel Ur Rahaman, Kaanapuli Ramkumar}

\emailAdd{upalab@iitk.ac.in, joydeep@iitk.ac.in, shakel@iitk.ac.in, kaanapuliramkumar08@gmail.com}
\affiliation{Indian Institute of Technology Kanpur, Kalyanpur, Kanpur 208016, Uttar Pradesh, India}

	
\preprint{}

\maketitle
	
\newpage
\section{Introduction}

     The precision era in particle physics began to dawn following the discovery of the Higgs boson. Since then, various experiments indicate that the Standard Model (SM) and the new physics are separated on the energy scale. In this situation, the Effective Field Theory (EFT)~\cite{Weinberg:1980wa,Georgi:1994qn,Manohar:2018aog,Cohen:2019wxr} emerges as the practical and economical working framework. Instead of dealing with all of the numerous feasible options for beyond Standard Model (BSM) scenarios, we can put them on the same footing by treating the SM as a low-energy effective theory. The Standard Model effective field theory (SMEFT) \cite{Brivio:2017vri,Isidori:2023pyp} works as a bridge between the unknown territory of new physics and the SM.
	
    EFT is primarily divided into two categories. The model-independent approach, also known as the bottom-up approach, involves augmenting the low-energy Lagrangian with higher mass dimensional operators while respecting the symmetry of the theory. Because of the redundancy caused by algebraic relations such as Fierz identities and integration by parts (IBP) satisfied by the operators, listing the higher dimensional operators becomes a challenging endeavour. The operators that are free from these redundancies are known to form Green's set \cite{Gherardi:2020det,Chala:2021cgt}. When considering the external states to be on-shell, the equation of motion or the freedom of field redefinition \cite{Georgi:1991ch,Barzinji:2018xvu,Criado:2018sdb,Banerjee:2022thk} must also be taken into account, imposing more constraints on the number of independent operators that form a basis. As we go higher and higher dimension the number of operators proliferates making it impossible to keep track of these redundancies. Several techniques and automated tools \cite{Henning:2015alf,Lehman:2015coa,Henning:2017fpj,Lehman:2015via,Fonseca:2019yya,Fonseca:2017lem,Gripaios:2018zrz,Criado:2019ugp,Marinissen:2020jmb,Banerjee:2020bym,Harlander:2023psl,Li:2021tsq,Li:2022tec} have been developed in order to construct higher dimensional operators \cite{BUCHMULLER1986621,Grzadkowski:2010es,Lehman:2014jma,Murphy:2020rsh,Li:2020gnx,Li:2020xlh,Liao:2020jmn,Anisha:2019nzx,Banerjee:2020jun}. In this bottom-up approach, the deviation from the low energy theory predictions is parametrised by the Wilson coefficients (WCs) of these effective operators by treating them as free parameters \cite{Hartland:2019bjb,Brivio:2019ius,Brivio:2017btx,Ellis:2020unq,Bagnaschi:2022whn,Ellis:2018gqa}. If some anomaly in the data is well explained by the effective operators, we can find UV completion using the straightforward symmetry base arguments, which aids in the model selection \cite{DasBakshi:2021xbl,Naskar:2022rpg,Cepedello:2022pyx,Guedes:2023azv,Gargalionis:2020xvt,Li:2022abx,Chakrabortty:2020mbc}.
	
    The alternate is the model dependent or the top-down approach. In this framework, we begin with an action $S[\Phi,\phi]$ for a UV complete model consisting of both heavy and light fields, which we denote as $\Phi$ and $\phi$ respectively. The corresponding low-energy effective action is obtained by expanding $\Phi$ around its classical minima ($\Phi_c$) and evaluating the path integral over the dynamic fluctuations ($\eta$), so the heavy field is written as, $\Phi = \Phi_c + \eta$. The effective action is obtained by,
{\small\begin{eqnarray}\label{eq:effective-action}
        e^{i\,S_{\text{eff}}[\phi]} &=& \int [\mathcal{D}\Phi]e^{i\,S[\Phi,\phi]}
        = \int [\mathcal{D}\eta]\, \exp\left[i\,\left(S[\Phi_c,\phi]\,+\,\frac{1}{2}\frac{\delta^2 S}{\delta \Phi^2}\bigg\vert_{\Phi=\Phi_c}\,\eta^2\,+\,\mathcal{O}(\eta^3)\right)\right],\nonumber\\
        S_{\text{eff}} &\approx& S[\Phi_c]\,+\,\frac{i}{2} \text{Tr} \log\left(\frac{\delta^2 S}{\delta \Phi^2}\bigg\vert_{\Phi=\Phi_c}\right)
        = S[\Phi_c]\,+\,\frac{i}{2} \text{Tr} \log\Big(D^2+M^2+U\Big),
\end{eqnarray}}
	where, $D$ is the covariant derivative, $M$ is the mass corresponding to the heavy field $\Phi$ and $U$ is the field-dependent term, functional of the light fields $\phi$. 
	The term $S[\Phi_c]$ in Eq.~\eqref{eq:effective-action} arises when the heavy fields are integrated out at the tree-level and can be simply obtained by substituting the heavy field with its classical solution in the action. The other term in the equation corresponds to the part of the effective action generated at the one-loop level. This procedure of integrating out the heavy field(s) is also known as the matching of the UV theory and the low energy effective theory at an energy scale $M$. In this approach, the WCs accompanying the effective operators are functions of the UV parameters. This framework provides the means to bring together a variety of UV models under the same umbrella and conduct a comparative analysis \cite{Anisha:2020ggj,Anisha:2021hgc,Bakshi:2020eyg}. 
	
	As one can infer from Eq.~\eqref{eq:effective-action}, the structure of the UV action is quite general and it does not depend on the specific choice of the UV theory. So computing the effective operators in terms of $D$ and $U$ with proper factors of $1/M$ should provide a universal formula for integrating out \cite{Henning:2014wua}. This is indeed the case and computation of the one-loop term in Eq.~\eqref{eq:effective-action}, has resulted in a master formula known as the universal one-loop effective action (UOLEA) \cite{Drozd:2015rsp,Ellis:2016enq,delAguila:2016zcb,Ellis:2017jns,Kramer:2019fwz,Angelescu:2020yzf,Ellis:2020ivx}. In addition, some automated tools have been devised to aid this process \cite{Bakshi:2018ics,Fuentes-Martin:2022jrf,Carmona:2021xtq,Cohen:2020qvb,Fuentes-Martin:2020udw,Criado:2017khh}. The UOLEA described in the aforementioned literature has been limited to computing operators up to mass dimension six (D6). In spite of the fact that it can be calculated up to any mass dimension, the process becomes very complicated as we go for the higher dimension. 

     There are multiple theoretical ways to integrate out the heavy fields to compute the low-energy effective Lagrangian: using functional methods \cite{Gaillard:1985uh,Cheyette:1987qz,Chan1985,Henning:2014wua,Henning:2016lyp,Dittmaier:2021fls,Fuentes-Martin:2016uol,Cohen:2020fcu}, using Feynman diagrams, and covariant diagram techniques \cite{Vandeven:1985, Zhang:2016pja}. Recently, in Ref.~\cite{Gersdorff:2023} the Heat-Kernel (HK) method is used to find covariant Feynman rules in the context of EFT. The HK method
     \cite{Minakshisundaram:1949xg,Minakshisundaram:1953xh,Hadamard2003,DeWitt:1964mxt,Seeley:1969,Schwinger:1951nm,Vassilevich:2003xt,Avramidi:2001ns,Avramidi:2015ch5,Kirsten:2001wz,Fulling:1989nb} has been instrumental to compute effective actions in quantum field theory and quantum gravity \cite{Avramidi:1994fx}. Using its intrinsic properties and background field method, the higher-order corrections in different space-time dimensions are computed in \cite{Avramidi:1991,Belkov:1996CA,Fliegner:1993wh,Fliegner:1997rk}. 
 
	As we commence to make progressively precise measurements of low energy observable, the dimension six terms will no longer be enough to explain the data, and much emphasis has been directed to mass dimension eight (D8) operators to keep up with the experimental precision \cite{Dawson:2021xei,Hays:2018zze,Corbett:2021eux,DasBakshi:2023htx,DasBakshi:2022mwk,Chala:2021pll,Alioli:2020kez,Degrande:2013kka,Hays:2020scx,Dawson:2023ebe,Dawson:2022cmu,Ellis:2019zex,Corbett:2023qtg,Ellis:2023zim,Degrande:2023iob,Banerjee:2023bzl}. Additionally, at the cross-section level, the interference term between the dimension eight and the renormalisable terms may become equivalent with the dimension six squared terms. It also provides leading order contributions to some physical processes such as light-by-light scattering that involves the neutral quartic gauge couplings. In the study of electroweak precision observable (EWPO),  the $\widehat{U}$-parameter receives its first contribution at dimension eight \cite{Grinstein:1991cd}.

	The primary goal of our paper is to extend the UOLEA to generate terms up to mass dimension eight. In this regard, we would also wish to advocate the advantages of the HK method.  The application of the Heat-Kernel method in this context is relatively new. We provide an extensive overview of the computation of the effective action and limit ourselves to the one-loop diagrams that involve either one heavy scalar or multiple degenerate heavy scalars. We  also use the conventional covariant diagram approach to validate some of our findings. We want to emphasise that the HK prescription is not just limited to  heavy scalar loops and can be applied to mixed loops regardless of the spin or mass (heavy or light) of the particles involved. We are aware that in order to complete the UOLEA, one needs also to take into consideration these contributions, so we have set this aside for future study. Heat-Kernel has the capacity to expand beyond one-loop,  making it more applicable to the study of precision physics.

     The paper is organised as follows. To start with we briefly review the relevant working principle of the Heat-Kernel method in Sec.~\ref{sec:heat-kernel-intro}. Then, in the following Sec.~\ref{sec:comp-hk-coeff-gen}, we discuss the computation of HK coefficients with one detailed example. In Sec.~\ref{sec: 1loop_HKC}, we define the connection between the one-loop effective action with the HK coefficients. In the following Sec.~\ref{sec:complete-result-from-HKM}, we discuss all possible independent structures of operators having up to mass dimension eight, accompanied by suitable numerical factors,  that appear after integrating out heavy scalars from one-loop diagrams. In the next Sec.~\ref{sec:verification}, we employ a new method to compute the effective operators, starting from the same UV action, based on the covariant diagram technique. This allows us to independently cross-check some of our results computed using the HK method. In Sec.~\ref{sec:uolea},  we write down the universal one-loop effective Lagrangian after collecting all the contributions computed in the earlier section and adding them up suitably. We also discuss a toy model to highlight the failure of naive power counting and the importance of different terms in our effective action. Then, we conclude  in Sec.~\ref{sec:conc}. 

\section{Heat Kernel: a brief review}
\label{sec:heat-kernel-intro}

The one-loop effective action in Eq.~\eqref{eq:effective-action} can be written as a spectral function of a second-order elliptic differential operator after Wick's rotation. This allows us to study the one-loop effective action using spectral analysis methods, one such example is the computation of the Heat-Kernel coefficients (HKC).

Any second-order elliptic partial differential operators can be written in terms of a covariant derivative and a scalar function as \cite{Avramidi:2015ch5},
\begin{equation}\label{eq:lap}
    \Delta= D^2 +U+M^2,
\end{equation}
where $D_\mu \equiv \partial_\mu-iA_\mu$ is the covariant derivative with $A_\mu$ being the connection, $U$ is a space dependent scalar function, and $M$ is a space independent scalar function. Later, $M$ will be mapped to the mass term of the heavy field(s) in the process of one-loop effective action computation using HKCs.

Let us consider $\lambda_n$ as the eigenvalues of the operator $\Delta$ corresponding to eigenvectors $\phi_n$. If $\Delta$ is a self-adjoint operator, the Heat-Kernel can be written as \cite{Kirsten:2001wz,Vassilevich:2003xt}
\begin{equation}
    K(t,x,y,\Delta)=\bra{y}e^{-t\Delta}\ket{x}=\sum_n e^{-t\Delta}\phi_n(x)\phi^\dagger_n(y).
\end{equation}
This HK satisfies the heat equation \cite{Avramidi:2015ch5,Vassilevich:2003xt}
\begin{equation}\label{eq:heat_eq}
    \left(\partial_t+\Delta_x\right)K(t,x,y,\Delta)=0,
\end{equation}
along with the initial condition
\begin{equation}\label{eq:initial_condition}
    K(0,x,y,\Delta)=\delta(x-y).
\end{equation}
In passing we would like to mention that $t>0$ is a parameter, not to be confused with time, while defining the heat equation.
In the absence of any interaction, i.e., for the free operator, $\Delta_0=\partial_\mu\partial^\mu+M^2$, the HK can be written as \cite{Avramidi:2015ch5,Fulling:1989nb}
\begin{equation}\label{eq:free_HK}
    K_0(t,x,y)=(4\pi t)^{-d/2}\ \mathrm{Exp}\left[\frac{z^2}{4t}-t\,M^2\right],
\end{equation}
where $z_\mu=(x-y)_\mu$ and $d$ is the dimension we are working in. From here on, we assume four-dimensional Euclidean space, i.e., $d=4$.

For any general Laplace type operator with a potential term within a gauge theory, see the operator in  Eq.~\eqref{eq:lap}, the HK can be written in terms of the free operator HK and an interaction part ($H$) as \cite{DeWitt:1964mxt,Avramidi:2015ch5}
\begin{equation}\label{eq:anzat}
     K(t,x,y,\Delta)=K_0(t,x,y)\ H(t,x,y,\Delta).
\end{equation}
In case of this potential term is bounded from below, the operator $\Delta$ is positive, and $H(t,x,y,\Delta)$ converges as $t\rightarrow\infty$. This allows one to write the interaction part in terms of a power law expansion in $t$ as \cite{DeWitt:1964mxt,Seeley:1969}
\begin{equation}\label{eq:interaction}
    H(t,x,y,\Delta)=\sum_k \frac{(-t)^k}{k\,!}b_k(x,y),
\end{equation}
where $b_k$ are the Heat-Kernel coefficients \footnote{Also known as the Hadamard - Minackshisundaram - De Witt - Seeley (HMDS) coefficients \cite{Hadamard2003,Minakshisundaram:1953xh,DeWitt:1964mxt,Seeley:1969}.} (HKCs).  These $b_k$ are polynomials of covariant derivative ($D_\mu$) and the scalar operator $U$. This method has no restrictions on the dimensions of the covariant derivative and scalar function in the elliptic operator and can be generalised to matrix-valued $U$ and $D_\mu$ \cite{Fulling:1989nb}.

\section{Computation of Heat-Kernel Coefficients: general method}
\label{sec:comp-hk-coeff-gen}
Blending  the ansatz depicted in Eq.~\eqref{eq:anzat} in heat equation, Eq.~\eqref{eq:heat_eq}, the interaction term satisfies \cite{Avramidi:2015ch5}
 \begin{equation}\label{eq:heat_eq_exp}
    \left(\partial_t+\frac{1}{t} z_\mu D^\mu +D^2 + U\right) H(t,x,y,\Delta)  =0.
\end{equation}
Then, further invoking the power law expansion in $t$, see Eq.~\eqref{eq:heat_eq_exp}, we obtain a recursion relation for the HKCs \cite{Belkov:1996CA}
\begin{equation}\label{eq:rec_rel}
    \left(k+z\cdot D\right)b_k=k\left(U+D^2\right)b_{k-1}.
\end{equation}
Here, $b_k$s for $k<0$ are considered to be 0. The combination of  the free heat kernel part  Eq.~\eqref{eq:free_HK}, and the initial condition Eq.~\eqref{eq:initial_condition} leads to $b_0(x,x)=I$, where $I$ is the identity matrix \cite{Avramidi:2015ch5,Fulling:1989nb}. This allows us to start the recursion relation with $k \geq 0$.

In a later section, Sec.~\ref{sec: 1loop_HKC}, we will see that the trace over the HKC at the coinciding limit, i.e., $\tr\ [ b_k (x,x) ]$, are related to the operators of the one-loop effective action where the scalar functions $U$ and $M^2$ will be mapped in to functional of light-fields and the mass term of the heavy scalar field. Thus, our primary focus is to compute the HKCs at the coincident limit $[b_k(x,x)]$. To do so, we use the following relations \cite{Belkov:1996CA},
\clearpage
\begin{align}
    & D_{\mu_1} D_{\mu_2}...D_{\mu_m} b_k\Big|_{z=0}= \frac{1}{m+k}\Big\{  k\,D_{\mu_1} D_{\mu_2}...D_{\mu_m}\ (U+D^2)\ b_{k-1} - T_{\mu_1\mu_2...\mu_m}b_k\Big\}\Big|_{z=0},\label{eq:derv_rec_rel}\\
    & \text{where, } T_{\mu_1\mu_2...\mu_m} = \{ D_{\mu_1} D_{\mu_2}...D_{\mu_m} (z\cdot D)\}
    \big|_{z=0} -m\,  D_{\mu_1} D_{\mu_2}...D_{\mu_m}, \label{eq:t_exp}\\
    & \hspace{.5cm}\Rightarrow\; T_{\mu_1\mu_2...\mu_m} = D_{\mu_1} T_{\mu_2...\mu_m} +  R_{\mu_2...\mu_m,\mu_1},\label{eq:t_rec}\\
    & \text{in the above equation, } R_{\mu_2...\mu_m,\mu_1}=[D_{\mu_2}...D_{\mu_m},D_{\mu_1}],\label{eq:r_exp}\\
    & \hspace{3cm}\Rightarrow\; R_{\mu_2\mu_3...\mu_m,\mu_1} = G_{\mu_2\mu_1} D_{\mu_3}...D_{\mu_m}+ D_{\mu_2} D_{\mu_3...\mu_m,\mu_1}\label{eq:r_rec},
\end{align}
where $G_{\mu\nu}=[D_\mu,D_\nu]$ is the stress tensor.
 
While solving the HKCs, the recursion relation Eq.~\eqref{eq:rec_rel} leads to terms involving derivatives of the HKCs, e.g.,  $D_\mu D_\nu...b_k(x,y)|_{z=0}$\footnote{It is important to note that for  HKC computation at coinciding point, one must not set $z=0$ in Eq.~\eqref{eq:rec_rel} as $\{D_\mu (z\cdot D) \,b_k(x,y)\}|_{z=0} = D_\mu\, b_k(x,y)|_{z=0} \neq 0$.}. To compute of one-loop effective Lagrangian systematically, we organise the operator classes of the form $\mathcal{O}(D^r U^s)$ and arrange them in a polynomial with different non-negative integer powers of the covariant derivative $(D)$ and the field-dependent term $(U)$. 

Operators of the form $\mathcal{O}(D^r U^s)$ appear in the HKC $b_n (x,y)$ where $n=r/2+s$. It is possible to extract the relevant part of the HKC $b_n$ for the operators of a specific class using Eqs.~\eqref{eq:derv_rec_rel}-\eqref{eq:r_rec} as
\begin{equation}\label{eq:HKC_class}
   \mathcal{O}(D^{r}U^s) \equiv [b_{r/2+s}] \llbracket U^s \rrbracket = \sum_{k=0}^{n=\frac{r}{2}+s} \frac{n!\,(n-1)!}{k!\,(2n-k)!}\Big\{k\ D^{2(n-k)} \{U\, b_{k-1}\llbracket U^{s-1} \rrbracket\} - T_{2(n-k)}\, b_k \llbracket U^s \rrbracket\Big\}_{z=0}.
\end{equation}
In the next sections, we will further use the following notations to express our results in a compact form
\begin{gather*}
b_k\llbracket U^s \rrbracket \equiv \text{terms of order}\ U^s\ \text{in the HKC}\ b_k; \qquad
[b_k] = b_k(x,x); \qquad G_{\mu\rho;\mu} \equiv J_\rho; \\
T_{2(k)} \equiv T_{\mu_1\mu_1...\mu_k\mu_k}; \qquad
D_{\mu_1...\mu_n} \equiv D_{\mu_1}...D_{\mu_n}; \qquad
D^{2(k)} \equiv D_{\mu_1\mu_1...\mu_k\mu_k}; \\
D_{\mu_1...\mu_n} (U) \equiv  U_{;\mu_n...\mu_1}, \qquad
D_{\mu_1...\mu_n} (G_{\rho\sigma}) \equiv G_{\rho\sigma;\mu_n...\mu_1}; \qquad
D_{\mu_1...\mu_n} (J_\rho) \equiv J_{\rho;\mu_n...\mu_1}.
\end{gather*}

Here, for the sake of the readers,  we systematically chalk out the necessary instructions to  calculate the operators of a particular class, $\mathcal{O}(D^r U^s) \subset$  HKC $b_n (x,y)$.

\begin{itemize}
    \item First, start with  Eq.~\eqref{eq:HKC_class} and note down all the equations for $\mathcal{O}(D^r U^{s-i})$ in a recursive way till $\mathcal{O}(D^r U^0)$. Each of these equations contains operators of the form  $D^{2(k)} (U\, b_j)$, and $T_{2(k)}$ with $k \in [0, r/2]$.
    \item  Then,  expand the operator $T_{2(k)}$,  using Eqs.~\eqref{eq:t_rec}-\eqref{eq:r_rec},  recursively until one is left with only field tensor $G_{\mu\nu}$ and the covariant derivatives $(D)$.
    \item Similarly expand the operator $D^{2(k)} (U\, b_j)$ such that the derivative acts either on $U$ or $b_j$. 
    
    \item[$\ast$] Following these steps, one ends up with terms that contain either HKCs of lower order or derivatives of HKCs.
    
    \item It is advised to start with operators of $\mathcal{O}(D^r U^0)$. Then, employing the initial condition $[b_0]=I$, evaluate the necessary derivatives of $b_0$ using Eq.~\eqref{eq:derv_rec_rel} and substitute them in the relation obtained for $\mathcal{O}(D^r U^0)$.
    \item Next work out another related class of operator $\mathcal{O}(D^r U^{i+1})$ and calculate the required derivatives of HKCs.
    \item[$\ast\;\ast$]  Repeat these  previously mentioned two steps until one reaches the desired class of operators $\mathcal{O}(D^r U^{s})$. Then, use the trace properties to note down the HKCs on a minimal basis.
\end{itemize}

Here, we want to explicitly mention a few subtleties while following the above-mentioned state of the art of calculation:
\begin{itemize}
    \item The order in which the coincidence limit is used is important especially when derivative operators are involved as 
    \[D_{\mu_1...\mu_n} b_k |_{z=0} \neq D_{\mu_1...\mu_n}([b_k]).\]
    \item Usage of the trace properties in intermediate steps of the recursive procedure is strictly forbidden.
    \item As HKCs ($b_k$) and its $n$\textsuperscript{th} derivative ($D^{n} b_k$) can be expressed in terms of the lower order HKCs $b_{k-i}$ and lower order derivatives $D^{(n-j)}b_{k}$, and $D^{(n-j)}b_{k-i}$. Thus, one may avoid reaching out to $b_0$ every time while calculating the HKC or its derivatives, and instead, it is advised to use results for the intermediate HKCs, if available already. 
\end{itemize}

\subsection{Computation of operators $\mathcal{O}(D^4 U^3)$: a detailed example}

Based on the methodology discussed in the previous section, we perform, here, an explicit calculation of the operators class $\mathcal{O}(D^4 U^3)$ for the sake of detailed demonstration. To compute $\mathcal{O}(D^4 U^3)$,  we obtain the following necessary relations using Eq.~\eqref{eq:HKC_class}.
\begin{flalign}    
    &[b_2] \llbracket U^0 \rrbracket=\mathcal{O}(D^4 U^0) =  -\frac{1}{12}\, \left\{T_{(4)}b_0\right\}_{z=0}, \label{eq:D4U0}\\
    &[b_3] \llbracket U^1 \rrbracket=\mathcal{O}(D^4 U^1) = U\,[b_2] \llbracket U^0 \rrbracket + \left\{\frac{1}{2} D^2 \left(U\,b_1 \right) + \frac{1}{10} \left( D^4 \left(U\,b_0 \right)- T_{(4)}b_1 \right)\right\}_{z=0}\llbracket U^1 \rrbracket,\label{eq:D4U1}\\
    &[b_4] \llbracket U^2 \rrbracket= \mathcal{O}(D^4 U^2) = U\,[b_3] \llbracket U^1 \rrbracket + \left\{\frac{3}{5} D^2 \left(U\,b_2 \right) + \frac{1}{10} \left( 2\,D^4 \left(U\,b_1 \right)- T_{(4)}b_2 \right)\right\}_{z=0} \llbracket U^2 \rrbracket,\label{eq:D4U2}\\
    &[b_5] \llbracket U^3 \rrbracket= \mathcal{O}(D^4 U^3) = U\,[b_4] \llbracket U^2 \rrbracket + \left\{\frac{2}{3} D^2 \left(U\,b_3 \right) + \frac{2}{21} \left( 3\,D^4 \left(U\,b_2 \right)- T_{(4)}b_3\right)\right\}_{z=0} \llbracket U^3 \rrbracket\label{eq:D4U3}.
\end{flalign}
Here, we set
\[T = 0\, ,\qquad T_{\mu} = 0\, ,\qquad T_{(2)}=T_{\mu \mu} = 0,\]
that are quite evident from Eq.~\eqref{eq:t_rec}. It is important to note that in the above equations, we have two different kinds of structures: (i) $D^4$ and $D^2$ act on $U\,b_k$, and (ii) $T_{(4)}$ acts on $b_k$. Thus, our initial aim is to calculate the explicit form of these operators, first. 

The actions of  $D^4$ and $D^2$ operators  on $(U\,b_k)$ are defined from Eqs.~\eqref{eq:D4U1}-\eqref{eq:D4U3}, as follows
\begin{equation}\label{eq:D2_op}
    D^2(U\,b_k) = U_{;\mu\mu}\, b_k + 2\,U_{;\mu} D_\mu b_k + U\,D^2b_k,
\end{equation}
\begin{equation}\label{eq:D2_op}
\begin{split}
    D^4(U\,b_k) =& U_{;\mu\mu\nu\nu}\, b_k + 2\, U_{;\mu\nu\nu} D_\mu b_k + 2\, U_{;\nu\nu\mu} D_\mu b_k + 2\,U_{;\mu} D_{\mu\nu\nu} b_k \\ & + 2\,U_{;\mu} D_{\nu\nu\mu}  b_k + 4\,U_{;\nu\mu} D_{\mu\nu}  b_k +  2\, U_{;\mu\mu} D^2  b_k + U\,D^4 b_k,
\end{split}
\end{equation}

Action of $T_{(4)}$ can be addressed in the following form derived from Eqs.~\eqref{eq:t_rec}-\eqref{eq:r_rec} 
\begin{equation}\label{eq:T_4}
\begin{split}
    T_{(4)}=T_{\mu\mu\nu\nu}&= D_\mu T_{\mu\nu\nu} + R_{\mu\nu\nu,\mu} =D_{\mu\mu} T_{\nu\nu} + D_\mu R_{\nu\nu,\mu}+ R_{\mu\nu\nu,\mu}\\
    &=D_\mu R_{\nu\nu,\mu}+ D_\mu R_{\nu\nu,\mu} + G_{\mu\mu}D_{\nu\nu} =2\,D_{\mu\nu} G_{\nu\mu}+ 2\,D_\mu G_{\nu\mu}D_\nu \\
    &= G_{\mu\nu} G_{\nu\mu} - 2\,G_{\mu\nu;\mu} D_\nu -2\,G_{\mu\nu}D_{\mu\nu} = -2\,(G_{\mu\nu})^2 - 2\,J_\nu D_\nu.
\end{split}
\end{equation}
In this derivation, we use the anti-symmetric nature of $G_{\mu\nu}$ and the following identity
\begin{equation}\label{eq: index_ch}
    X_{;\nu\mu} = X_{;\mu\nu} + G_{\mu\nu}X - X\, G_{\mu\nu},
\end{equation}
where $X$ is any arbitrary tensor. This leads to our finding
\[2\,D_{\mu\nu} G_{\mu\nu} = (G_{\mu\nu})^2.\]
Now, we are ready to demonstrate the explicit computation of operators $\mathcal{O}(D^4 U^3)$ belonging to the HKC $b_5$.

\subsubsection*{\underline{ \Large$\mathcal{O}(D^4 U^0)$}}
\vskip 0.2cm 
$T_{(4)}$ operator contains a derivative  acting on HKCs. Hence, to calculate $[b_2]_{U^0}$, we, first, need to calculate $ D_\nu b_0 |_{z=0}$. From Eq. \eqref{eq:derv_rec_rel} we find
\begin{equation}
    \begin{split}
        D_\nu b_0|_{z=0} = -T_\nu b_0|_{z=0} = 0.
    \end{split}
\end{equation}
This provides $\mathcal{O}(D^4U^0)$, directly from Eq.~\eqref{eq:D4U0}, as
\begin{equation}
\begin{split}
    [b_2] \llbracket U^0 \rrbracket=\mathcal{O}(D^4 U^0) &= \frac{1}{6}\, \Big\{(G_{\mu\nu})^2 b_0 + J_\nu D_\nu b_0\Big\}_{z=0} = \frac{1}{6}\, (G_{\mu\nu})^2.
\end{split}
\end{equation}
\subsubsection*{\underline{ \Large$\mathcal{O}(D^4 U^1)$}}
\vskip 0.2cm
Next, to compute $\mathcal{O} (D^4 U^1)$ we  calculate the necessary derivatives of HKCs using Eqs.~\eqref{eq:derv_rec_rel}-\eqref{eq:r_rec}.

\begin{equation}
    \begin{split}
        D^2 b_0\Big|_{z=0} = D_{\mu\mu} b_0\Big|_{z=0} &= -\frac{1}{2} T_{\mu\mu} b_0\Big|_{z=0} = 0.
    \end{split}
\end{equation}

\begin{equation}
        D_{\mu\nu} b_0\Big|_{z=0} = -\frac{1}{2} T_{\mu\nu} b_0\Big|_{z=0} = -\frac{1}{2}\, \{D_\mu T_{\nu} + R_{\nu,\mu}\} b_0 \Big|_{z=0} = \frac{1}{2}\, G_{\mu\nu}.
\end{equation}

\begin{multline}
        D_{\mu\nu\nu} b_0\Big|_{z=0} = -\frac{1}{3} T_{\mu\nu\nu} b_0\Big|_{z=0} = -\frac{1}{3}\, \{D_\mu T_{\nu\nu} + R_{\nu\nu,\mu}\} b_0 \Big|_{z=0} = -\frac{1}{3}\, \{D_\nu G_{\nu\mu} + G_{\nu\mu} D_\nu\} b_0 \Big|_{z=0},\\ 
        = -\frac{1}{3}\, \{G_{\nu\mu;\nu} + 2\,G_{\nu\mu} D_\nu\} b_0 \Big|_{z=0} -\frac{1}{3}\,J_\mu. \qquad \qquad \quad
\end{multline}

\begin{multline}
        D_{\nu\nu\mu} b_0\Big|_{z=0} = -\frac{1}{3} T_{\nu\nu\mu} b_0\Big|_{z=0} = -\frac{1}{3}\, \{D_\nu T_{\nu\mu} + R_{\nu\mu,\nu}\} b_0 \Big|_{z=0} = -\frac{2}{3}\, D_\nu G_{\mu\nu} b_0 \Big|_{z=0},\\
         = \frac{2}{3}\, \{G_{\nu\mu;\nu} + \,G_{\nu\mu} D_\nu\} b_0 \Big|_{z=0} = \frac{2}{3}\,J_\mu. \qquad \qquad
\end{multline}

\begin{equation}
        D^4 b_0\Big|_{z=0} = D_{\mu\mu\nu\nu} b_0\Big|_{z=0} = -\frac{1}{4} T_{(4)} b_0\Big|_{z=0} = \frac{1}{2}\, (G_{\mu\nu})^2.
\end{equation}

\begin{equation}
        [b_1] = \{U+D^2\}b_0\Big|_{z=0} = U.
\end{equation}

\begin{equation}
    \begin{split}
        D_\mu b_1\Big|_{z=0} = \frac{1}{2}\{D_\mu(U+D^2)b_0-T_\mu b_1\}\Big|_{z=0} & = \frac{1}{2} \Big\{U_{;\mu}b_0 + U\,D_\mu b_0 + D_{\mu\nu\nu} b_0\Big\}_{z=0}, \\
        & = \frac{1}{2} U_{;\mu} - \frac{1}{6} J_\mu.
    \end{split}
\end{equation}

\begin{equation}
    \begin{split}
        D_{\mu\mu} b_1\Big|_{z=0} \llbracket U^0 \rrbracket &= \frac{1}{3} \{D_{\mu\mu}(U+D^2)b_0 - T_{\mu\mu} b_1\}\Big|_{z=0} \llbracket U^0 \rrbracket = \frac{1}{3} D_{\mu\mu\nu\nu}b_0\Big|_{z=0} \llbracket U^0 \rrbracket = \frac{1}{6} (G_{\mu\nu})^2.
    \end{split}
\end{equation}
Assembling all the contributions, computed here, in Eq.~\eqref{eq:D4U1} we find
\begin{equation}
    \begin{split}
        [b_3] \llbracket U^0 \rrbracket =\mathcal{O}(D^4 U^1) =& \frac{3}{10} U\,(G_{\mu\nu})^2 +\frac{1}{5}(G_{\mu\nu})^2 U - \frac{1}{10} U_{;\mu}J_{;\mu} + \frac{1}{10} J_{;\mu}U_{;\mu}\\& + \frac{1}{10} U_{;\mu\mu\nu\nu} + \frac{2}{10} U_{;\nu\mu}G_{\mu\nu}.
    \end{split}
\end{equation}

\subsubsection*{\underline{ \Large$\mathcal{O}(D^4 U^2)$}}
\vskip 0.2cm
Following the similar path, we calculate the derivatives of HKCs required for the computation of operators $\mathcal{O}(D^4 U^2)$.

\begin{equation}
    \begin{split}
        D_{\mu\mu} b_1|_{z=0} \llbracket U^1 \rrbracket &= \frac{1}{3} \{D^2(U+D^2)b_0 - T_{\mu\mu} b_1\}|_{z=0} \llbracket U^1 \rrbracket,\\
        &= \frac{1}{3} \{U_{;\mu\mu} b_0 + 2\,U_{;\mu} D_\mu b_0 + U\,D^2 b_0\}|_{z=0} \llbracket U^1 \rrbracket,\\
        &= \frac{1}{3} U_{;\mu\mu}.
    \end{split}
\end{equation}

\begin{equation}
    \begin{split}
        D_{\mu\nu} b_1|_{z=0} \llbracket U^1 \rrbracket &= \frac{1}{3} \{D_{\mu\nu}(U+D^2)b_0 - T_{\mu\nu} b_1\}|_{z=0} \llbracket U^1 \rrbracket,\\
        &= \frac{1}{3} \{U_{;\nu\mu} b_0 + U_{;\mu} D_\nu b_0 + U_{;\nu} D_\mu b_0 + U\,D_{\mu\nu} b_0 - G_{\nu\mu}b_1\}|_{z=0} \llbracket U^1 \rrbracket,\\
        &= \frac{1}{3}\{ U_{;\nu\mu} + \frac{1}{2}U\,G_{\mu\nu} +  G_{\mu\nu}U\},\\
    \end{split}
\end{equation}

\begin{equation}
    \begin{split}
        D_{\mu\nu\nu} b_1|_{z=0} \llbracket U^1 \rrbracket &= \frac{1}{4} \{D_{\mu\nu\nu}(U+D^2)b_0 - T_{\mu\nu\nu} b_1\}|_{z=0} \llbracket U^1 \rrbracket,\\
        &= \frac{1}{4} \{D_{\mu}(U_{;\nu\nu} b_0 + 2\,U_{;\nu}D_\nu b_0 + U\,D^2 b_0) - D_\nu G_{\nu\mu} b_1 - G_{\nu\mu} D_\nu b_1\}|_{z=0} \llbracket U^1 \rrbracket,\\
        &= \frac{1}{4} \{U_{;\nu\nu\mu} b_0 + U_{;\nu\nu} D_\mu b_0 + 2\,U_{;\nu\mu}D_\nu b_0 + 2\,U_{;\nu}D_{\mu\nu} b_0 + U_{;\mu}\,D^2 b_0 \\ & \quad + U\,D_{\mu\nu\nu} b_0 - G_{\nu\mu;\nu} b_1 - 2\, G_{\nu\mu}D_\nu b_1\}|_{z=0} \llbracket U^1 \rrbracket,\\
        &= \frac{1}{4} \{U_{;\nu\nu\mu} + U_{;\nu}G_{\mu\nu} -\frac{1}{3} U\,J_{\mu} -J_\mu U + G_{\mu\nu} U_{;\nu}\}.
    \end{split}
\end{equation}

\begin{equation}
    \begin{split}
        D_{\nu\nu\mu} b_1|_{z=0} \llbracket U^1 \rrbracket &= \frac{1}{4} \{D_{\nu\nu\mu}(U+D^2)b_0 - T_{\nu\nu\mu} b_1\}|_{z=0} \llbracket U^1 \rrbracket,\\
        &= \frac{1}{4} \{D_{\nu\nu}(U_{\mu}b_0 + U\,D_\mu b_0) - 2\,D_\nu G_{\mu\nu} b_1\}|_{z=0} \llbracket U^1 \rrbracket,\\
        &= \frac{1}{4} \{U_{;\mu\nu\nu}b_0 + U_{\mu} D^2 b_0 + 2\,U_{\mu\nu} D_\nu b_0 + U_{\nu\nu}\,D_\mu b_0 + U\,D_{\nu\nu\mu} b_0 \\ &\quad +2\, U_{;\nu}\,D_{\nu\mu} b_0 + 2\, G_{\nu\mu;\nu} b_1 + 2\, G_{\nu\mu} D_\nu b_1\}|_{z=0} \llbracket U^1 \rrbracket,\\
        &= \frac{1}{4} \{U_{;\mu\nu\nu} + \frac{2}{3}U\,J_{\mu} + U_{;\nu}G_{\nu\mu} + 2\, J_\mu U + G_{\nu\mu} U_{;\nu}\}.
    \end{split}
\end{equation}

\begin{equation}
    \begin{split}
        D^4\,b_1|_{z=0} \llbracket U^1 \rrbracket &= \frac{1}{5} \{D_{\mu\mu\nu\nu}(U+D^2)b_0 - T_{\mu\mu\nu\nu} b_1\}|_{z=0} \llbracket U^1 \rrbracket,\\
        &= \frac{1}{5} \{U_{;\mu\mu\nu\nu} b_0 + 2\, U_{;\mu\nu\nu} D_\mu b_0 + 2\, U_{;\nu\nu\mu} D_\mu b_0 + 2\,U_{;\mu} D_{\mu\nu\nu} b_0 + 2\,U_{;\mu} D_{\nu\nu\mu}  b_0 \\ & \quad + 4\,U_{;\nu\mu} D_{\mu\nu}  b_0 +  2\, U_{;\mu\mu} D^2  b_0 + U\,D^4 b_0 + 2\, J_\mu D_\mu b_1 + 2\, (G_{\mu\nu})^2 b_1\}|_{z=0} \llbracket U^1 \rrbracket,\\
        &= \frac{1}{5} \{U_{;\mu\mu\nu\nu} + \frac{2}{3}\,U_{;\mu} J_{\mu} + 2\,U_{;\nu\mu} G_{\mu\nu}  + \frac{1}{2} U\,(G_{\mu\nu})^2 + J_\mu U_{;\mu} + 2\, (G_{\mu\nu})^2 U\}.
    \end{split}
\end{equation}

\begin{equation}
    \begin{split}
        [b_2] \llbracket U^1,U^2 \rrbracket &= \{U+D^2\}b_1|_{z=0} \llbracket U^1,U^2 \rrbracket = U^2 + \frac{1}{3} U_{;\mu\mu}.
    \end{split}
\end{equation}

\begin{equation}
    \begin{split}
        D_\mu b_2|_{z=0} \llbracket U^1,U^2 \rrbracket &= \frac{1}{3}\{2\,D_\mu(U+D^2)b_1-T_\mu b_2\}_{z=0} \llbracket U^1,U^2 \rrbracket,\\
        &= \frac{2}{3} \{U_{;\mu}b_1 + U\,D_\mu b_1 + D_{\mu\nu\nu} b_1\}_{z=0} \llbracket U^1,U^2 \rrbracket,\\
        &= \frac{2}{3} \{U_{;\mu} U + \frac{1}{2} U\,U_{;\mu} -\frac{1}{4} U\,J_\mu + \frac{1}{4} U_{;\nu\nu\mu} + \frac{1}{4}U_{;\nu}G_{\mu\nu} - \frac{1}{4}J_\mu U + \frac{1}{4}G_{\mu\nu} U_{;\nu}\}.
    \end{split}
\end{equation}

\begin{equation}
    \begin{split}
        D_{\mu\mu} b_2|_{z=0} \llbracket U^1 \rrbracket &= \frac{1}{4}\{2\,D_{\mu\mu}(U+D^2)b_1-T_{\mu\mu} b_2\}_{z=0} \llbracket U^1 \rrbracket,\\
        &= \frac{1}{2} \{U_{;\mu\mu} b_1 + 2\,U_{;\mu} D_\mu b_1 + U\,D^2 b_1 + D_{\mu\mu\nu\nu} b_1\}_{z=0} \llbracket U^1 \rrbracket,\\
        &= \frac{1}{2} \{ -\frac{1}{5}\,U_{;\mu} J_\mu + \frac{4}{15}U\,(G_{\mu\nu})^2 + \frac{1}{5}U_{;\mu\mu\nu\nu} +  \frac{2}{5}\,U_{;\nu\mu} G_{\mu\nu}+ \frac{2}{5}\, (G_{\mu\nu})^2 U\}.
    \end{split}
\end{equation}

Again, we collect all these contributions and with the help of Eq.~\eqref{eq:D4U2}, we note the following equation.

\begin{equation}
    \begin{split}
        [b_4] \llbracket U^2 \rrbracket =\mathcal{O}(D^4 U^2) =& \frac{1}{5} U_{;\mu\mu\nu} U_{;\nu} +\frac{3}{10} U_{;\mu} U_{;\nu\nu\mu} + \frac{1}{5} U_{;\mu\nu\nu} U_{;\mu} + \frac{4}{15} (U_{;\mu\nu})^2 +\frac{1}{3} (U_{;\mu\mu})^2  \\ 
        &  +\frac{1}{5} U_{;\mu\mu\nu\nu} U +\frac{1}{10} U_{;\mu} U_{;\mu\nu\nu} +\frac{1}{5} U\,U_{;\mu\mu\nu\nu}  +\frac{2}{15} J_\mu U_{;\mu U }  \\
        & -\frac{2}{15} U\,U_{;\mu}J_\mu +\frac{1}{5}U\,J_\mu U_{;\mu} - \frac{1}{6} U_{;\mu}U\,J_\mu -\frac{1}{10} U_{;\mu}J_\mu U \\ 
        & + \frac{1}{15} J_\mu U\, U_{;\mu} +\frac{1}{5} U^2 (G_{\mu\nu})^2 +\frac{4}{15} U(G_{\mu\nu})^2U+\frac{2}{15} (U\, G_{\mu\nu})^2 \\ 
        & +\frac{1}{15} G_{\mu\nu} U^2 G_{\mu\nu} +\frac{2}{15} (G_{\mu\nu} U)^2 + \frac{1}{5} U_{;\mu} U_{;\nu} G_{\mu\nu} + \frac{1}{5} U_{;\mu} G_{\mu\nu} U_{;\nu}\\ & +\frac{1}{5} (G_{\mu\nu})^2 U^2. 
    \end{split}
\end{equation}

\subsubsection*{\underline{ \Large$\mathcal{O}(D^4 U^3)$}}
\vskip 0.2cm
We repeat the same task, one more time. We focus on the computation of the relevant derivatives of HKCs to calculate the operators $\mathcal{O}(D^4 U^3)$.

\begin{equation}
    \begin{split}
        D_{\mu\mu} b_2|_{z=0} \llbracket U^2 \rrbracket &= \frac{1}{4} \{2\,D^2(U+D^2)b_1 - T_{\mu\mu} b_2\}|_{z=0} \llbracket U^2 \rrbracket,\\
        &= \frac{1}{2} \{U_{;\mu\mu} b_1 + 2\,U_{;\mu} D_\mu b_1 + U\,D^2 b_1\}|_{z=0} \llbracket U^2 \rrbracket,\\
        &= \frac{1}{2} \{U_{;\mu\mu} U + U_{;\mu} U_{;\mu} + \frac{1}{3}U\,U_{;\mu\mu}\}.
    \end{split}
\end{equation}

\begin{equation}
    \begin{split}
        D_{\mu\nu} b_2|_{z=0} \llbracket U^2 \rrbracket &= \frac{1}{4} \{2\,D_{\mu\nu}(U+D^2)b_1 - T_{\mu\nu} b_2\}|_{z=0} \llbracket U^2 \rrbracket,\\
        &= \frac{1}{4} \{2\,U_{;\nu\mu} b_1 + 2\,U_{;\mu} D_\nu b_1 + 2\,U_{;\nu} D_\mu b_1 +2\, U\,D_{\mu\nu} b_1 - G_{\nu\mu}b_2\}|_{z=0} \llbracket U^2 \rrbracket,\\
        &= \frac{1}{4} \{2\,U_{;\nu\mu} U + U_{;\mu} U_{;\nu} + U_{;\nu} U_{;\mu} + \frac{2}{3} U\,U_{;\nu\mu} + \frac{1}{3} U^2\,G_{\mu\nu}+ \frac{2}{3} U\,G_{\mu\nu}U \\
        &\quad + G_{\mu\nu}U^2\}.
    \end{split}
\end{equation}

\begin{equation}
    \begin{split}
        D_{\mu\nu\nu} b_2|_{z=0} \llbracket U^2 \rrbracket &= \frac{1}{5} \{2\,D_{\mu\nu\nu}(U+D^2)b_1 - T_{\mu\nu\nu} b_2\}|_{z=0} \llbracket U^2 \rrbracket,\\
        &= \frac{2}{5} \{D_{\mu}(U_{;\nu\nu} b_1 + 2\,U_{;\nu}D_\nu b_1 + U\,D^2 b_1) - \frac{1}{2}D_\nu G_{\nu\mu} b_2\\ 
        & \quad - \frac{1}{2} G_{\nu\mu} D_\nu b_2\}|_{z=0} \llbracket U^2 \rrbracket,\\
        &= \frac{2}{5} \{U_{;\nu\nu\mu} b_1 + U_{;\nu\nu} D_\mu b_1 + 2\,U_{;\nu\mu}D_\nu b_1 + 2\,U_{;\nu}D_{\mu\nu} b_1 + U_{;\mu}\,D^2 b_1 \\ & \quad + U\,D_{\mu\nu\nu} b_1 - \frac{1}{2} G_{\nu\mu;\nu} b_2 - G_{\nu\mu}D_\nu b_2\}|_{z=0} \llbracket U^2 \rrbracket,\\
        &= \frac{2}{5} \{U_{;\nu\nu\mu} U + \frac{1}{2} U_{;\nu\nu} U_{;\mu} + U_{;\nu\mu}U_{;\nu} + \frac{2}{3}\,U_{;\nu}U_{;\nu\mu} + \frac{1}{3}\,U_{;\nu}U\,G_{\mu\nu} \\ 
        & \quad + \frac{2}{3}\,U_{;\nu}G_{\mu\nu}U  + \frac{1}{3}U_{;\mu}U_{;\nu\nu} + \frac{1}{4}U\,U_{;\nu\nu\mu} + \frac{1}{4}U\,U_{;\nu}G_{\mu\nu} - \frac{1}{12}U^2\,J_{\mu}\\ 
        & \quad - \frac{1}{4}U\,J_{\mu}U + \frac{1}{4}U\,G_{\mu\nu}U_{;\nu} -\frac{1}{2} J_\mu U^2 + \frac{2}{3} G_{\mu\nu}U_{;\nu}U + \frac{1}{3} G_{\mu\nu}U\,U_{;\nu} \}.
    \end{split}
\end{equation}

\begin{equation}
    \begin{split}
        D_{\nu\nu\mu} b_2|_{z=0} \llbracket U^2 \rrbracket &= \frac{1}{5} \{2\,D_{\nu\nu\mu}(U+D^2)b_1 - T_{\nu\nu\mu} b_2\}|_{z=0} \llbracket U^2 \rrbracket,\\
        &= \frac{2}{5} \{D_{\nu\nu}(U_{\mu}b_1 + U\,D_\mu b_1) - D_\nu G_{\mu\nu} b_2\}|_{z=0} \llbracket U^2 \rrbracket,\\
        &= \frac{2}{5} \{U_{;\mu\nu\nu}b_1 + U_{\mu} D^2 b_1 + 2\,U_{\mu\nu} D_\nu b_1 + U_{\nu\nu}\,D_\mu b_1  \\ 
        &\quad + U\,D_{\nu\nu\mu} b_1 +2\, U_{;\nu}\,D_{\nu\mu} b_1 + G_{\nu\mu;\nu} b_2 + G_{\nu\mu} D_\nu b_2\}|_{z=0} \llbracket U^2 \rrbracket,\\
        &= \frac{2}{5} \{U_{;\mu\nu\nu} U + \frac{1}{3}U_{\mu} U_{;\nu\nu} +  U_{\mu\nu} U_{;\nu} + \frac{1}{2}U_{\nu\nu} U_{;\mu}+ \frac{1}{4}U\,U_{;\mu\nu\nu} + J_\mu U^2 \\
        &\quad +\frac{1}{4} U\,U_{;\nu}G_{\nu\mu} + \frac{1}{6} U^2 J_\mu + \frac{1}{2}U\,J_\mu U +\frac{1}{4} U\,G_{\nu\mu}U_{;\nu} +\frac{2}{3}\, U_{;\nu}\,U_{;\mu\nu}\\ 
        &\quad  +\frac{1}{3}\, U_{;\nu}U\,G_{\nu\mu} +\frac{2}{3}\, U_{;\nu}G_{\nu\mu}U + \frac{2}{3} G_{\nu\mu} U_{;\nu}U + \frac{1}{3} G_{\nu\mu}U\, U_{;\nu}\}.
    \end{split}
\end{equation}

\begin{equation}
    \begin{split}
        D^4\,b_2|_{z=0} \llbracket U^2 \rrbracket &= \frac{1}{6} \{2\,D_{\mu\mu\nu\nu}(U+D^2)b_1 - T_{\mu\mu\nu\nu} b_2\}|_{z=0} \llbracket U^2 \rrbracket,\\
        &= \frac{1}{3} \{U_{;\mu\mu\nu\nu} b_1 + 2\, U_{;\mu\nu\nu} D_\mu b_1 + 2\, U_{;\nu\nu\mu} D_\mu b_1 + 2\,U_{;\mu} D_{\mu\nu\nu} b_1 + 2\,U_{;\mu} D_{\nu\nu\mu}  b_1 \\ 
        & \quad + 4\,U_{;\nu\mu} D_{\mu\nu}  b_1 +  2\, U_{;\mu\mu} D^2  b_1 + U\,D^4 b_1 + J_\mu D_\mu b_2 + (G_{\mu\nu})^2 b_2\}|_{z=0} \llbracket U^2 \rrbracket,\\
        &= \frac{1}{3} \{ U_{;\mu\mu\nu\nu} U + \frac{1}{5} U\,U_{;\mu\mu\nu\nu} + \frac{2}{3}\,U_{;\mu\mu}U_{;\nu\nu} + U_{;\nu\nu\mu} U_{;\mu} + U_{;\mu\nu\nu} U_{;\mu} \\
        & \quad + \frac{2}{5} U\,U_{;\mu}J_\mu -\frac{4}{15}U\,(G_{\mu\nu})^2 U + (G_{\mu\nu})^2 U^2 +\frac{1}{5}U\,J_\mu U_{;\mu} - \frac{2}{15}(U\,G_{\mu\nu})^2 \\
       & \quad  - \frac{1}{10}U^2 (G_{\mu\nu})^2 + \frac{1}{2} U_{;\mu} U_{;\nu\nu\mu} +\frac{1}{2} U_{;\mu}J_\mu U +\frac{1}{6} U_{;\mu}U\,J_\mu + \frac{1}{2}U_{;\mu}U_{;\mu\nu\nu}\\
       & \quad +\frac{4}{3}(U_{;\mu\nu})^2 +\frac{1}{3}G_{\mu\nu}U^2\,G_{\mu\nu} + \frac{2}{3}(G_{;\mu\nu}U)^2 +\frac{2}{3} J_\mu U_{;\mu} U +\frac{1}{3} J_\mu U U_{;\mu} \}.
    \end{split}
\end{equation}

\begin{equation}
    \begin{split}
        [b_3] \llbracket U^2,U^3 \rrbracket &= \{U+D^2\}b_2|_{z=0} \llbracket U^2,U^3 \rrbracket,\\
        &= U^3 + \frac{1}{2} U\,U_{;\mu\mu} + \frac{1}{2} U_{;\mu\mu} U + \frac{1}{2} (U_{;\mu})^2.\\
    \end{split}
\end{equation}

\begin{equation}
    \begin{split}
        D_\mu b_3|_{z=0} \llbracket U^2,U^3 \rrbracket &= \frac{1}{4}\{3\,D_\mu(U+D^2)b_2-T_\mu b_3\}_{z=0} \llbracket U^2,U^3 \rrbracket,\\
        &= \frac{3}{4} \{U_{;\mu}b_2 + U\,D_\mu b_2 + D_{\mu\nu\nu} b_2\}_{z=0} \llbracket U^2,U^3 \rrbracket,\\
        &= \frac{3}{4} \{U_{;\mu} U^2 + \frac{2}{3} U\,U_{;\mu}U + \frac{1}{3} U^2U_{;\mu} -\frac{1}{5} U^2J_\mu - \frac{4}{15} UJ_\mu U\\
        & \quad+ \frac{4}{15} U\,U_{;\nu\nu\mu} + \frac{4}{15}U\,U_{;\nu}G_{\mu\nu} + \frac{4}{15} U\,G_{\mu\nu} U_{;\nu} + \frac{7}{15} U_{;\mu} U_{;\nu\nu}\\
        & \quad + \frac{2}{5} U_{;\nu\nu\mu} U + \frac{1}{5} U_{;\nu\nu} U_{;\mu} + \frac{2}{5} U_{;\nu\mu}U_{;\nu} + \frac{4}{15}\,U_{;\nu}U_{;\nu\mu} -\frac{1}{5} J_\mu U^2\\ 
        & \quad  + \frac{2}{15}\,U_{;\nu}U\,G_{\mu\nu} + \frac{4}{15}\,U_{;\nu}G_{\mu\nu}U + \frac{4}{15} G_{\mu\nu}U_{;\nu}U + \frac{2}{15} G_{\mu\nu}U\,U_{;\nu} \}.
    \end{split}
\end{equation}

\begin{equation}
    \begin{split}
        D_{\mu\mu} b_3|_{z=0} \llbracket U^2 \rrbracket &= \frac{1}{5}\{3\,D_{\mu\mu}(U+D^2)b_2 -T_{\mu\mu} b_3\}_{z=0} \llbracket U^2 \rrbracket,\\
        &= \frac{3}{5} \{U_{;\mu\mu} b_2 + 2\,U_{;\mu} D_\mu b_2 + U\,D^2 b_2 + D_{\mu\mu\nu\nu} b_2\}_{z=0} \llbracket U^2 \rrbracket,\\
        &= \frac{1}{5} \{  \frac{1}{10}\,U\,U_{;\mu} J_\mu  + \frac{1}{2}U\,U_{;\mu\mu\nu\nu} +  \frac{1}{6}\,(U\,G_{\mu\nu})^2 +\frac{1}{3}G_{\mu\nu}U^2\,G_{\mu\nu}\\
        &\quad + \frac{1}{2}U\,J_\mu U_{;\mu} + \frac{3}{2}U_{;\mu}U_{;\nu\nu\mu} + U_{;\mu}G_{\mu\nu} U_{;\nu} + U_{;\mu} U_{;\nu} G_{\mu\nu} +\frac{4}{3}(U_{;\mu\nu})^2 \\
        & \quad + U_{;\mu\mu\nu\nu} U  + \frac{2}{3}\,U_{;\mu\mu}U_{;\nu\nu} + U_{;\nu\nu\mu} U_{;\mu} + U_{;\mu\nu\nu} U_{;\mu} + \frac{1}{3}\, U(G_{\mu\nu})^2 U \\
        & \quad + (G_{\mu\nu})^2 U^2  -\frac{1}{2} U_{;\mu}J_\mu U -\frac{5}{6} U_{;\mu}U\,J_\mu + \frac{1}{2}U_{;\mu}U_{;\mu\nu\nu} +\frac{1}{3} J_\mu U U_\mu \\
       & \quad   + (U_{;\mu\mu})^2  + \frac{2}{3}(G_{;\mu\nu}U)^2 +\frac{2}{3} J_\mu U_\mu U \}.
    \end{split}
\end{equation}

Finally, we combine all the above-computed expressions and put them in Eq.~\eqref{eq:D4U3} to find the operators of the class $\mathcal{O}(D^4 U^3)$. Note that, at this stage, all the evaluated operator structures are not independent. We employ the trace properties and a few identities\footnote{Under trace, as cyclic permutations are equivalent, a commutator is zero. Total derivatives can be written as a commutator, i.e., $(D_\mu\;U)=[D_\mu,U]$ and hence are zero under a trace.  Along with the identity given in Eq.~\eqref{eq: index_ch} we further use the Bianchi identity, $ G_{\rho\sigma;\mu}+G_{\sigma\mu;\rho}+G_{\mu\rho;\sigma} = 0$, to simplify terms.} that simplify the HKCs and allow us to write them in terms of independent operators. Finally, we find the independent operators of the form $\mathcal{O}(D^4 U^3)$ as
 \begin{equation}
     \begin{split}
       [b_5] \llbracket U^3 \rrbracket = \mathcal{O}(D^4 U^3) = &\ U^3 (G_{\mu\nu})^2 + \frac{2}{3} U^2 G_{\mu\nu} U\,G_{\mu\nu} + \frac{1}{3} U^2 J_\mu U_{;\mu} + \frac{1}{3} U\,G_{\mu\nu}U_{;\mu}U_{;\nu}  \\
    & + \frac{1}{3} U\,U_{;\mu}U_{;\nu}\,G_{\mu\nu} - \frac{1}{3} U^2 U_{;\mu}J_\mu + U\,U_{;\mu\mu}U_{;\nu\nu} + \frac{2}{3} U_{;\mu\mu} (U_{;\nu})^2.
     \end{split}
 \end{equation}


\subsection{Relevant Coefficients at Coincidence point}
\label{subsec:heat-kernel-result}

The necessary and relevant HKCs ($[b_k]$) computed at the coincidence point can be written in a compact form as
\begin{align}
    &\tr [b_0]=\tr I,& \label{eq:HKC_0}\\ 
    &\tr [b_1]=\tr U, \label{eq:HKC_1}\\ 
    &\tr [b_2]=\tr \Big[U^2+\frac{1}{6}\,(G_{\mu\nu})^2\Big], \label{eq:HKC_2}\\
    &\tr [b_3]=\tr \Big[U^3-\frac{1}{2} (U_{;\mu})^2+\frac{1}{2}U\,G_{\mu\nu}G_{\mu\nu}  -\frac{1}{10}(J_\nu)^2+\frac{1}{15}\,G_{\mu\nu}\,G_{\nu\rho}\,G_{\rho\mu} \Big],\label{eq:HKC_3} \\ 
    &\tr [b_4]=\tr \Big[U^4+ U^2 U_{;\mu\mu} + \frac{4}{5}U^2 (G_{\mu\nu})^2 + \frac{1}{5} (U\,G_{\mu\nu})^2 +  \frac{1}{5} (U_{;\mu\mu})^2 -\frac{2}{5} U\, U_{;\nu}\,J_{\nu} \notag\\ 
    &\quad\quad\quad - \frac{2}{5} U(J_\mu)^2 +\frac{2}{15} U_{;\mu\mu} (G_{\rho\sigma})^2 +\frac{4}{15} U\,G_{\mu\nu}G_{\nu\rho} G_{\rho\mu} +\frac{8}{15} U_{;\nu\mu}\, G_{\rho\mu} G_{\rho\nu} \label{eq:HKC_4}\\ 
    &\quad\quad\quad +\frac{1}{35}(J_{\mu;\nu})^2 + \frac{16}{105}G_{\mu\nu}J_{\mu}J_{\nu}+ \frac{1}{420} (G_{\mu\nu}G_{\rho\sigma})^2 +\frac{17}{210}(G_{\mu\nu})^2(G_{\rho\sigma})^2 \notag\\
    &\quad\quad\quad + \frac{1}{105} G_{\mu\nu}G_{\nu\rho}G_{\rho\sigma}G_{\sigma\mu} +\frac{2}{35}(G_{\mu\nu}G_{\nu\rho})^2+\frac{16}{105} J_{\nu;\mu} G_{\nu\sigma}G_{\sigma\mu} \Big], \notag \\
    &\tr [b_5] \llbracket U^5,U^4,U^3,U^2 \rrbracket=\tr \Big[ U^5 + 2\,U^3 U_{;\mu\mu} + U^2(U_{;\mu})^2 + U^3 (G_{\mu\nu})^2 + \frac{2}{3} U^2 G_{\mu\nu} U\,G_{\mu\nu} \notag \\
    & \quad\quad\quad\quad\quad\quad\quad\quad\quad\quad- \frac{1}{3} U^2 U_{;\mu}J_\mu + \frac{1}{3} U^2 J_\mu U_{;\mu} + \frac{1}{3} U\,G_{\mu\nu}U_{;\mu}U_{;\nu} + \frac{1}{3} U\,U_{;\mu}U_{;\nu}\,G_{\mu\nu}  \notag \\
    &\quad\quad\quad\quad\quad\quad\quad\quad\quad\quad + U\,U_{;\mu\mu}U_{;\nu\nu} + \frac{2}{3} U_{;\mu\mu} (U_{;\nu})^2 + \mathcal{O}(D^6 U^2)\Big], \label{eq:HKC_5} \\
    &\tr [b_6] \llbracket U^6,U^5,U^4 \rrbracket =\tr \Big[U^6 + 3\,U^4U_{;\mu\mu}+2\,U^3(U_{;\mu})^2 + \frac{12}{7}U^2 U_{;\nu\mu}U_{;\mu\nu} + \frac{17}{14} (U_{;\mu}U_{;\nu})^2  \notag\\
    &\quad\quad\quad\quad\quad\quad\quad\quad\quad\quad + \frac{9}{7} U\,U_{;\nu\mu}U\,U_{;\mu\nu} +\frac{26}{7} U_{;\nu\mu} U_{;\mu}U_{;\nu}U + \frac{18}{7} U_{;\nu\mu} U_{;\mu}U\,U_{;\nu} \notag\\
    &\quad\quad\quad\quad\quad\quad\quad\quad\quad\quad +\frac{26}{7} U_{;\nu\mu} U\,U_{;\mu}U_{;\nu} + \frac{9}{7} (U_{;\mu})^2(U_{;\nu})^2  + \frac{18}{7} G_{\mu\nu}U\,U_{;\mu}U_{;\nu}U \notag\\
    &\quad\quad\quad\quad\quad\quad\quad\quad\quad\quad  + \frac{5}{7} U^4(G_{\mu\nu})^2+ \frac{8}{7} U^3G_{\mu\nu}U\,G_{\mu\nu} + \frac{18}{7} G_{\mu\nu}U_{;\mu}U^2U_{;\nu}  \notag\\
    &\quad\quad\quad\quad\quad\quad\quad\quad\quad\quad + \frac{9}{14} (U^2G_{\mu\nu})^2 +  \frac{26}{7} G_{\mu\nu}U_{;\mu}U\,U_{;\nu}U  +  \frac{8}{7} G_{\mu\nu}U\,U_{;\mu}U\,U_{;\nu} \notag\\
    &\quad\quad\quad\quad\quad\quad\quad\quad\quad\quad  + \frac{24}{7} G_{\mu\nu}U_{;\mu}U_{;\nu}U^2 -  \frac{2}{7} G_{\mu\nu}U^2U_{;\mu}U_{;\nu}\Big], \label{eq:HKC_7}\\
    &\tr [b_7] \llbracket U^7,U^6 \rrbracket=\tr \Big[U^7 - 5\, U^4 (U_{;\mu})^2-8\,U^3U_{;\mu}U\,U_{;\mu} -\frac{9}{2} (U^2 U_{;\mu})^2 \Big],\label{eq:HKC_7} \\
    &\tr [b_8] \llbracket U^8 \rrbracket=\tr \Big[U^8\Big].\label{eq:HKC_8}
\end{align}
\section{One-loop effective Lagrangian and Heat-Kernel Coefficients} \label{sec: 1loop_HKC}
The one-loop effective Lagrangian obtained from Eq. \eqref{eq:effective-action} in the Euclidean space is given by
\begin{equation}
    \mathcal{L}_{\eff}= c_s \tr \log (-P^2+U+M^2),
\end{equation}
where $P_\mu=iD^E_\mu$, with $D_\mu^E$ being the derivative operator in the Euclidean signature\footnote{From now onwards we will drop the superscript $E$ and will use $D$ uniformly.}, and $c_s=+1/2$ and $+1$ for $\phi$ being a real scalar and complex scalar background respectively. In the Minkowski signature, the d'Alembertian operator is a hyperbolic second-order partial differential operator for which the Heat-Kernel expansion (HKE) is not convergent. Hence, by performing Wick's rotation to Euclidean space, the second-order partial differential operator is transformed into an elliptical one for which a convergent HKE is well-defined. 
\noindent
The following identity,
\begin{equation}
    \ln \lambda = -\int_0^\infty \frac{dt}{t} e^{-t\lambda},
\end{equation}
helps to recast the one-loop effective action  in terms of the HK as
\begin{equation}
    \mathcal{L}_{\eff}=c_s \tr \int_0^\infty \frac{dt}{t} K(t,x,x,\Delta).
\end{equation}
Employing  the ansatz, noted in Eq.~\eqref{eq:anzat}, the $\mathcal{L}_{\eff}$ can be expressed in terms of the coincident limit HKCs ($[b_k]$) as
\begin{equation}
\begin{split}
    \mathcal{L}_{\eff}&=c_s \int_0^\infty \frac{dt}{t} (4\pi t)^{-d/2}\ e^{-t\,M^2} \sum_k \frac{(-t)^k}{k\,!} \tr [b_k]\\
    &=\frac{c_s}{(4\pi)^{d/2}} \sum_k \frac{(-1)^k}{k!}\int_0^\infty dt\ t^{k-1-d/2}\ e^{-t\,M^2} \tr [b_k].   
\end{split}
\end{equation}
A suitable change in variable $t\,M^2 \to \tau^2$, the above integral reads as
\begin{equation}
     \mathcal{L}_{\eff}=\frac{c_s}{(4\pi)^{d/2}}\sum_k M^{d-2k}\frac{(-1)^k}{k!}\ 2\int_0^\infty d\tau\ \tau^{2(k-d/2)-1}\ e^{-\tau^2} \tr [b_k]. 
\end{equation}
Note that the integral over  $\tau$ mimics the integral representation of gamma function $\Gamma[z]$
\begin{equation}
    \Gamma[z]=2\int_0^\infty d\tau\ \tau^{2z-1}\ e^{-\tau^2},
\end{equation}
and that eases out writing down a compact form of the  one-loop effective Lagrangian as
\begin{equation}\label{eq:heat_ker_eff}
     \mathcal{L}_{\eff}=\frac{c_s}{(4\pi)^{d/2}}\sum_{k=0}^\infty M^{d-2k}\frac{(-1)^k}{k!}\ \Gamma[k-d/2] \tr [b_k]. 
\end{equation}

\subsection{Effective Contributions to Renormalisable Lagrangian}
It is evident from Eq.~\eqref{eq:heat_ker_eff}, that for $k\leq d/2$ the  Gamma function has simple poles. Thus, for such cases, we need to renormalise the theory employing dimensional regularisation, and $\overline{MS}$ renormalisation scheme.

We are working with 4-dim Euclidean space. Assuming $d=4-\epsilon$, we find
\begin{equation}
    \Gamma[k-d/2]=\frac{(\epsilon/2-3+k)!}{(\epsilon/2-1)!}\,\Gamma[\epsilon/2].
\end{equation}
In case of 4-dim, $d\to4$, i.e., $\epsilon\rightarrow 0$, the Gamma function has simple poles as $\Gamma[\epsilon/2]=2/\epsilon-\gamma_E+\mathcal{O}(\epsilon)$. In that scenario, the divergent part of the one-loop effective Lagrangian can be written as
\begin{equation}
    \mathcal{L}^{(k)}_{div}=\frac{c_s}{(4\pi)^{2-\epsilon/2}} M^{d-2k}\frac{(-1)^k}{k!}\ \frac{(\epsilon/2-3+k)!}{(\epsilon/2-1)!}\,\Gamma[\epsilon/2]\ \tr [b_k],
\end{equation}
with $k=0,1,2$. 
These three cases are explicitly demonstrated below.
\subsubsection*{\underline{\Large $k=0$}}

\begin{equation}
\begin{split}
     \mathcal{L}^{(0)}_{div}&=\frac{c_s}{(4\pi)^{2-\epsilon/2}} M^{d}\ \frac{(\epsilon/2-3)!}{(\epsilon/2-1)!}\,\Gamma[\epsilon/2]\ \tr[b_0]\\
    &=\frac{c_s}{(4\pi)^{2-\epsilon/2}} M^{4-\epsilon}\ \frac{1}{(\epsilon/2-1)(\epsilon/2-2)}\,\Gamma[\epsilon/2]\ \tr[b_0]\\
    &=c_s\left(\frac{M^2}{4\pi}\right)^2\left(\frac{4\pi}{M^2}\right)^{\epsilon/2}\ \frac{1}{(\epsilon/2-1)(\epsilon/2-2)}\,\Gamma[\epsilon/2]\ \tr [b_0].\\
\end{split}
\end{equation}
Taylor expansion in limit $\epsilon \rightarrow 0$ leads to
\begin{equation}
    \mathcal{L}^{(0)}_{div}=\frac{c_s}{(4\pi)^2}\ M^4\ \frac{1}{2}\,\left(\frac{2}{\epsilon}-\gamma_E-\ln\left[\frac{M^2}{4\pi}\right]+3/2\right)\ \tr[b_0].
\end{equation}
Employing $\overline{MS}$ scheme, we can write the finite part as
\begin{equation}
    \mathcal{L}^{(0)}_{\eff}=\frac{c_s}{(4\pi)^2}\ M^4\ \left[-\frac{1}{2}\,\left(\ln\left[\frac{M^2}{\mu^2}\right]-3/2\right)\, \tr [b_0]\right].
\end{equation}

\subsubsection*{\underline{\Large $k=1$}}

\begin{equation}
\begin{split}
    \mathcal{L}^{(1)}_{div}&=-\frac{c_s}{(4\pi)^{2-\epsilon/2}} M^{2-\epsilon}\ \frac{(\epsilon/2-2)!}{(\epsilon/2-1)!}\,\Gamma[\epsilon/2]\ \tr[b_1]\\
    &=-c_s\,\left(\frac{m}{4\pi}\right)^{2} \left(\frac{4\pi}{M^2}\right)^{\epsilon/2}\ \frac{1}{(\epsilon/2-1)}\,\Gamma[\epsilon/2]\ \tr[b_1].\\
\end{split}
\end{equation}
Taylor expansion in limit $\epsilon \rightarrow 0$ leads to
\begin{equation}
    \mathcal{L}^{(1)}_{div}=-\frac{c_s}{(4\pi)^2}\ M^2\ (-1)\,\left(\frac{2}{\epsilon}-\gamma_E-\ln\left[\frac{M^2}{4\pi}\right]+1\right)\ \tr[b_1].
\end{equation}
Employing $\overline{MS}$ scheme, we can write the finite part as
\begin{equation}
    \mathcal{L}^{(1)}_{\eff}=\frac{c_s}{(4\pi)^2}\ M^2\ \left[-\left(\ln\left[\frac{M^2}{\mu^2}\right]-1\right)\, \tr[b_1]\right].
\end{equation}

\subsubsection*{\underline{\Large $k=2$}}

\begin{equation}
\begin{split}
    \mathcal{L}^{(2)}_{div} =\frac{c_s}{(4\pi)^{2-\epsilon/2}} M^{-\epsilon}\frac{1}{2}\ \frac{(\epsilon/2-1)!}{(\epsilon/2-1)!}\,\Gamma[\epsilon/2]\ \tr[b_2] =\frac{c_s}{(4\pi)^{2}} \left(\frac{4\pi}{M^2}\right)^{\epsilon/2}\frac{1}{2} \,\Gamma[\epsilon/2]\ \tr[b_2].
\end{split}
\end{equation}
Taylor expansion in limit $\epsilon \rightarrow 0$ leads to
\begin{equation}
    \mathcal{L}^{(2)}_{div}=\frac{c_s}{(4\pi)^2}\ M^0\,\frac{1}{2}\left(\frac{2}{\epsilon}-\gamma_E-\ln\left[\frac{M^2}{4\pi}\right]\right)\ \tr[b_2].
\end{equation}
Employing $\overline{MS}$ scheme, we can write the finite part as
\begin{equation}
    \mathcal{L}^{(1)}_{\eff}=\frac{c_s}{(4\pi)^2}\ M^0\,\frac{1}{2}\left[-\left(\ln\left[\frac{M^2}{\mu^2}\right]\right)\, \tr[b_2]\right].
\end{equation}

\subsection*{Renormalised one-loop effective Lagrangian}
After renormalising the effective Lagrangian for three cases, $k=0,1,2$ we collect all the finite parts, and the  one-loop effective contributions to renormalisable part of the Lagrangian that contains operators up to mass dimension four can be written as
\begin{multline}\label{eq:finite}
    \mathcal{L}_{\eff}^{\text{ren}}=\frac{c_s}{(4\pi)^{2}} \tr \Bigg\{ M^4\ \left[-\frac{1}{2}\,\left(\ln\left[\frac{M^2}{\mu^2}\right]-3/2\right)\, [b_0]\right] + M^2\ \left[-\left(\ln\left[\frac{M^2}{\mu^2}\right]-1\right)\, [b_1]\right]\\
     + M^0\ \left[-\frac{1}{2}\left(\ln\left[\frac{M^2}{\mu^2}\right]\right)\, [b_2]\right] \Bigg\}.
\end{multline}
\section{Pure Heavy Scalar loop UOLEA up to D8}
\label{sec:complete-result-from-HKM}

In order to facilitate the readability of our result, we organise the effective operators according to the number of appearances of its constituents, i.e., covariant derivatives $(P)$, and the light field-dependent functional $(U)$.  Here, if an operator is composed of $i$ number of covariant derivatives and $j$ number of $U$, the corresponding quantities are represented by the integrer-superscripts $(i,j)$. The Lorentz invariance allows $i$ to be only even integer.  The effective Lagrangian up to mass dimension eight can be written as 
\begin{eqnarray}\label{eq:eff-Lag}
 \L_\eff = \frac{c_s}{(4\pi)^2} \sum_{i,j} \L^{(i,j)}_\eff = \frac{c_s}{(4\pi)^2} \sum_{i,j,k} \mathcal C_k^{(i,j)} O_k(P^i\,U^j).\nonumber
\end{eqnarray}
Here, $\mathcal  C_k^{(i,j)}$ is the coefficient associated with different operator $O_k(P^i\,U^j)$. The $k$ sums over the number of operators in each class with $i,j=[0,8]$. To keep in agreement with the  literature, we use the following notation $P_\mu = i D_\mu$. With the help of the HKCs computed at the coincident point, see Subsec. \ref{subsec:heat-kernel-result}, and using  Eq.~\eqref{eq:finite}, here, we catalogue the operators associated with the one-loop effective Lagrangian. 

\subsection{\Large $O(P^8\;U^j)$}

\begin{multline*}\label{eq:Lag-P8}
    \L^{(8,0)}_\eff = \frac{1}{M^4} \frac{1}{24}\tr \bigg[ \frac{17}{210}[P_\mu,P_\nu] [P_\mu,P_\nu] [P_\rho,P_\sigma][P_\rho,P_\sigma] \nn\\
     + \frac{2}{35}[P_\mu,P_\rho][P_\rho,P_\nu][P_\mu,P_\sigma][P_\sigma,P_\nu]
     + \frac{1}{105} [P_\mu,P_\nu][P_\nu,P_\rho] [P_\rho,P_\sigma][P_\sigma,P_\mu] \nn\\
     + \frac{1}{420} [P_\mu,P_\nu][P_\rho,P_\sigma][P_\mu,P_\nu][P_\rho,P_\sigma] 
     + \frac{1}{35}[P_\mu, [P_\rho,[P_\rho,P_\nu]]] [P_\mu, [P_\sigma,[P_\sigma,P_\nu]]] \nn\\
     + \frac{16}{105} [P_\mu, [P_\rho,[P_\rho,P_\nu]] [P_\nu,P_\sigma][P_\sigma,P_\mu]
     + \frac{16}{105}[P_\mu,P_\nu][P_\sigma,[P_\sigma,P_\mu]][P_\rho,[P_\rho,P_\nu]] \bigg],
\end{multline*}
\begin{gather} 
    \mathcal C^{(8,0)}_1 = \frac{1}{M^4}\frac{17}{5040}, \quad 
    \mathcal C^{(8,0)}_2 = \frac{1}{M^4}\frac{1}{420}, \quad
    \mathcal C^{(8,0)}_3 = \frac{1}{M^4}\frac{1}{2520}, \quad
    \mathcal C^{(8,0)}_4 = \frac{1}{M^4}\frac{1}{10080}, \nn \\
    \mathcal C^{(8,0)}_5 = \frac{1}{M^4}\frac{1}{840}, \quad
    \mathcal C^{(8,0)}_6 = \frac{1}{M^4}\frac{2}{315}, \quad
    \mathcal C^{(8,0)}_7 = \frac{1}{M^4}\frac{2}{315}. 
\end{gather}
\subsection{\Large  $O(P^6\;U^j)$}


\begin{gather}
    \L^{(6,0)}_\eff = \frac{1}{M^2} \frac{1}{6}\tr \bigg[ \frac{1}{15}[P_\mu,P_\nu][P_\nu,P_\rho] [P_\rho,P_\mu] -\frac{1}{10} [P_\mu, [P_\mu, P_\nu]] [P_\rho,[P_\rho,P_\nu]\bigg], \nn \\
    \mathcal C^{(6,0)}_1 = \frac{1}{M^2}\frac{1}{90}, \quad
    \mathcal C^{(6,0)}_2 = - \frac{1}{M^2}\frac{1}{60}.
\end{gather}


\begin{multline*}
    \L^{(6,1)}_\eff = \frac{1}{M^4} \frac{1}{24}\tr \bigg[-\frac{4}{15} U\,[P_\mu,P_\nu][P_\nu,P_\rho] [P_\rho,P_\mu] + \frac{2}{5} U[P_\rho,[P_\rho,P_\mu]][P_\sigma,[P_\sigma,P_\mu]] \\
     -\frac{2}{15} [P_\mu, [P_\mu, U]] [P_\rho,P_\sigma][P_\rho,P_\sigma]  +\frac{8}{15} [P_\mu, [P_\nu, U]]\, [P_\rho,P_\mu] [P_\nu,P_\rho] \bigg],
\end{multline*}
\begin{equation} \label{eq:Lag-P6U}
    \mathcal C^{(6,1)}_1 = -\frac{1}{M^4}\frac{1}{90}, 
    \mathcal C^{(6,1)}_2 = +\frac{1}{M^4}\frac{1}{60}, 
    \mathcal C^{(6,1)}_3 = -\frac{1}{M^4}\frac{1}{180}, 
    \mathcal C^{(6,1)}_4 = +\frac{1}{M^4}\frac{1}{45}.
\end{equation}

\begin{equation}\label{eq:Lag-P6U2}
\begin{split}
    \L^{(6,2)}_\eff =\ \tr \Big[&  \mathcal C^{(6,2)}_1\, U^2\,[P_{\mu},P_{\nu}]\,[P_{\nu},P_{\alpha}][P_{\alpha},P_{\mu}] \,+\, \mathcal C^{(6,2)}_2\, U\,[P_{\mu},P_{\nu}]\,U\,[P_{\nu},P_{\alpha}][P_{\alpha},P_{\mu}]  \\
&+ \, \mathcal C^{(6,2)}_3\, U^2\,[P_{\mu},[P_{\mu},P_{\nu}]]\,[P_{\alpha},[P_{\alpha},P_{\nu}]] \,+\, \mathcal C^{(6,2)}_4\, U\,[P_{\mu},[P_{\mu},P_{\nu}]]\,U\,[P_{\alpha},[P_{\alpha},P_{\nu}]]  \\
&+ \, \mathcal C^{(6,2)}_5\,  U\,[P_{\mu},[P_{\mu},U]]\,[P_{\nu},P_{\alpha}]\,[P_{\nu},P_{\alpha}] + \, \mathcal C^{(6,2)}_{6} [P_{\mu},[P_{\mu},U]]\,U\,[P_{\nu},P_{\alpha}]\,[P_{\nu},P_{\alpha}]   \\
&+ \, \mathcal C^{(6,2)}_7\,  U\,[P_{\mu},U]\,[P_{\nu}[P_{\nu},P_{\alpha}]\,[P_{\mu},P_{\alpha}] + \, \mathcal C^{(6,2)}_{8} U\,[P_{\mu},P_{\alpha}]\,[P_{\nu}[P_{\nu},P_{\alpha}]\,[P_{\mu},U]   \\
&+\, \mathcal C^{(6,2)}_9\,  U\,[P_{\mu},U]\,[P_{\mu},P_{\nu}]\,[P_{\alpha}[P_{\alpha},P_{\nu}] +\, \mathcal C^{(6,2)}_{10} U\,[P_{\alpha}[P_{\alpha},P_{\nu}]\,[P_{\mu},P_{\nu}]\,[P_{\mu},U]   \\
&+ \mathcal C^{(6,2)}_{11}\, U\,[P_{\mu},P_{\nu}]\,[P_{\alpha},[P_{\alpha},U]\,[P_{\mu},P_{\nu}] \,+\, \mathcal C^{(6,2)}_{12}\, \,[P_{\mu},U]\,[P_{\mu},U]\,[P_{\nu},P_{\alpha}]\,[P_{\nu},P_{\alpha}]  \\
&+\, \mathcal C^{(6,2)}_{13}\,  \,[P_{\mu},U]\,[P_{\nu},[P_{\nu},P_\alpha]]\,U\,[P_{\mu},P_{\alpha}] +\, \mathcal C^{(6,2)}_{14} \,[P_{\mu},P_{\alpha}]\,U\,[P_{\nu},[P_{\nu},P_\alpha]]\,[P_{\mu},U]   \\
&+\, \mathcal C^{(6,2)}_{15}\, \,[P_{\mu},U]\,[P_{\nu},U]\,[P_{\mu},P_{\alpha}]\,[P_{\alpha},P_{\nu}] +\, \mathcal C^{(6,2)}_{16}\, \,[P_{\mu},U]\,[P_{\nu},U]\,[P_{\nu},P_{\alpha}]\,[P_{\alpha},P_{\mu}]  \\
&+\, \mathcal C^{(6,2)}_{17}\,  \,[P_{\mu},U]\,[P_{\mu},P_{\nu}]\,[P_{\alpha},U]\,[P_{\alpha},P_{\nu}] + \, \mathcal C^{(6,2)}_{18} \,[P_{\mu},U]\,[P_{\alpha},P_{\nu}]\,[P_{\alpha},U]\,[P_{\mu},P_{\nu}]   \\
&+\, \mathcal C^{(6,2)}_{19}\, \,[P_{\mu},U]\,[P_{\nu},P_{\alpha}]\,[P_{\mu},U]\,[P_{\nu},P_{\alpha}] 
\,+\, \mathcal C^{(6,2)}_{20}\,  \,[P_{\mu},[P_{\mu},U]]\,[P_{\nu},U]\,[P_{\alpha},[P_{\alpha},P_{\nu}]]  \\
&+\, \mathcal C^{(6,2)}_{21} \,[P_{\mu},[P_{\mu},U]]\,[P_{\alpha},[P_{\alpha},P_{\nu}]]\,[P_{\nu},U]  \,+\, \mathcal C^{(6,2)}_{22}\, \,[P_{\mu},U]\,[P_{\nu},U]\,[P_{\mu},[P_{\alpha},[P_{\alpha},P_{\nu}]]]  \\
&+ \, \mathcal C^{(6,2)}_{23} \,[P_{\mu},U]\,[P_{\mu},[P_{\alpha},[P_{\alpha},P_{\nu}]]]\,[P_{\nu},U]
+\, \mathcal C^{(6,2)}_{24}\,[P_{\mu},[P_{\nu},[P_{\nu},U]]]\,[P_{\mu},[P_{\alpha},[P_{\alpha},U]]]\\
&+ \, \mathcal C^{(6,2)}_{25}[P_\alpha,P_\mu][P_\mu,P_\beta]U\,[P_\alpha,[P_\beta,U]] + \, \mathcal C^{(6,2)}_{26}[P_\alpha,P_\mu][P_\mu,P_\beta][P_\alpha,[P_\beta,U]] U\,  \Big],
\end{split}
\end{equation}

\subsection{\Large  $O(P^4\;U^j)$}

\begin{eqnarray}\label{eq:Lag-P4U0}
    \L^{(4,0)}_\eff = - M^0 \frac{1}{12} \ln\left[\frac{M^2}{\mu^2}\right] \tr \bigg[ \, [P_\mu,P_\nu][P_\mu,P_\nu]\bigg], \;\Rightarrow\; \mathcal C^{(4,0)}_1 =- M^0 \frac{1}{12} \ln\left[\frac{M^2}{\mu^2}\right].
\end{eqnarray}


\begin{eqnarray}\label{eq:Lag-P4U1}
    \L^{(4,1)}_\eff = - \frac{1}{M^2} \frac{1}{12} \tr \bigg[U [P_\mu,P_\nu][P_\mu,P_\nu]\bigg], \;\Rightarrow\;  \mathcal C^{(4,1)}_1 =- \frac{1}{M^2} \frac{1}{12}.
\end{eqnarray}


\begin{multline*}
    \L^{(4,2)}_\eff = \frac{1}{M^4} \frac{1}{24}\tr  \bigg[ \frac{4}{5}U^2 [P_\mu,P_\nu][P_\mu,P_\nu] + \frac{1}{5} U\,[P_\mu,P_\nu]\, U\,[P_\mu,P_\nu]\\
    +  \frac{1}{5} [P_\mu,[P_\mu,U]] [P_\nu,[P_\nu,U]] -\frac{2}{5} U\, [P_\nu,U]\,[P_\rho,[P_\rho,P_\nu]]  \bigg],
\end{multline*}
\begin{equation} \label{eq:Lag-P4U2}
    \mathcal C^{(4,2)}_1 = \frac{1}{M^4}\frac{1}{30}, 
    \mathcal C^{(4,2)}_2 = \frac{1}{M^4}\frac{1}{120}, 
    \mathcal C^{(4,2)}_3 = \frac{1}{M^4}\frac{1}{120}, 
    \mathcal C^{(4,2)}_4 = -\frac{1}{M^4}\frac{1}{60}.
\end{equation}

\begin{multline*}
    \L^{(4,3)}_\eff = \frac{1}{M^6} \frac{1}{60}\tr  \bigg[ - U^3 [P_\mu,P_\nu][P_\mu,P_\nu] - \frac{2}{3} U^2 [P_\mu,P_\nu]\,U\,[P_\mu,P_\nu]\\
    + \frac{1}{3} U^2 [P_\mu,U][P_\rho,[P_\rho,P_\mu]] - \frac{1}{3} U^2 [P_\rho,[P_\rho,P_\nu]] [P_\nu,U] \\ 
    - \frac{1}{3} U\,[P_\mu,P_\nu][P_\mu,U][P_\nu,U] - \frac{1}{3} U\,[P_\mu,U][P_\nu,U][P_\mu,P_\nu]\\ 
    - U\,[P_\mu,[P_\mu,U]] [P_\nu,[P_\nu,U]] - \frac{2}{3} [P_\mu,[P_\mu,U]] [P_\nu,U][P_\nu,U] \bigg],
\end{multline*}
\begin{align}\label{eq:Lag-P4U3}
    \mathcal C^{(4,3)}_1 &= -\frac{1}{M^6}\frac{1}{60}, & 
    \mathcal C^{(4,3)}_2 &= -\frac{1}{M^6}\frac{1}{90}, &
    \mathcal C^{(4,3)}_3 &= \frac{1}{M^6}\frac{1}{180}, &
    \mathcal C^{(4,3)}_4 &= -\frac{1}{M^6}\frac{1}{180}, \nn \\
    \mathcal C^{(4,3)}_5 &= -\frac{1}{M^6}\frac{1}{180}, &
    \mathcal C^{(4,3)}_6 &= -\frac{1}{M^6}\frac{1}{180}, &
    \mathcal C^{(4,3)}_7 &= -\frac{1}{M^6}\frac{1}{60}, &
    \mathcal C^{(4,3)}_8 &= -\frac{1}{M^6}\frac{1}{90}.
\end{align}

\begin{equation*}
\begin{split}
    \L^{(4,4)}_\eff = \frac{1}{M^8} \frac{1}{120}\tr & \bigg[ \frac{12}{7}U^2[P_\mu,[P_\nu,U]][P_\nu,[P_\mu,U]] + \frac{9}{7} U[P_\mu,[P_\nu,U]]U[P_\nu,[P_\mu,U]] \\
    & +\frac{26}{7} [P_\mu,[P_\nu,U]] [P_\mu,U][P_\nu,U]U + \frac{18}{7} [P_\mu,[P_\nu,U]] [P_\mu,U]U[P_\nu,U] \\
    & +\frac{26}{7} [P_\mu,[P_\nu,U]] U[P_\mu,U][P_\nu,U]  + \frac{17}{14} [P_\mu,U][P_\nu,U] [P_\mu,U][P_\nu,U] \\
    & + \frac{9}{7} [P_\mu,U][P_\mu,U] [P_\nu,U][P_\nu,U] + \frac{5}{7} U^4[P_\mu,P_\nu][P_\mu,P_\nu] \\
    & + \frac{8}{7} U^3[P_\mu,P_\nu]U[P_\mu,P_\nu] + \frac{9}{14} U^2[P_\mu,P_\nu]U^2[P_\mu,P_\nu] \\
    & + \frac{18}{7} [P_\mu,P_\nu][P_\mu,U]U^2[P_\nu,U] + \frac{18}{7} [P_\mu,P_\nu]U[P_\mu,U][P_\nu,U]U \\
    & +  \frac{8}{7} [P_\mu,P_\nu]U[P_\mu,U]U[P_\nu,U] +  \frac{26}{7} [P_\mu,P_\nu][P_\mu,U]U[P_\nu,U]U\\
    & + \frac{24}{7} [P_\mu,P_\nu][P_\mu,U][P_\nu,U]U^2 -  \frac{2}{7} [P_\mu,P_\nu]U^2[P_\mu,U][P_\nu,U]\bigg],
\end{split}
\end{equation*}
\begin{align}\label{eq:Lag-P4U4}
    \mathcal C^{(4,4)}_1 &= \frac{1}{M^8}\frac{1}{70}, & 
    \mathcal C^{(4,4)}_2 &= \frac{1}{M^8}\frac{3}{280}, &
    \mathcal C^{(4,4)}_3 &= \frac{1}{M^8}\frac{13}{420}, &
    \mathcal C^{(4,4)}_4 &= \frac{1}{M^8}\frac{3}{140}, \nn \\
    \mathcal C^{(4,4)}_5 &= \frac{1}{M^8}\frac{13}{420}, &
    \mathcal C^{(4,4)}_6 &= \frac{1}{M^8}\frac{17}{1680}, &
    \mathcal C^{(4,4)}_7 &= \frac{1}{M^8}\frac{3}{280}, &
    \mathcal C^{(4,4)}_8 &= \frac{1}{M^8}\frac{1}{168}, \nn \\
    \mathcal C^{(4,4)}_9 &= \frac{1}{M^8}\frac{1}{105}, & 
    \mathcal C^{(4,4)}_{10} &= \frac{1}{M^8}\frac{3}{560}, &
    \mathcal C^{(4,4)}_{11} &= \frac{1}{M^8}\frac{3}{140}, &
    \mathcal C^{(4,4)}_{12} &= \frac{1}{M^8}\frac{3}{140}, \nn \\
    \mathcal C^{(4,4)}_{13} &= \frac{1}{M^8}\frac{1}{105}, &
    \mathcal C^{(4,4)}_{14} &= \frac{1}{M^8}\frac{13}{420}, &
    \mathcal C^{(4,4)}_{15} &= \frac{1}{M^8}\frac{1}{35}, &
    \mathcal C^{(4,4)}_{16} &= -\frac{1}{M^8}\frac{1}{420}.
\end{align}
\vspace{.3cm}
\subsection{\Large $O(P^2\;U^j)$}

\begin{eqnarray}\label{eq:Lag-P2U3}
    \L^{(2,2)}_\eff = \frac{1}{M^2} \frac{1}{12}\tr \bigg[ -[P_\mu,U] [P_\mu,U]]  \bigg], \;\Rightarrow\;
    \mathcal C^{(2,2)}_1 = -\frac{1}{M^{2}}\frac{1}{12}.
\end{eqnarray}


\begin{eqnarray}\label{eq:Lag-P2U3}
    \L^{(2,3)}_\eff = \frac{1}{M^4} \frac{1}{24}\tr \bigg[ -U^2 [P_\mu,[P_\mu,U]]  \bigg], \;\Rightarrow\; \mathcal C^{(2,3)}_1 = -\frac{1}{M^{4}}\frac{1}{24}.
\end{eqnarray}


\begin{equation*}
    \L^{(2,4)}_\eff = \frac{1}{M^6} \frac{1}{60}\tr \bigg[  2\,U^3 [P_\mu,[P_\mu,U]] + U^2[P_\mu,U][P_\mu,U] \bigg],
\end{equation*}
\begin{equation}
    \mathcal C^{(2,4)}_1 = -\frac{1}{M^{8}}\frac{1}{40}, \quad
    \mathcal C^{(2,4)}_2 = -\frac{1}{M^{8}}\frac{4}{60}.
\end{equation}

\begin{equation*}
    \L^{(2,5)}_\eff = \frac{1}{M^8} \frac{1}{120}\tr \bigg[-3\,U^4[P_\mu,[P_\mu,U]]-2\,U^3[P_\mu,U][P_\mu,U]\bigg],
\end{equation*}
\begin{equation}
    \mathcal C^{(2,5)}_1 = -\frac{1}{M^{8}}\frac{1}{40}, \quad
    \mathcal C^{(2,5)}_2 = -\frac{1}{M^{8}}\frac{4}{60}.
\end{equation}

\begin{multline*}
    \L^{(2,6)}_\eff = \frac{1}{M^{10}} \frac{1}{210}\tr  \bigg[-5\, U^4 [P_\mu,U][P_\mu,U] - 8\,U^3 [P_\mu,U]\, U\,[P_\mu,U]\\
    - \frac{9}{2} U^2 [P_\mu,U]U^2 [P_\mu,U] \bigg],
\end{multline*}
\begin{equation}\label{eq:Lag-P2U6}
    \mathcal C^{(2,6)}_1 = -\frac{1}{M^{10}}\frac{1}{42}, \quad
    \mathcal C^{(2,6)}_2 = -\frac{1}{M^{10}}\frac{4}{105}, \quad
    \mathcal C^{(2,6)}_3 = -\frac{1}{M^{10}}\frac{3}{140}.
\end{equation}
\vspace{.3cm}
\subsection{\Large  $O(P^0\;U^j)$}

\begin{eqnarray}
    \L^{(0,0)}_\eff = M^4\ \frac{1}{2}\,\tr \left[\frac{3}{2}-\ln\left[\frac{M^2}{\mu^2}\right]\right], \;\Rightarrow\; \mathcal C^{(0,0)}_1 = M^4\ \frac{1}{2}\,\tr \left[\frac{3}{2}-\ln\left[\frac{M^2}{\mu^2}\right]\right].
\end{eqnarray}


\begin{eqnarray}
    \L^{(0,1)}_\eff = M^2\,\left(1-\ln\left[\frac{M^2}{\mu^2}\right]\right) \tr \left[\, U\right], \;\Rightarrow\;  \mathcal C^{(0,1)}_1 = M^2\ \left(1-\ln\left[\frac{M^2}{\mu^2}\right]\right).
\end{eqnarray}


\begin{eqnarray}
    \L^{(0,2)}_\eff = - M^0\,\ln\left[\frac{M^2}{\mu^2}\right] \, \tr \left[\, U^2\right], \;\Rightarrow\; \mathcal C^{(0,2)}_1 = - M^0\ \ln\left[\frac{M^2}{\mu^2}\right].
\end{eqnarray}


\begin{eqnarray}
    \L^{(0,3)}_\eff =  -\frac{1}{M^2}\frac{1}{6} \,\text{tr}\,\left[U^3\right], \;\Rightarrow\;  \mathcal C^{(0,3)}_1 = -\frac{1}{M^2}\frac{1}{6}.
\end{eqnarray}


\begin{eqnarray}
    \L^{(0,4)}_\eff = \frac{1}{M^4}\frac{1}{24} \,\text{tr}\,\left[ U^4\right], \;\Rightarrow\; \mathcal C^{(0,4)}_1 = \frac{1}{M^4}\frac{1}{24}.
\end{eqnarray}


\begin{eqnarray}
    \L^{(0,5)}_\eff = -\frac{1}{M^6} \frac{1}{60} \,\text{tr}\,\left[ U^5\right], \;\Rightarrow\; \mathcal C^{(0,5)}_1 = -\frac{1}{M^6} \frac{1}{60}.
\end{eqnarray}


\begin{eqnarray}
    \L^{(0,6)}_\eff = \frac{1}{M^8} \frac{1}{120} \,\text{tr}\,\left[ U^6\right], \;\Rightarrow\; \mathcal C^{(0,6)}_1 = \frac{1}{M^8} \frac{1}{120}.
\end{eqnarray}


\begin{eqnarray}
    \L^{(0,7)}_\eff = - \frac{1}{M^{10}}\frac{1}{210} \,\text{tr}\,\left[ U^7\right], \;\Rightarrow\; \mathcal C^{(0,7)}_1 = - \frac{1}{M^{10}}\frac{1}{210}.
\end{eqnarray}


\begin{eqnarray}
    \L^{(0,8)}_\eff = \frac{1}{M^{12}}\frac{1}{336} \,\text{tr}\,\left[ U^8\right] \;\Rightarrow\; \mathcal C^{(0,8)}_1 = \frac{1}{M^{12}}\frac{1}{336}.
\end{eqnarray}

\section{Validation using Covariant Diagram}
\label{sec:verification}


We present the one-loop effective action up to dimension eight after integrating out heavy degenerate scalars in Sec.~\ref{sec:complete-result-from-HKM}. We agree with the results so far available in the existing literature \cite{Henning:2014wua,Drozd:2015rsp,Ellis:2016enq}.  To cross-check the new results computed in the earlier section, we employ a new method of computation based on covariant diagram techniques, see Refs.~\cite{Vandeven:1985, Zhang:2016pja} for details. A brief review of this method has been given in Appendix~\ref{app:cov-diag-review}. We focus on computing only relevant one-loop diagrams that can generate the results depicted in the previous section.


\subsection{Method of Covariant diagram}

Here, we illustrate some aspects of the covariant diagram method that are pertinent to our objectives. Eq.~\eqref{eq:one-loop-Lagrangian}, given in Appendix~\ref{app:cov-diag-review}, reduces to Eq.~\eqref{eq:Lag-one-loop} up to additive constant while integrating out heavy scalar field or multiple degenerate heavy scalar fields \cite{Zhang:2016pja}.
\begin{eqnarray}\label{eq:Lag-one-loop}
    \mathcal{L}_{\text{eff}}\,[\phi] = -ic_s\,\text{tr}\,\sum_{n=1}^{\infty}\,\frac{1}{n}\int\frac{d^d\,q}{(2\pi)^d}\,\big[(q^2-M^2)^{-1}\,(2q.P-P^2+U)\big]^n.
\end{eqnarray}
This method allows one to map each integral of order $n$ into a number of covariant diagrams consisting of $n$ number of heavy propagators $1/(q^2-M^2)$ and along with all permissible combinations of $2q.P$, $-P^2$, and $U$ as vertex insertions. This automatically respects the covariant nature of the functional matching. The most generic form of the loop integrals at $n^{th}$ order with $2n_c$ number of $2q.P$ vertices that can appear in the process, is  given as
\begin{eqnarray}\label{Eq:generic-on-loop}
    \int \frac{d^d q}{(2\pi)^d}\,\frac{q^{\mu_1}\cdots q^{\mu_{2n_c}}}{(q^2-M^2)^{n}}\,\equiv \,g^{\mu_1\cdots \mu_{2n_c}} \,\mathcal{I}[q^{2n_c}]^{n}.
\end{eqnarray}
The completely symmetric tensor $g^{\mu_1\cdots \mu_{2n_c}}$, in Eq.~\eqref{Eq:generic-on-loop}, takes care of all possible contractions among the $P_{\mu}$'s. We present the explicit expressions for all the relevant and necessary master integrals `` $\mathcal{I}$ " for our calculation in the Appendix~\ref{app:master-integral}. 


\subsection{A detailed case analysis: $O(P^2\,U^6)$}

The covariant diagram method maintains a record of all CDE terms and maps the integral at each order to one-loop diagrams. To exemplify we focus on the $O(P^2U^6)$ that appears at $n=8$. Adhering to the ideas discussed in Ref.~\cite{Zhang:2016pja}, we provide all possible diagrams for this operator class in Table~\ref{table:P2U6-diags}.

\begin{table}[!h]
			\centering \scriptsize
			\renewcommand{\arraystretch}{2.6}
			\adjustbox{width = 0.8\textwidth}{\begin{tabular}{|c|c|c|c|}
				\hline
				\textsf{\quad Diagram}$\quad$&
				\textsf{\quad $O(P^2U^6)$ structure}$\quad$&
				\textsf{\quad Values}$\quad$\\
				\hline
                    \hline

                \begin{minipage}{0.1\textwidth}
                  \includegraphics[width=\linewidth, height=1.5cm]{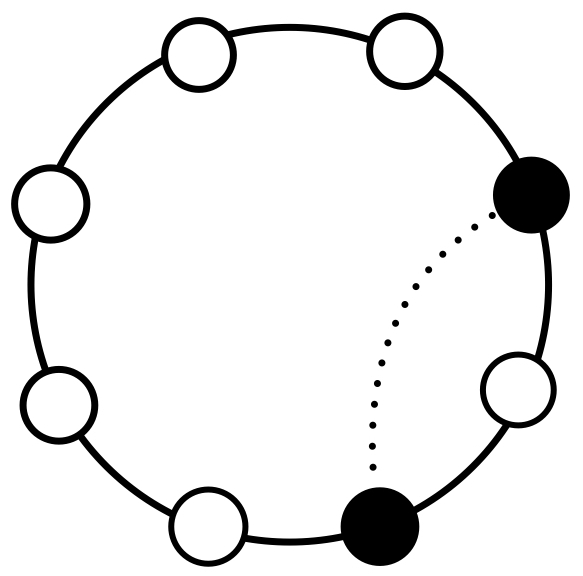}
                \end{minipage}&
                tr$\big(P_{\mu}UP_{\mu}UUUUU\big)$
                &$-ic_s\,2^2\,\mathcal{I}[q^2]^8$
               \\

             \hline

                 \begin{minipage}{0.1\textwidth}
                  \includegraphics[width=\linewidth, height=1.5cm]{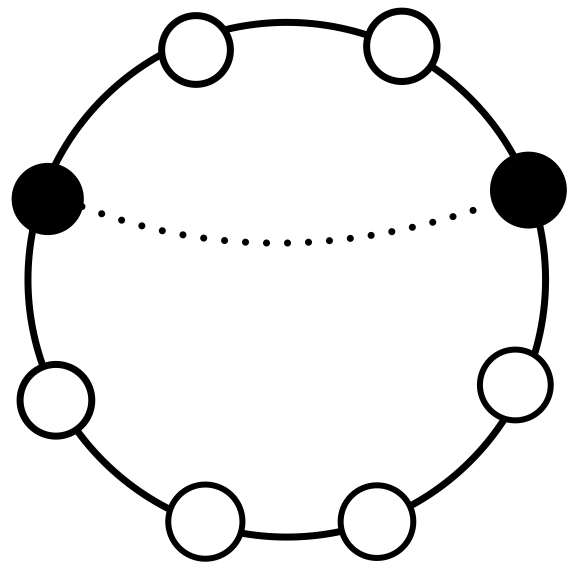}
                \end{minipage}&
                tr$\big(P_{\mu}UUP_{\mu}UUUU\big)$
                &$-ic_s\,2^2\,\mathcal{I}[q^2]^8$
               \\
               \hline

                 \begin{minipage}{0.1\textwidth}
                  \includegraphics[width=\linewidth, height=1.5cm]{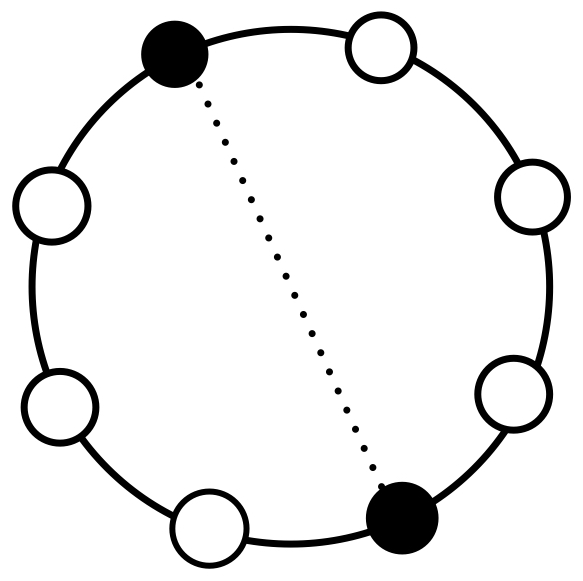}
                \end{minipage}&
                tr$\big(P_{\mu}UUUP_{\mu}UUU\big)$
                &$-i\frac{c_s}{2}\,2^2\,\mathcal{I}[q^2]^8$
               \\
               \hline
				
			\end{tabular}}
			\caption{\small All possible diagrams at the level of $\mathcal{O}(P^2U^6)$ containing two $P_{\mu}$'s that are contracted among themselves and six $U$'s. In the second column, we present their corresponding operator structures with open covariant derivatives. Their values are given in the third column. }
			\label{table:P2U6-diags}
		\end{table}

\begin{wrapfigure}[10]{l}{4cm}
     \parbox{4cm}{
     \centering
     \includegraphics[scale=0.13]{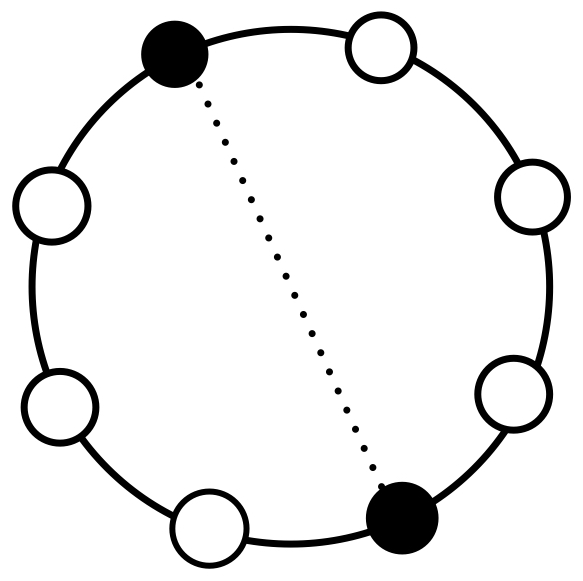}
     \caption{\small Representative diagram at $\mathcal{O}(P^2U^6)$.}\label{fig:symm-P2U6}}
\end{wrapfigure}

In order to discuss how to compute these diagrams, we will concentrate on Fig.~\ref{fig:symm-P2U6}. Since in this category only two $P_{\mu}$'s are present, they must be contracted among themselves to form a Lorentz-invariant structure. Following the conventions of \cite{Zhang:2016pja}, we denote the vertices containing $P_{\mu}$'s and $U$'s with black and white blobs on the loop diagrams respectively and the contraction of the $P_{\mu}$'s is shown with a dotted line. The structure that is unique to the diagram then can be read off starting from one particular blob and going clockwise until we exhaust all the vertices present in the diagram. Following the rule, the structure corresponding to this diagram can then be written as $\text{tr}\,\big(P_{\mu}UUUP_{\mu}UUU\big)$.
  Each of the black blobs corresponds to a $2q.P$ vertex factor, which leaves an additional factor of $2$ along with $P_{\mu}$'s when the loop momentum $q_{\mu}$'s are taken inside the integral $\mathcal{I}[q^{2n_c}]$. When the diagram exhibits an $N$-fold rotational symmetry, we divide the total value of the loop-integral with a factor of $N$\footnote{This division is necessary to eliminate the overcounting when different operator structures under trace give rise to the same diagram.}. Thus,  the contribution from this diagram reads as: $-i\frac{c_s}{2}\,2^2\,\mathcal{I}[q^2]^8 = i\frac{c_s}{2}\,2^2\times\big(\frac{i}{16\pi^2}\frac{1}{M^{10}}\frac{12}{7!}\big)$\footnote{It should be noted that if a specific diagram and its mirror image cannot be superimposed onto each other even after rotation under trace (see e.g. the second diagram in Table~\ref{table:P6U-diags}), these diagrams are connected via Hermitian conjugation. The conjugate diagrams receive exactly similar contributions as their parent diagrams after the expansion of the covariant structures, so we avoid writing them separately.}. 
 
 Along with this, two more independent diagrams can arise at this level. Table~\ref{table:P2U6-diags} contains all the diagrams, their corresponding structures with open covariant derivatives, and individual contributions. The number of possible structures appearing at each class implies that the same number of independent covariant structures must be present where $P_{\mu}$'s only appear through commutators. To verify the results obtained in Eq.~\eqref{eq:Lag-P2U6}, first, we start with three distinct forms of the effective operators given in that equation. Then we expand the commutators to encompass all the diagrams within this class as
\begin{multline}\label{eq:P2U6-final-structures}
C^{(2,6)}_1\,\text{tr}\big(U^4\,[P_{\mu},U]\,[P_{\mu},U]\big)\,+\,C^{(2,6)}_2\,\text{tr}\big(U^3\,[P_{\mu},U]\,U\,[P_{\mu},U]\big)\\
+\,C^{(2,6)}_3\,\text{tr}\big(U^2\,[P_{\mu},U]\,U^2\,[P_{\mu},U]\big)
= \,\,(2C^{(2,6)}_1-C^{(2,6)}_2)\,\text{tr}\,\big(P_{\mu}UP_{\mu}UUUUU\big)\\+\,(2C^{(2,6)}_2-2C^{(2,6)}_3-C^{(2,6)}_1)\,\text{tr}\,\big(P_{\mu}UUP_{\mu}UUUU\big)\\
+\,(2C^{(2,6)}_3-C^{(2,6)}_2)\,\text{tr}\,\big(P_{\mu}UUUP_{\mu}UUU\big)\,+\,\text{tr}\big(\cdots P^2\cdots\big)\text{terms}.
\end{multline}  
The contraction of two adjacent $2q.P$ vertices and the contribution from $(-P^2)$ vertices can produce $\text{tr}\big(\cdots P^2\cdots\big)$ terms with the diagrams where the adjacent $P_{\mu}$'s are contracted. It is not necessary to consider these diagrams separately since their coefficients are functions of the same $C_k^{(i,j)}$'s which can be determined from other diagrams. The coefficients of the structures in the RHS of Eq.~\eqref{eq:P2U6-final-structures}, correspond to values of the diagrams given in the third column of Table~\ref{table:P2U6-diags},
\begin{eqnarray}
    2C^{(2,6)}_1-C^{(2,6)}_2 &=& -\frac{c_s}{16\pi^2}\frac{1}{M^{10}}\frac{48}{7!},\nonumber\\
    2C^{(2,6)}_2-C^{(2,6)}_1-2C^{(2,6)}_3 &=& -\frac{c_s}{16\pi^2}\frac{1}{M^{10}}\frac{48}{7!},\nonumber\\
    2C^{(2,6)}_3-C^{(2,6)}_2 &=& -\frac{c_s}{16\pi^2}\frac{1}{M^{10}}\frac{24}{7!}.
\end{eqnarray}
Finally, we find the coefficients associated with the operators given in Eq.~\eqref{eq:P2U6-final-structures} as
\begin{equation}
    C^{(2,6)}_1 = -\frac{c_s}{16\pi^2}\frac{1}{M^{10}}\frac{1}{42},\,\hspace{.7cm} C^{(2,6)}_2 = -\frac{c_s}{16\pi^2}\frac{1}{M^{10}}\frac{4}{105},\hspace{.7cm} C^{(2,6)}_3 = -\frac{c_s}{16\pi^2}\frac{1}{M^{10}}\frac{3}{140},
\end{equation}
comparing with Eq.~\eqref{eq:Lag-P2U6}, we can infer $\mathcal{C}^{(2,6)}_k =(c_s/(16\pi^2)) C^{(2,6)}_k$, which validates our findings.

\subsection{Coefficients obtained using Covaraint diagrams}
\label{subsec:wc-cov-diags}
 
\subsubsection*{$\bullet$\quad\boldsymbol{$O(P^8)$}}

Starting with the covariant structures derived in Eq.~\eqref{eq:Lag-P8}, we expand them to find the contributions to each of the diagrams of this class. The diagrams and their corresponding values are given in Table~\ref{table:P8-diags},
\begin{align}\label{eq:P8-final-result}
 & C^{(8,0)}_1\,\text{tr}\Big([P_{\mu},P_{\nu}]\,[P_{\mu},P_{\nu}][P_{\alpha},P_{\beta}][P_{\alpha},P_{\beta}]\Big)\,+  \,C^{(8,0)}_2\,\text{tr}\Big([P_{\mu},P_{\nu}]\,[P_{\nu},P_{\rho}][P_{\mu},P_{\sigma}][P_{\sigma},P_{\rho}]\Big)  \nn \\ 
+& \,C^{(8,0)}_3\,\text{tr}\Big([P_{\mu},P_{\nu}]\,[P_{\nu},P_{\rho}][P_{\rho},P_{\sigma}][P_{\sigma},P_{\mu}]\Big)
+  \,C^{(8,0)}_4\,\text{tr}\Big([P_{\mu},P_{\nu}]\,[P_{\rho},P_{\sigma}][P_{\mu},P_{\nu}][P_{\rho},P_{\sigma}]\Big)  \nn \\
+ & \,C^{(8,0)}_5\,\text{tr}\Big([P_{\mu},[P_{\alpha},[P_{\alpha},P_{\nu}]]]\,[P_{\mu},[P_{\rho},[P_{\rho},P_{\nu}]]]\Big) \nn\\
+& \,C^{(8,0)}_6\,\text{tr}\Big([P_{\mu},[P_{\alpha},[P_{\alpha},P_{\nu}]]]\,[P_{\nu},P_{\rho}][P_{\rho},P_{\mu}]\Big) \nn \\
+ & \,C^{(8,0)}_7\,\text{tr}\Big([P_{\alpha},[P_{\alpha},P_{\mu}]]\,P_{\beta},[P_{\beta},P_{\nu}]]\,[P_{\mu},P_{\nu}]\Big)  \nn \\ 
\nn
 \supset & \quad \Big(4C^{(8,0)}_1\,+\,2C^{(8,0)}_3\,+\,2C^{(8,0)}_6\,-\,4C^{(8,0)}_7 \Big)\,\text{tr} \Big(P_{\mu}P_{\nu}P_{\mu}P_{\rho}P_{\sigma}P_{\nu}P_{\rho}P_{\sigma}\Big) \nn \\
 + & \,\Big(2C^{(8,0)}_2\,-\,2C^{(8,0)}_6\,+\,8C^{(8,0)}_5 \Big) \,\text{tr} \Big(P_{\mu}P_{\nu}P_{\rho}P_{\nu}P_{\mu}P_{\sigma}P_{\rho}P_{\sigma}\Big) \nn \\
+ & \,\Big( 2C^{(8,0)}_2\,-\,4C^{(8,0)}_3\,-\,4C^{(8,0)}_6\,+\,4C^{(8,0)}_7 \Big) \,\text{tr} \Big(P_{\mu}P_{\nu}P_{\mu}P_{\rho}P_{\nu}P_{\sigma}P_{\rho}P_{\sigma}\Big) \nn \\
- & \,\Big(4C^{(8,0)}_2\,-\,2C^{(8,0)}_6 \Big) \,\text{tr} \Big(P_{\mu}P_{\nu}P_{\mu}P_{\rho}P_{\sigma}P_{\nu}P_{\rho}P_{\sigma}\Big) \nn\\
+ &  \,\Big(C^{(8,0)}_2\,-\,8C^{(8,0)}_4\Big) \,\text{tr} \Big(P_{\mu}P_{\nu}P_{\rho}P_{\sigma}P_{\mu}P_{\nu}P_{\sigma}P_{\rho}\Big) \nn \\ +&\,\Big(C^{(8,0)}_3\,+\,4C^{(8,0)}_4\Big) \,\text{tr} \,\Big(P_{\mu}P_{\nu}P_{\rho}P_{\mu}P_{\sigma}P_{\rho}P_{\nu}P_{\sigma}\Big)
+\,4C^{(8,0)}_4\,\text{tr}\,\Big(P_{\mu}P_{\nu}P_{\rho}P_{\sigma}P_{\mu}P_{\nu}P_{\rho}P_{\sigma}\Big).
\end{align} 
Equating the coefficients of the structures on the RHS of the Eq.~\eqref{eq:P8-final-result} with the values in Table~\ref{table:P8-diags}, we find,

\begin{align}
    C^{(8,0)}_1 &= \frac{c_s}{16\pi^2}\frac{1}{M^4}\frac{17}{5040}, &
    C^{(8,0)}_2 &= \frac{c_s}{16\pi^2}\frac{1}{M^4}\frac{1}{420}, &
    C^{(8,0)}_3 &= \frac{c_s}{16\pi^2}\frac{1}{M^4}\frac{1}{2520}, \nn \\
    C^{(8,0)}_4 &= \frac{c_s}{16\pi^2}\frac{1}{M^4}\frac{1}{10080}, &
    C^{(8,0)}_5 &= \frac{c_s}{16\pi^2}\frac{1}{M^4}\frac{1}{840}, &
    C^{(8,0)}_6 &= \frac{c_s}{16\pi^2}\frac{1}{M^4}\frac{2}{315},\nn \\ 
    & & C^{(8,0)}_7 &= \frac{c_s}{16\pi^2}\frac{1}{M^4}\frac{2}{315}.
\end{align}
From Eq.~\eqref{eq:Lag-P8}, one can check $ \mathcal{C}^{(8,0)}_k = (c_s/(16\pi^2))C^{(8,0)}_k$, which validates our findings.

\subsubsection*{$\bullet$\quad\boldsymbol{$O(P^6\,U)$}}
The structures derived in Eq.~\eqref{eq:Lag-P6U} are expanded to map them back to the diagrams of this class given in Table.~\ref{table:P6U-diags}, 
\begin{align}
& C^{(6,1)}_1\,\text{tr}\Big(U\,[P_{\mu},P_{\nu}]\,[P_{\nu},P_{\rho}]\,[P_{\rho},P_{\mu}]\Big)\,+\,C^{(6,1)}_2\,\text{tr}\Big(U\,[P_{\nu},[P_{\nu},P_{\mu}]]\,[P_{\rho},[P_{\rho},P_{\mu}]]\Big) \nn \\
+ & \,C^{(6,1)}_3\,\text{tr}\Big([P_{\alpha},[P_{\alpha},U]\,[P_{\mu},P_{\nu}]\,[P_{\mu},P_{\nu}]]\Big)
+\,C^{(6,1)}_4\,\text{tr}\Big([P_{\beta},[P_{\gamma},U]]\,[P_{\gamma},P_{\alpha}]\,[P_{\alpha},P_{\beta}]\Big) \nn \\ \supset & \,\,-\Big( C^{(6,1)}_1 +4C^{(6,1)}_3 +C^{(6,1)}_4\Big) \,\text{tr}\Big(P_{\mu}P_{\nu}P_{\rho}P_{\mu}P_{\rho}P_{\nu}U\Big)\,\nn\\
+&\,\Big(C^{(6,1)}_1 + C^{(6,1)}_4\Big) \,\text{tr}\Big(P_{\mu}P_{\nu}P_{\rho}P_{\mu}P_{\rho}P_{\nu}U\Big) \nn \\
+& \,\Big(C^{(6,1)}_1+4C^{(6,1)}_2-2C^{(6,1)}_4\Big) \,\text{tr}\Big(P_{\mu}P_{\nu}P_{\mu}P_{\rho}P_{\nu}P_{\rho}U\Big)\,-\,C^{(6,1)}_1\,\text{tr}\Big(P_{\mu}P_{\nu}P_{\rho}P_{\mu}P_{\nu}P_{\rho}U\Big).
\end{align}  
 
We equate the coefficients of the structures with the open covariant derivatives with their corresponding values and reproduce the results given in Eq.~\eqref{eq:Lag-P6U},
\begin{align}
    C^{(6,1)}_1 &= -\frac{c_s}{16\pi^2}\frac{1}{M^4}\frac{1}{90}, &
    C^{(6,1)}_2 &= +\frac{c_s}{16\pi^2}\frac{1}{M^4}\frac{1}{60},&
    C^{(6,1)}_3 &= -\frac{c_s}{16\pi^2}\frac{1}{M^4}\frac{1}{180},\nn \\
    & & C^{(6,1)}_4 &= +\frac{c_s}{16\pi^2}\frac{1}{M^4}\frac{1}{45}.
\end{align}
With an appropriate scale factor, one can compare these coefficients to the previously obtained coefficients shown in Eq.~\eqref{eq:Lag-P6U}, the relation reads, $\mathcal{C}^{(6,1)}_k = (c_s/(16\pi^2))C^{(6,1)}_k$.

\subsubsection*{$\bullet$\quad\boldsymbol{$O(P^4\,U^3)$}}

We start with the structures obtained in Eq.~\eqref{eq:Lag-P4U3}. From Table~\ref{table:P4U3-diags}, it can be seen that there are six independent diagrams that can arise in this category. Therefore, we assume that the covariant structures that have distinct Hermitian conjugates (h.c.'s), both of them should have the same coefficients to maintain the overall Hermiticity of the Lagrangian. In this way, we encounter exactly six independent variables that can be solved from six associated diagrams.
\begin{align}
& C^{(4,3)}_1\,\text{tr}\Big(U^3\,[P_{\mu},P_{\nu}]\,[P_{\mu},P_{\nu}]\,\Big)\,+\,C^{(4,3)}_2\,\text{tr}\Big(U^2\,[P_{\mu},P_{\nu}]\,U\,[P_{\mu},P_{\nu}]\Big)\nn\\ 
&+C^{(4,3)}_3\, \Bigg\{\,\text{tr}\Big(U^2\,[P_{\mu},U]\,[P_{\nu},[P_{\nu},P_{\mu}]]\Big)
-\text{tr}\Big([P_{\mu},U]\,U^2\,[P_{\nu},[P_{\nu},P_{\mu}]]\Big)\Bigg\}\nn\\
&+\,C^{(4,3)}_4\,\Bigg\{\text{tr}\Big([P_{\mu},U]\,[P_{\nu},U]\,U\,[P_{\mu},P_{\nu}]\Big)+\text{tr}\Big(U\,[P_{\mu},U]\,[P_{\nu},U]\,[P_{\mu},P_{\nu}]\Big) \Bigg\}\nn \\
&+\,C^{(4,3)}_5\,\text{tr}\Big(U\,[P_{\mu}[P_{\mu},U]]\,[P_{\nu}[P_{\nu},U]]\Big)+\,C^{(4,3)}_6\,\text{tr}\Big([P_{\mu}[P_{\mu},U]]\,[P_{\nu},U][P_{\nu},U]\Big) \nn \\\supset & \,\,\Big(2 C^{(4,3)}_1 +4C^{(4,3)}_3 \Big) \,\text{tr}\Big(UUUP_{\nu}P_{\mu}P_{\nu}P_{\mu}\Big)\,+\,\Big(2C^{(4,3)}_2 - 2C^{(4,3)}_4\Big) \,\text{tr}\Big(UUP_{\mu}P_{\nu}UP_{\mu}P_{\nu}\Big) \nn \\
+& \Big(-2C^{(4,3)}_2+2C^{(4,3)}_4+2C^{(4,3)}_6\Big) \,\text{tr}\Big(UUP_{\mu}P_{\nu}UP_{\nu}P_{\mu}\Big)\,-\,2C^{(4,3)}_3\,\text{tr}\Big(UUP_{\mu}UP_{\nu}P_{\mu}P_{\nu}\Big)\nn\\
+&2C^{(4,3)}_4\,\text{tr}\Big(UP_{\mu}UP_{\nu}UP_{\mu}P_{\nu}\Big)-\Big(2C^{(4,3)}_4-4C^{(4,3)}_5+4C^{(4,3)}_6\Big) \,\text{tr}\Big(UP_{\mu}UP_{\mu}P_{\nu}UP_{\nu}\Big).
\end{align}  
It is evident from the covariant structures that the operators that have distinct h.c.'s, both yield identical contributions to the mirror-symmetric diagrams, while the conjugate operators produce the mirror images of the diagrams that are not mirror-symmetric.
\begin{align}
    C^{(4,3)}_1 & = \frac{c_s}{16\pi^2}\,\mathcal{C}^{(4,3)}_1= -\frac{c_s}{16\pi^2}\frac{1}{M^6}\frac{1}{60}, \nn\\
    C^{(4,3)}_2 &= \frac{c_s}{16\pi^2}\,\mathcal{C}^{(4,3)}_2= -\frac{c_s}{16\pi^2}\frac{1}{M^6}\frac{1}{90},\nn\\
    C^{(4,3)}_3 &= \frac{c_s}{16\pi^2}\,\mathcal{C}^{(4,3)}_3 = -\frac{c_s}{16\pi^2}\,\mathcal{C}^{(4,3)}_4 = \frac{c_s}{16\pi^2}\frac{1}{M^6}\frac{1}{180}, \nn\\
    C^{(4,3)}_4 &= \frac{c_s}{16\pi^2}\,\mathcal{C}^{(4,3)}_5 = \frac{c_s}{16\pi^2}\,\mathcal{C}^{(4,3)}_6 = -\frac{c_s}{16\pi^2}\frac{1}{M^6}\frac{1}{180} \nn\\
    C^{(4,3)}_5 &=  \frac{c_s}{16\pi^2}\,\mathcal{C}^{(4,3)}_7=  -\frac{c_s}{16\pi^2}\frac{1}{M^6}\frac{1}{60},\nn\\
    C^{(4,3)}_6 &=  \frac{c_s}{16\pi^2}\,\mathcal{C}^{(4,3)}_8=  - \frac{c_s}{16\pi^2}\frac{1}{M^6}\frac{1}{90}.
\end{align}
The coefficients match exactly with those obtained in Eq.~\eqref{eq:Lag-P4U3}.

\subsubsection*{$\bullet$\quad\boldsymbol{$O(P^6\,U^2)$}}
Here, to begin with, we consider the operator structures given in Eq.~\eqref{eq:Lag-P6U2} derived using the Heat-Kernel method. Noting the possible independent covariant diagrams, allowed for this class (see Tables~\ref{tab:P6U2-diags-1} and \ref{tab:P6U2-diags-2}), we anticipate that there should be at most seventeen independent covariant operators (excluding the h.c.'s). Regarding the h.c.'s, we follow the similar prescription discussed in the case of $O(P^4\,U^3)$. We express the last two operators in Eq.~\eqref{eq:Lag-P6U2} (i.e., $[P_\alpha,P_\mu][P_\mu,P_\beta]U\,[P_\alpha,[P_\beta,U]]$, and $[P_\alpha,P_\mu][P_\mu,P_\beta][P_\alpha,[P_\beta,U]] U$) in terms of other ones.
\begin{align}\label{eq:cov-p6u2}
& C^{(6,2)}_1\,\text{tr}\Big(U^2\,[P_{\mu},P_{\nu}]\,[P_{\nu},P_{\alpha}][P_{\alpha},P_{\mu}]\Big)\,+\,C^{(6,2)}_2\,\text{tr}\Big(U\,[P_{\mu},P_{\nu}]\,U\,[P_{\nu},P_{\alpha}][P_{\alpha},P_{\mu}]\Big)\nn \\
+ &\,C^{(6,2)}_3\,\text{tr}\Big(U^2\,[P_{\mu},[P_{\mu},P_{\nu}]]\,[P_{\alpha},[P_{\alpha},P_{\nu}]]\Big)\,+\,C^{(6,2)}_4\,\text{tr}\Big(U\,[P_{\mu},[P_{\mu},P_{\nu}]]\,U\,[P_{\alpha},[P_{\alpha},P_{\nu}]]\Big)\nn\\
+ &\,C^{(6,2)}_5\,\Bigg\{\text{tr}\Big(U\,[P_{\mu},[P_{\mu},U]]\,[P_{\nu},P_{\alpha}]\,[P_{\nu},P_{\alpha}]\Big)+\text{tr}\Big([P_{\mu},[P_{\mu},U]]\,U\,[P_{\nu},P_{\alpha}]\,[P_{\nu},P_{\alpha}]\Big)\Bigg\}\nn\\
+ &\,C^{(6,2)}_6\,\Bigg\{\text{tr}\Big(U\,[P_{\mu},U]\,[P_{\nu}[P_{\nu},P_{\alpha}]\,[P_{\mu},P_{\alpha}]\Big)+\text{tr}\Big(U\,[P_{\mu},P_{\alpha}]\,[P_{\nu}[P_{\nu},P_{\alpha}]\,[P_{\mu},U]\Big)\Bigg\}\nn\\
+&\,C^{(6,2)}_7\,\Bigg\{\text{tr}\Big(U\,[P_{\mu},U]\,[P_{\mu},P_{\nu}]\,[P_{\alpha}[P_{\alpha},P_{\nu}]\Big)+\text{tr}\Big(U\,[P_{\alpha}[P_{\alpha},P_{\nu}]\,[P_{\mu},P_{\nu}]\,[P_{\mu},U]\Big)\Bigg\}\nn\\
+&C^{(6,2)}_8\,\text{tr}\Big(U\,[P_{\mu},P_{\nu}]\,[P_{\alpha},[P_{\alpha},U]\,[P_{\mu},P_{\nu}]\Big)\,+\,C^{(6,2)}_9\,\text{tr}\Big(\,[P_{\mu},U]\,[P_{\mu},U]\,[P_{\nu},P_{\alpha}]\,[P_{\nu},P_{\alpha}]\Big)\nn\\
+&\,C^{(6,2)}_{10}\,\Bigg\{\text{tr}\Big(\,[P_{\mu},U]\,[P_{\nu},[P_{\nu},P_{\alpha}]]\,U\,[P_{\mu},P_{\alpha}]\Big)+\text{tr}\Big(\,[P_{\mu},P_{\alpha}]\,U\,[P_{\nu},[P_{\nu},P_{\alpha}]]\,[P_{\mu},U]\Big)\Bigg\}\nn\\
+&\,C^{(6,2)}_{11}\,\text{tr}\Big(\,[P_{\mu},U]\,[P_{\nu},U]\,[P_{\mu},P_{\alpha}]\,[P_{\alpha},P_{\nu}]\Big)+\,C^{(6,2)}_{12}\,\text{tr}\Big(\,[P_{\mu},U]\,[P_{\nu},U]\,[P_{\nu},P_{\alpha}]\,[P_{\alpha},P_{\mu}]\Big)\nn\\
+&\,C^{(6,2)}_{13}\,\Bigg\{\text{tr}\Big(\,[P_{\mu},U]\,[P_{\mu},P_{\nu}]\,[P_{\alpha},U]\,[P_{\alpha},P_{\nu}]\Big)+\text{tr}\Big(\,[P_{\mu},U]\,[P_{\alpha},P_{\nu}]\,[P_{\alpha},U]\,[P_{\mu},P_{\nu}]\Big)\Bigg\}\nn\\
+&\,C^{(6,2)}_{14}\,\text{tr}\Big(\,[P_{\mu},U]\,[P_{\nu},P_{\alpha}]\,[P_{\mu},U]\,[P_{\nu},P_{\alpha}]\Big)
\,+\,C^{(6,2)}_{15}\,\Bigg\{\text{tr}\Big(\,[P_{\mu},[P_{\mu},U]]\,[P_{\nu},U]\,[P_{\alpha},[P_{\alpha},P_{\nu}]]\Big)\nn\\
-&\text{tr}\Big(\,[P_{\mu},[P_{\mu},U]]\,[P_{\alpha},[P_{\alpha},P_{\nu}]]\,[P_{\nu},U]\Big)\Bigg\}\,+\,C^{(6,2)}_{16}\,\Bigg\{\text{tr}\Big(\,[P_{\mu},U]\,[P_{\nu},U]\,[P_{\mu},[P_{\alpha},[P_{\alpha},P_{\nu}]]]\Big)\nn\\
-&\text{tr}\Big(\,[P_{\mu},U]\,[P_{\mu},[P_{\alpha},[P_{\alpha},P_{\nu}]]]\,[P_{\nu},U]\Big)\Bigg\}
+2\,C^{(6,2)}_{17}\,\text{tr}\Big(\,[P_{\mu},[P_{\nu},[P_{\nu},U]]]\,[P_{\mu},[P_{\alpha},[P_{\alpha},U]]]\Big) \nn\\
\supset & -(C_1+C_{11})\,\text{tr}\,\Big(P_{\mu}P_{\nu}P_{\rho}UUP_{\mu}P_{\nu}P_{\rho}\Big)+\,(C_1\,+\,4C_3\,-\,4C_7\,-C_{12})\,\text{tr}\,\Big(P_{\mu}P_{\nu}P_{\mu}UUP_{\rho}P_{\nu}P_{\rho}\Big)\nn\\
-&(C_1\,+\,2C_9\,+\,4C_6)\,\text{tr}\,\Big(P_{\mu}P_{\nu}P_{\rho}UUP_{\rho}P_{\mu}P_{\nu}\Big)\nn\\
+&(C_1+2C_6+C_{11}+2C_{16})\,\text{tr}\,\Big(P_{\mu}P_{\nu}P_{\rho}UUP_{\nu}P_{\mu}P_{\rho}\Big)\nn\\
+&(4C_4-4C_{10}+2C_{13})\,\text{tr}\,\Big(UP_{\mu}P_{\nu}P_{\mu}UP_{\rho}P_{\nu}P_{\rho}\Big)+4C_{14}\,\text{tr}\,\Big(UP_{\mu}P_{\nu}P_{\rho}UP_{\mu}P_{\nu}P_{\rho}\Big)\nn\\
-&(2C_{14}-C_{13})\,\text{tr}\,\Big(UP_{\mu}P_{\nu}P_{\rho}UP_{\nu}P_{\mu}P_{\rho}\Big)-(4C_8+4C_{13})\,\text{tr}\,\Big(UP_{\mu}P_{\nu}P_{\rho}UP_{\rho}P_{\mu}P_{\nu}\Big)\nn\\
+&(4C_8+2C_{13}+8C_{17})\,\text{tr}\,\Big(UP_{\mu}P_{\nu}P_{\rho}UP_{\rho}P_{\nu}P_{\mu}\Big)\nn\\
-&(\frac{1}{2}C_2+\frac{1}{2}C_{11}+2C_{14})\,\text{tr}\,\Big(UP_{\mu}P_{\nu}UP_{\rho}P_{\mu}P_{\nu}P_{\rho}\Big)\nn\\
+&(C_2-C_{12}+4C_{14})\,\text{tr}\,\Big(UP_{\mu}P_{\nu}UP_{\rho}P_{\nu}P_{\mu}P_{\rho}\Big)\nn\\
+&(C_2+C_{11}+2C_{10}+2C_{16}-2C_{13})\,\text{tr}\,\Big(UP_{\mu}P_{\nu}UP_{\mu}P_{\rho}P_{\nu}P_{\rho}\Big)\nn\\
-&(C_2+2C_{10}-C_{12}-2C_{13}-4C_{15}+2C_{16})\,\text{tr}\,\Big(UP_{\mu}P_{\nu}UP_{\nu}P_{\rho}P_{\mu}P_{\rho}\Big)\nn\\
+&2C_{11}\,\text{tr}\,\Big(P_{\mu}P_{\nu}P_{\rho}UP_{\mu}UP_{\nu}P_{\rho}\Big)\nn\\
+&(2C_6+2C_9-C_{12}-4C_{15}+2C_{16}-4C_5-2C_7)\,\text{tr}\,\Big(P_{\mu}UP_{\mu}UP_{\nu}P_{\rho}P_{\nu}P_{\rho}\Big)\nn\\
-&(2C_6+C_{11}-C_{12}-2C_7+2C_{16})\,\text{tr}\,\Big(P_{\mu}UP_{\nu}UP_{\rho}P_{\mu}P_{\rho}P_{\nu}\Big)\nn\\
-&(2C_{11}+4C_{16})\,\text{tr}\,\Big(P_{\mu}P_{\nu}P_{\rho}P_{\nu}P_{\mu}UP_{\rho}U\Big)
\end{align}  
The coefficients in the LHS of the Eq.~\eqref{eq:cov-p6u2} can be obtained from the values of the loop diagrams, 
\begin{align}
    C^{(6,2)}_1 &= \frac{c_s}{16\pi^2} \frac{1}{M^6}\frac{1}{210}, &
    C^{(6,2)}_2 &= \frac{c_s}{16\pi^2}\frac{1}{M^6}\frac{2}{315}, &
    C^{(6,2)}_3 &= -\frac{c_s}{16\pi^2}\frac{1}{M^6}\frac{1}{105}, \nn \\
    C^{(6,2)}_4 &= -\frac{c_s}{16\pi^2}\frac{1}{M^6}\frac{1}{140}, &
    C^{(6,2)}_5 &= \frac{c_s}{16\pi^2}\frac{1}{M^6}\frac{1}{105}, &
    C^{(6,2)}_6 &= -\frac{c_s}{16\pi^2}\frac{1}{M^6}\frac{1}{210},\nn \\ 
    C^{(6,2)}_7 &= -\frac{c_s}{16\pi^2}\frac{1}{M^6}\frac{1}{105},& 
    C^{(6,2)}_8 &= \frac{c_s}{16\pi^2}\frac{1}{M^6}\frac{1}{315},&
    C^{(6,2)}_9 &= \frac{c_s}{16\pi^2}\frac{1}{M^6}\frac{11}{1260},\nn \\
    C^{(6,2)}_{10} &= -\frac{c_s}{16\pi^2}\frac{1}{M^6}\frac{1}{126},& 
    C^{(6,2)}_{11} &= -\frac{c_s}{16\pi^2}\frac{1}{M^6}\frac{1}{630},&
    C^{(6,2)}_{12} &= \frac{c_s}{16\pi^2}\frac{1}{M^6}\frac{1}{126},\nn \\
    C^{(6,2)}_{13} &= -\frac{c_s}{16\pi^2}\frac{1}{M^6}\frac{1}{420},& 
    C^{(6,2)}_{14} &= -\frac{c_s}{16\pi^2}\frac{1}{M^6}\frac{1}{2520},&
    C^{(6,2)}_{15} &= -\frac{c_s}{16\pi^2}\frac{1}{M^6}\frac{1}{315},\nn \\
    C^{(6,2)}_{16} &= \frac{c_s}{16\pi^2}\frac{1}{M^6}\frac{1}{630},&
    C^{(6,2)}_{17} &= -\frac{c_s}{16\pi^2}\frac{1}{M^6}\frac{1}{840}.
\end{align}

\section{Universal One-loop Effective Lagrangian up to $D8$}
\label{sec:uolea}
Relying on the computation based on the Heat-Kernel method and supported by the covariant diagram technique, we exhaustively compute all possible operator structures that can emerge after integrating out degenerate heavy scalars at the one-loop considering only heavy propagators in the loop. We collect all such terms and provide the universal one-loop effective Lagrangian up to dimension eight. This effective action does not necessarily depend on either the specific UV or low energy theories, and in that sense it is universal. 
\begin{align}\label{eq:finite}
    \mathcal{L}_{\eff}^{d \leq 8} = & \mathcal{L}_{\eff}^{\text{ren}} + \cfrac{c_s}{(4\pi)^{2}} \Bigg[ \L^{(8,0)}_\eff  +  \L^{(6,0)}_\eff + \L^{(6,1)}_\eff + \L^{(6,2)}_\eff + \L^{(4,0)}_\eff + \L^{(4,1)}_\eff + \L^{(4,2)}_\eff + \L^{(4,3)}_\eff + \L^{(4,4)}_\eff  \nn \\
    & + \L^{(2,0)}_\eff +  \L^{(2,1)}_\eff + \L^{(2,2)}_\eff + \L^{(2,3)}_\eff + \L^{(2,4)}_\eff + \L^{(2,5)}_\eff + \L^{(2,6)}_\eff + \L^{(0,0)}_\eff + \L^{(0,1)}_\eff  \nn \\ 
    & + \L^{(0,2)}_\eff  +  \L^{(0,3)}_\eff +  \L^{(0,4)}_\eff + \L^{(0,5)}_\eff  + \L^{(0,6)}_\eff + \L^{(0,7)}_\eff + \L^{(0,8)}_\eff \Bigg] \nn \\
    = & \cfrac{c_s}{(4\pi)^{2}}   M^4\ \left[-\frac{1}{2}\,\left(\ln\left[\frac{M^2}{\mu^2}\right]-\frac{3}{2}\right)\right ]  + \cfrac{c_s}{(4\pi)^{2}}  \tr   \,\Bigg\{ M^2\ \Bigg[-\left(\ln\left[\frac{M^2}{\mu^2}\right]-1\right)\, U\Bigg]  \nn \\
    & + M^0\ \frac{1}{2}  \Bigg[- \ln\left[\frac{M^2}{\mu^2}\right] \, U^2 -\frac{1}{6} \ln\left[\frac{M^2}{\mu^2}\right] \, (G_{\mu\nu})^2\Bigg] \nn\\
    & + \frac{1}{M^2} \frac{1}{6}  \,\Bigg[ -U^3 - \frac{1}{2} (P_\mu U)^2-\frac{1}{2}U\,(G_{\mu\nu})^2  - \frac{1}{10}(J_\nu)^2   + \frac{1}{15}\,G_{\mu\nu}\,G_{\nu\rho}\,G_{\rho\mu} \Bigg]  \nn\\
    &+ \frac{1}{M^4} \frac{1}{24} \, \Bigg[U^4 - U^2 (P^2 U) + \frac{4}{5}U^2 (G_{\mu\nu})^2 + \frac{1}{5} (U\,G_{\mu\nu})^2 +  \frac{1}{5} (P^2 U)^2  \nn\\ 
    & \hspace{2cm} -\frac{2}{5} U\, (P_\mu U)\,J_{\mu} + \frac{2}{5} U(J_\mu)^2 - \frac{2}{15} (P^2 U) (G_{\rho\sigma})^2 +\frac{1}{35}(P_\nu J_{\mu})^2  \nn\\ 
    & \hspace{2cm} - \frac{4}{15} U\,G_{\mu\nu}G_{\nu\rho} G_{\rho\mu} - \frac{8}{15} (P_\mu P_\nu U)\, G_{\rho\mu} G_{\rho\nu} + \frac{16}{105}G_{\mu\nu}J_{\mu}J_{\nu}   \nn\\
    & \hspace{2cm} + \frac{1}{420} (G_{\mu\nu}G_{\rho\sigma})^2 +\frac{17}{210}(G_{\mu\nu})^2(G_{\rho\sigma})^2 +\frac{2}{35}(G_{\mu\nu}G_{\nu\rho})^2 \nn\\
    & \hspace{2cm} + \frac{1}{105} G_{\mu\nu}G_{\nu\rho}G_{\rho\sigma}G_{\sigma\mu} +\frac{16}{105} (P_\mu J_{\nu}) G_{\nu\sigma}G_{\sigma\mu} \Bigg]  \nn\\
    & + \frac{1}{M^6} \frac{1}{60}  \,\Bigg[ -U^5 + 2\,U^3 (P^2 U) + U^2(P_\mu U)^2 - \frac{2}{3} U^2 G_{\mu\nu} U\,G_{\mu\nu}  - U^3 (G_{\mu\nu})^2  \nn \\
    & \hspace{2cm} + \frac{1}{3} U^2 (P_\mu U)J_\mu - \frac{1}{3} U\,(P_\mu U)(P_\nu U)\,G_{\mu\nu}   - \frac{1}{3} U^2 J_\mu (P_\mu U)  \nn\\
    & \hspace{2cm} -  \frac{1}{3} U\,G_{\mu\nu}(P_\mu U)(P_\nu U) - U\,(P^2 U)^2  -  \frac{2}{3} (P^2 U) (P_\nu U)^2  - \frac{1}{7} ((P_\mu U)G_{\mu\alpha})^2 \nn\\
    & \hspace{2cm} +\frac{2}{7} U^2 G_{\mu\nu}G_{\nu\alpha}G_{\alpha\mu}+\frac{8}{21}U\,G_{\mu\nu}U\,G_{\nu\alpha}G_{\alpha\mu}-\frac{4}{7}U^2(J_\mu)^2 -\frac{3}{7} (U\,J_\mu)^2 \nn \\
    & \hspace{2cm} +\frac{4}{7}U\,(P^2U)(G_{\mu\nu})^2 +\frac{4}{7}(P^2U)U(G_{\mu\nu})^2 -\frac{2}{7}U\,(P_\mu U)J_\nu G_{\mu\nu} \nn \\
    & \hspace{2cm} -\frac{2}{7}(P_\mu U)U\,G_{\mu\nu} J_\nu -\frac{4}{7}U\,(P_\mu U)G_{\mu\nu} J_\nu -\frac{4}{7}(P_\mu U)U\, J_\nu G_{\mu\nu} \nn \\
    & \hspace{2cm} +\frac{4}{21}U\,G_{\mu\nu}(P^2U)G_{\mu\nu}  +\frac{11}{21}(P_\alpha U)^2(G_{\mu\nu})^2 - \frac{10}{21}(P_\mu U)J_\nu U\, G_{\mu\nu}  \nn \\
    & \hspace{2cm} - \frac{10}{21}(P_\mu U) G_{\mu\nu} U \,J_\nu - \frac{2}{21} (P_\mu U)(P_\nu U)G_{\mu\alpha}G_{\alpha\nu} + \frac{10}{21} (P_\nu U)(P_\mu U)G_{\mu\alpha}G_{\alpha\nu}  \nn \\
    & \hspace{2cm}-\frac{1}{7} (G_{\alpha\mu}(P_\mu U))^2 - \frac{1}{42} ((P_\alpha U)G_{\mu\nu})^2 -\frac{1}{14} (P_\mu P^2 U)^2 -\frac{4}{21} (P^2U) (P_\mu U)J_\mu \nn \\
    & \hspace{2cm} +\frac{4}{21} (P_\mu U)(P^2U)J_\mu +\frac{2}{21} (P_\mu U) (P_\nu U)(P_\mu J_{\nu}) - \frac{2}{21} (P_\nu U) (P_\mu U)(P_\mu J_{\nu}) \Bigg]  \nn\\
    & + \frac{1}{M^8} \frac{1}{120}  \,\Bigg[U^6 - 3\,U^4 (P^2 U) - 2\,U^3(P_\nu U)^2 + \frac{12}{7}U^2 (P_\mu P_\nu U)(P_\nu P_\mu U)  \nn\\
    & \hspace{.5cm}  +\frac{26}{7} (P_\mu P_\nu U) U\,(P_\mu U)(P_\nu U) +\frac{26}{7} (P_\mu P_\nu U) (P_\mu U)(P_\nu U)U  + \frac{9}{7} (P_\mu U)^2(P_\nu U)^2  \nn\\
    & \hspace{2cm} + \frac{9}{7} U\,(P_\mu P_\nu U)U\,(P_\nu P_\mu U)  + \frac{17}{14} ((P_\mu U)(P_\nu U))^2 + \frac{8}{7} U^3G_{\mu\nu}U\,G_{\mu\nu}  \nn\\
    & \hspace{2cm} + \frac{5}{7} U^4(G_{\mu\nu})^2 + \frac{18}{7} G_{\mu\nu}(P_\mu U)U^2(P_\nu U) + \frac{9}{14} (U^2G_{\mu\nu})^2   \nn\\
    & \hspace{2cm} + \frac{18}{7} G_{\mu\nu}U\,(P_\mu U)(P_\nu U)U + \frac{18}{7} (P_\mu P_\nu U) (P_\mu U)U\,(P_\nu U)   \nn\\
    & \hspace{2cm} +  \Bigg( \frac{8}{7} G_{\mu\nu}U\,(P_\mu U)U\,(P_\nu U) +  \frac{26}{7} G_{\mu\nu}(P_\mu U)U\,(P_\nu U)U \Bigg) \nn\\
    & \hspace{2cm} +  \Bigg( \frac{24}{7} G_{\mu\nu}(P_\mu U)(P_\nu U)U^2 - \frac{2}{7} G_{\mu\nu}U^2(P_\mu U)(P_\nu U)\Bigg)\Bigg] \nn\\
    & + \frac{1}{M^{10}} \frac{1}{210}  \,\Bigg[-U^7 - 5\, U^4 (P_\nu U)^2 - 8\,U^3(P_\mu U)U(P_\mu U)  -\frac{9}{2} (U^2 (P_\mu U))^2 \Bigg] \nn\\
    & + \frac{1}{M^{12}} \frac{1}{336} \,\Bigg[U^8\Bigg] \Bigg\}.
\end{align}
Here, we consider that the tensors $G_{\mu\nu}$ and $J_\mu$ are functions of $P$: $G_{\mu\nu} = [P_\mu,P_\nu]$, and $J_\mu = P_\nu G_{\nu\mu} = [P_\nu,[P_\nu,P_\mu]]$. Please note that the hermitian conjugates are already fed in the above expression such that the effective Lagrangian is self-hermitian.
We agree with the effective action up to dimension six computed using the functional method \cite{Henning:2014wua} and covariant diagram \cite{Zhang:2016pja}.

\subsection{Dimension of $U$ vs Dimension of Operator}
\label{subsec:examples}

We note that the dimension of scalar functional $U$ does not always reflect the dimension of the emerged operator consisting of light fields $(\phi)$. As $U$ is defined as a double functional derivative of action {\it w.r.to.} the heavy fields $(\Phi_i)$, its mass dimension (in a four-dimensional space-time case) is always $+2$.  Thus, $U$ may contain a single light scalar field accompanied by a non-zero mass-dimensional coupling. In that case, identification of the operator's mass dimension by naive power counting of $U$ may be misleading. For example, the structure  $\mathcal{O}(P^{2m}U^{n})$ has mass dimension $2(m+n)$, with $m,n \in \mathbb{Z}^{+}$. But an operator of the mass dimension $2(m+n)$ can also be generated from all the structures up to $\mathcal{O}(P^{2m}U^{2n})$. We demonstrate this notion through a simple toy example where the scalar potential takes the following form 
\begin{multline}
    V_{\text{scalar}} \supset \lambda\,(\phi^{\dagger}\phi)^2 + \lambda_{1}\,(\Phi_1^{\dagger}\Phi_1)^2 + \lambda_{2}\,(\Phi_2^{\dagger}\Phi_2)^2 + \lambda_{3}\,(\phi^{\dagger}\phi)(\Phi_1^{\dagger}\Phi_1) \\
    + \lambda_{4}\,(\phi^{\dagger}\phi)(\Phi_2^{\dagger}\Phi_2) + \lambda_{5}\,(\Phi_1^{\dagger}\Phi_1)(\Phi_2^{\dagger}\Phi_2)+\big(\kappa\,\phi^{\dagger}(\Phi_1^{\dagger}\Phi_2)+\text{h.c.}\big).
\end{multline}

\noindent The explicit structure of $U$ can be obtained through, 
\begin{equation}
U[\phi] =
    \begin{bmatrix}
       \cfrac{\delta^2\,V_{\text{scalar}}}{\delta\Phi_1^{\dagger}\delta\Phi_1} & \cfrac{\delta^2\,V_{\text{scalar}}}{\delta\Phi_1^{\dagger}\delta\Phi_2}\\
       \cfrac{\delta^2\,V_{\text{scalar}}}{\delta\Phi_2^{\dagger}\delta\Phi_1} & \cfrac{\delta^2\,V_{\text{scalar}}}{\delta\Phi_2^{\dagger}\delta\Phi_2}
\end{bmatrix}
\rule[-30pt]{1pt}{70pt}_{\,\substack{\Phi_1=\Phi_{1,c}[\phi],\\ \Phi_2=\Phi_{2,c}[\phi]}}.
\end{equation}
As the Lagrangian does not contain terms linear in each of the heavy fields, their classical solutions correspond to $\Phi_{1,c}=\Phi_{2,c}=0$. So the final form of $U$ can be written as
\begin{equation}
U[\phi] =
    \begin{bmatrix}
       \lambda_3 \,(\phi^{\dagger}\phi) & \kappa\,\phi^{\dagger}\\
       \kappa^{*}\,\phi & \lambda_4 \,(\phi^{\dagger}\phi)
\end{bmatrix}.
\end{equation}
Now, we compute the total contributions to the dimension eight operator $(\phi^{\dagger}\phi)^4$
\begin{eqnarray}
    \mathcal{L}_{\text{eff}}^{(0,4)} &\supset& \frac{1}{24\,M^4}\,(\lambda_3^4+\lambda_4^4)\,(\phi^{\dagger}\phi)^4 \sim \mathcal{O}(U^4),\nn\\
     \mathcal{L}_{\text{eff}}^{(0,5)} &\supset& -\frac{1}{12\,M^6} \,(|\kappa|^2\lambda_3^3+|\kappa|^2\lambda_3^2\lambda_4+|\kappa|^2\lambda_3\lambda_4^2+|\kappa|^2\lambda_4^3)\,(\phi^{\dagger}\phi)^4 \sim \mathcal{O}(U^5),\nn\\
     \mathcal{L}_{\text{eff}}^{(0,6)} &\supset& \frac{1}{40\,M^8} \,(3|\kappa|^4\lambda_3^2+4|\kappa|^4\lambda_3\lambda_4+3|\kappa|^4\lambda_4^2)\,(\phi^{\dagger}\phi)^4 \sim \mathcal{O}(U^6),\nn\\
     \mathcal{L}_{\text{eff}}^{(0,7)} &\supset& -\frac{1}{30\,M^{10}}\,(|\kappa|^6\lambda_3+|\kappa|^6\lambda_4)\,(\phi^{\dagger}\phi)^4 \sim \mathcal{O}(U^7),\nn\\
     \mathcal{L}_{\text{eff}}^{(0,8)} &\supset& \frac{1}{168\,M^{12}}\,|\kappa|^8\,(\phi^{\dagger}\phi)^4 \sim \mathcal{O}(U^8).
\end{eqnarray}
As the mass dimension of $U$ is $+2$, it was expected to have a solitary contribution from $U^4$ to the dimension eight operators $(\phi^{\dagger}\phi)^4$. But, that is not certainly true and we find contributions to the same operator from $U^{5,6,7,8}$ all the structures. As an example, one can find a similar effect when the SM is extended by another Higgs Doublet (2HDM) and a gauge singlet scalar and low energy theory is SMEFT.

\section{Conclusions}
\label{sec:conc}
Effective Field Theory (EFT) has drawn much attention in recent times by its virtue. We are working to achieve more and more precision in EFT calculations. Also, higher dimensional effective operators have their own signatures and impact to crack the degeneracy among different UV theories as well. At this point, the computation of dimension eight effective operators will have a very large impact on our ongoing analysis. In this paper, we have stepped in that direction. We have enabled the Heat-Kernel (HK) method to compute the dimension eight one-loop effective Lagrangian after integrating out either a heavy or multiple degenerate heavy scalars. Our results do not depend on the form of either the UV or low energy theories. In this paper, we have computed the contributions from the loops that consist of only heavy scalar propagators. We have also employed the covariant diagram method to cross-check part of our results. These two methods complement each other to validate our results. We are in the process of extending this result by including heavy fermions and light-heavy propagator mixing in future works.

\section*{Acknowledgements}
We acknowledge the useful discussions with Diptarka Das and Nilay Kundu.  The authors would also like to acknowledge the initial discussions with Priyank Kaushik.

\appendix

\section{Review of covariant diagram method}
\label{app:cov-diag-review}

In this section, we present a brief review of the development of the covariant diagram representation starting from the original gauge-covariant functional form of the effective action. The topic has been greatly discussed in Refs.~\cite{Zhang:2016pja,Vandeven:1985}. As shown in Eq.~\eqref{eq:effective-action}, the one-loop part of the effective action for a field can be given as,
\begin{equation}
    \Delta S_{\text{eff}} = ic_s\, \text{Tr} \log(-P^2+M^2+U),
\end{equation}
 here, $c_s = +1/2$, or $+1 $ depending on whether the heavy field is a real scalar or complex scalar. The trace ``Tr" can then be evaluated by taking an integral over the momentum eigenstate basis,
\begin{eqnarray}
    \int d^dx\,\mathcal{L}_{\text{eff}}[\phi] 
    &=& ic_s\,\int\,\frac{d^d q}{(2\pi)^d}\bra{q}\text{tr} \log(-\widehat{P}^2+M^2+U)\ket{q}\nonumber\\
    &=& ic_s\int d^dx\int\frac{d^d q}{(2\pi)^d}\braket{q}{x}\bra{x}\text{tr} \log(-\widehat{P}^2+M^2+U)\ket{q}\nonumber\\
    &=& ic_s\int d^dx\int\frac{d^d q}{(2\pi)^d}\,e^{i\,q.x}\,\text{tr} \log(-P^2+M^2+U)\,e^{-i\,q.x}.
\end{eqnarray}
By following a straightforward manipulation of introducing the completeness relation
for the basis of the spatial eigenstates, we find the following form of the effective one-loop Lagrangian, 
\begin{align}\label{eq:one-loop-Lagrangian}
    \mathcal{L}_{\text{eff}}[\phi] &= ic_s\int\frac{d^d q}{(2\pi)^d}\,\text{tr} \log(-P^2+M^2+U)_{P\to P-q}\nonumber\\
    &= ic_s\int\frac{d^d q}{(2\pi)^d}\text{tr} \log(-P^2-q^2+2q.P+M^2+U)\nonumber\\
    \begin{split}
    &= ic_s\int\frac{d^d q}{(2\pi)^d}\text{tr} \bigg\{\log(-q^2+M^2) \\ 
    &\qquad\qquad\qquad+\log\big[1-(q^2-M^2)^{-1}(-P^2+2q.P+U)\big]\bigg\}.
    \end{split}
 \end{align} 
After performing the momentum integral, the first term in Eq.~\eqref{eq:one-loop-Lagrangian} reduces to a constant, while the second term can be expanded in an infinite series as shown in Eq.~\eqref{eq:Lag-one-loop}.

\clearpage
\newpage

\subsection{Covariant loop diagrams, their structures and values}\label{app:wc-cov-diags}
In this subsection, we present all the diagrams that can contribute to dimension eight interactions at each order of $P^{2n}U^{m}$ with possible contractions among $P_{\mu}$'s and note down the corresponding operator structures containing open covariant derivatives ($P_{\mu}$'s) and value for the loops. 

\subsubsection*{\Large$\mathbf{O(P^4\,U^3)}$}

\begin{table}[!h]
			\centering \scriptsize
			\renewcommand{\arraystretch}{2.6}
			\adjustbox{width = 0.75\textwidth}{\begin{tabular}{|c|c|c|c|}
				\hline
				\textsf{\quad Diagram}$\quad$&
				\textsf{\quad $O(P^4U^3)$ structure}$\quad$&
				\textsf{\quad Value}$\quad$\\
				\hline
                    \hline

                \begin{minipage}{0.09\textwidth}
                  \includegraphics[width=\linewidth, height=1.4cm]{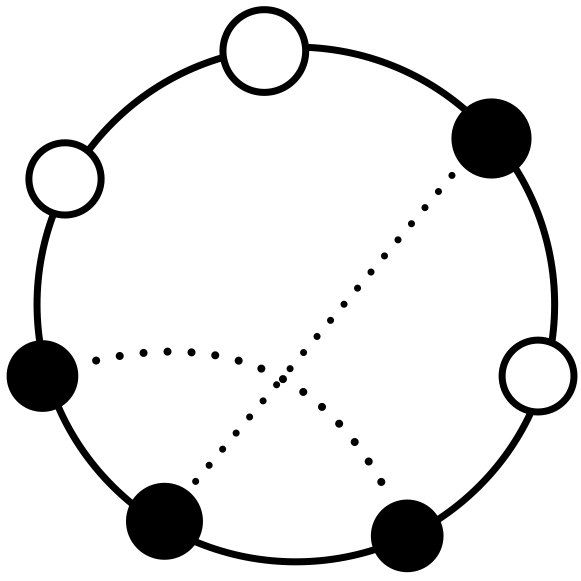}
                \end{minipage}&
                tr$\big(P_{\mu}UP_{\nu}P_{\mu}P_{\nu}UU\big)$
                &$-ic_s\,2^4\mathcal{I}[q^4]^7$
               \\

             \hline

                 \begin{minipage}{0.09\textwidth}
                  \includegraphics[width=\linewidth, height=1.4cm]{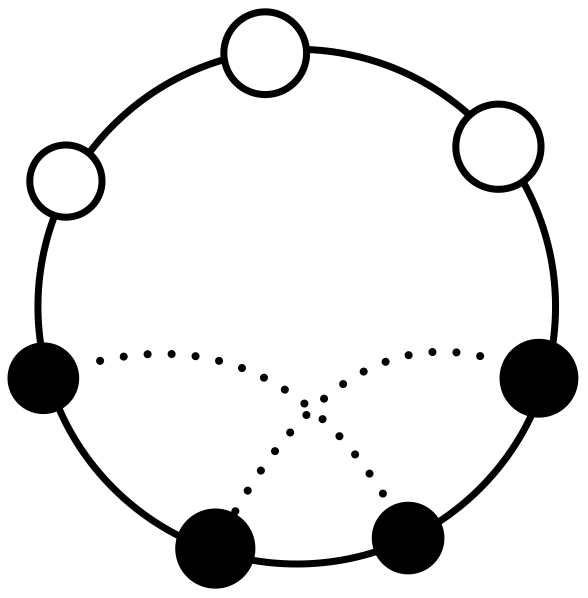}
                \end{minipage}&
                tr$\big(UUUP_{\mu}P_{\nu}P_{\mu}P_{\nu}\big)$
                &$-ic_s\,2^4\mathcal{I}[q^4]^7$
               \\
               \hline

                 \begin{minipage}{0.09\textwidth}
                  \includegraphics[width=\linewidth, height=1.4cm]{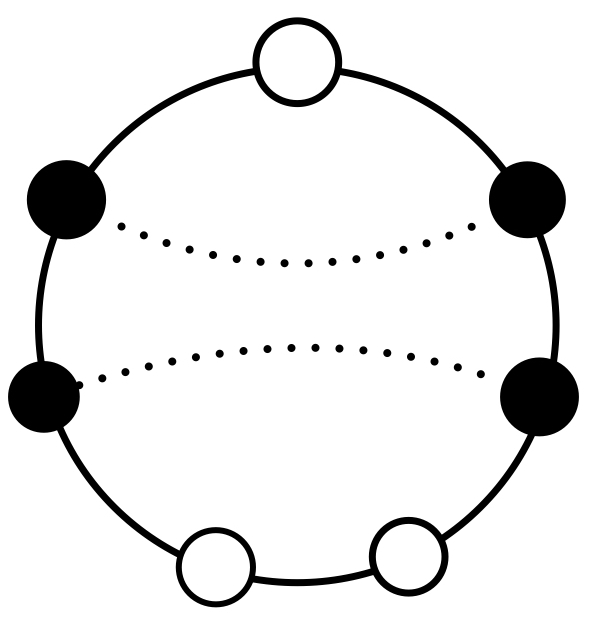}
                \end{minipage}&
                tr$\big(UP_{\mu}P_{\nu}UUP_{\nu}P_{\mu}\big)$
                &$-ic_s\,2^4\mathcal{I}[q^4]^7$
               \\
               \hline

                   \begin{minipage}{0.09\textwidth}
                  \includegraphics[width=\linewidth, height=1.4cm]{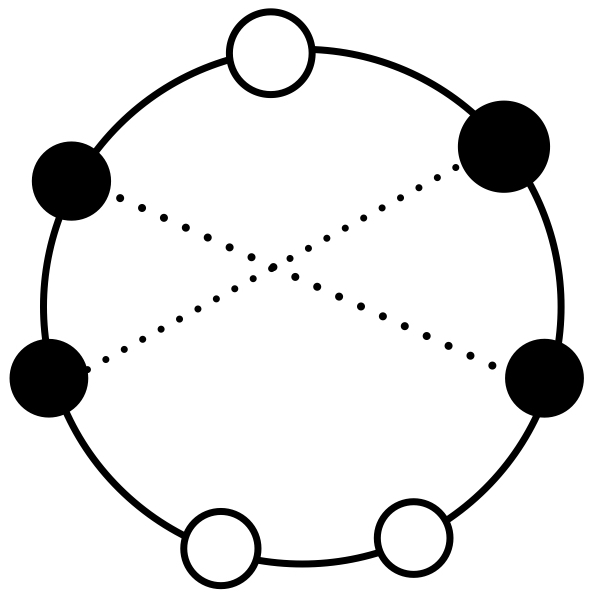}
                \end{minipage}&
                tr$\big(UP_{\mu}P_{\nu}UUP_{\mu}P_{\nu}\big)$
                &$-ic_s\,2^4\mathcal{I}[q^4]^7$
               \\
               \hline

                   \begin{minipage}{0.09\textwidth}
                  \includegraphics[width=\linewidth, height=1.4cm]{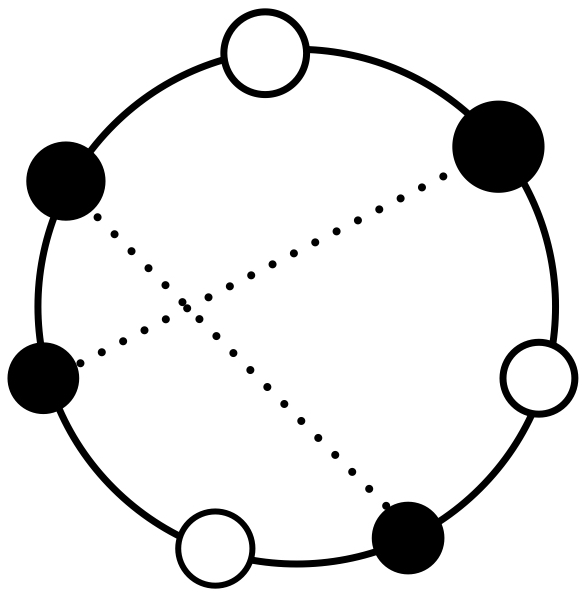}
                \end{minipage}&
                tr$\big(P_{\mu}UP_{\nu}UP_{\mu}P_{\nu}U\big)$
                &$-ic_s\,2^4\mathcal{I}[q^4]^7$
               \\
               \hline

                \begin{minipage}{0.09\textwidth}
                  \includegraphics[width=\linewidth, height=1.4cm]{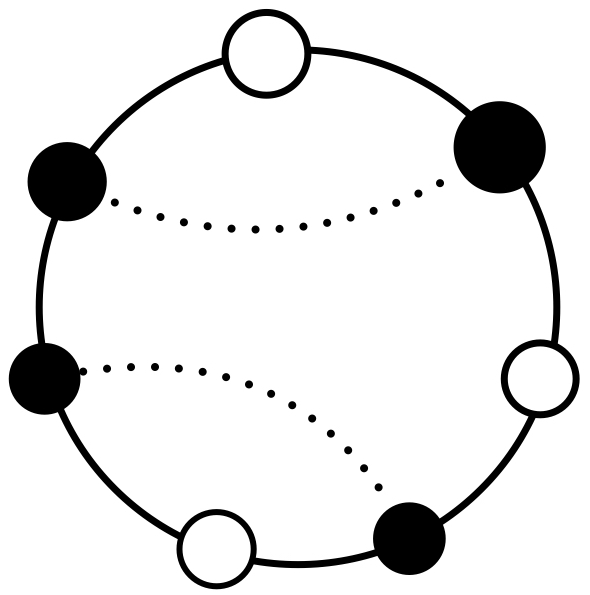}
                \end{minipage}&
                tr$\big(P_{\mu}UP_{\nu}UP_{\nu}P_{\mu}U\big)$
                &$-ic_s\,2^4\mathcal{I}[q^4]^7$
               \\
               \hline
               
			\end{tabular}}
			\caption{\small All possible diagrams at the level of $O(P^4U^3)$. In the second column, we present their corresponding operator structures with open covariant derivatives. Their values are given in the third column.}
			\label{table:P4U3-diags}
		\end{table} 

\clearpage
\newpage
\subsubsection*{\Large$\mathbf{O(P^8)}$}
\begin{table}[!h]
			\centering \scriptsize
			\renewcommand{\arraystretch}{2.6}
			\adjustbox{width = 0.75\textwidth}{\begin{tabular}{|c|c|c|c|}
				\hline
				\textsf{\quad Diagram}$\quad$&
				\textsf{\quad $O(P^8)$ structure}$\quad$&
				\textsf{\quad Value}$\quad$\\
				\hline
                    \hline

                \begin{minipage}{0.1\textwidth}
                  \includegraphics[width=\linewidth, height=1.5cm]{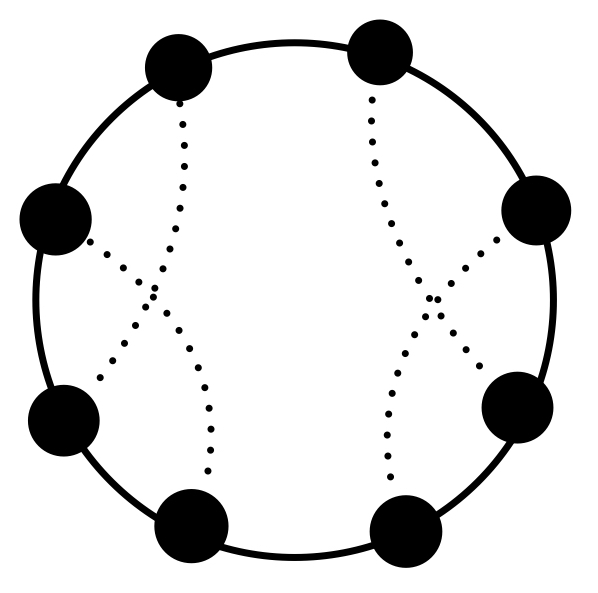}
                \end{minipage}&
                tr$\big(P_{\mu}P_{\nu}P_{\mu}P_{\nu}P_{\rho}P_{\sigma}P_{\rho}P_{\sigma}\big)$
                &
                $-i\frac{c_s}{2}\,2^8\mathcal{I}[q^8]^8$
               \\

             \hline

                 \begin{minipage}{0.1\textwidth}
                  \includegraphics[width=\linewidth, height=1.5cm]{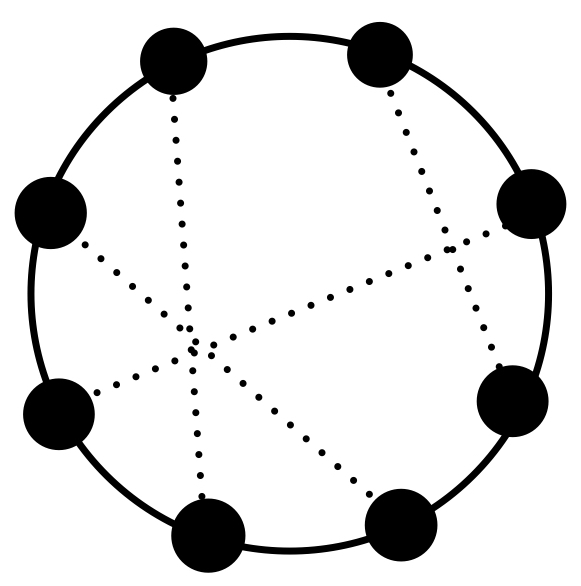}
                \end{minipage}&
                tr$\big(P_{\mu}P_{\nu}P_{\mu}P_{\rho}P_{\sigma}P_{\nu}P_{\rho}P_{\sigma}\big)$
                &$-ic_s\,2^8\mathcal{I}[q^8]^8$
               \\
               \hline

                 \begin{minipage}{0.1\textwidth}
                  \includegraphics[width=\linewidth, height=1.5cm]{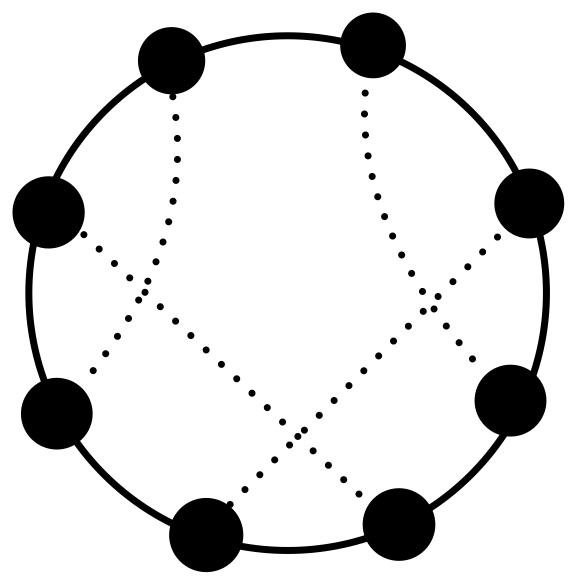}
                \end{minipage}&
                tr$\big(P_{\mu}P_{\nu}P_{\mu}P_{\rho}P_{\nu}P_{\sigma}P_{\rho}P_{\sigma}\big)$
                &$-ic_s\,2^8\mathcal{I}[q^8]^8$
               \\
               \hline

                \begin{minipage}{0.1\textwidth}
                  \includegraphics[width=\linewidth, height=1.5cm]{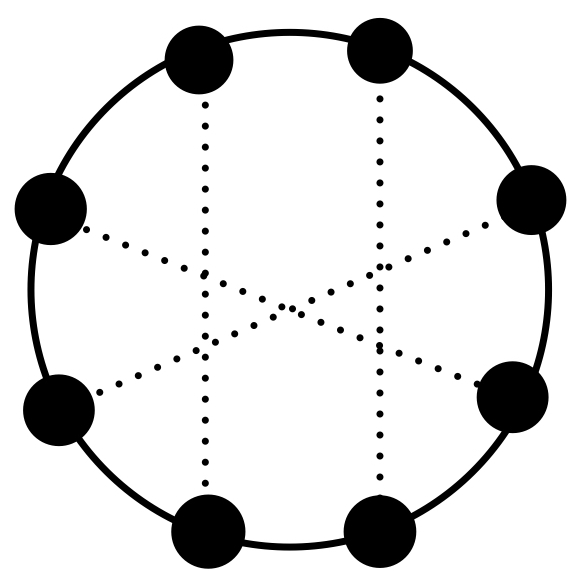}
                \end{minipage}&
                tr$\big(P_{\mu}P_{\nu}P_{\rho}P_{\sigma}P_{\mu}P_{\nu}P_{\sigma}P_{\rho}\big)$
                &$-i\frac{c_s}{2}\,2^8\mathcal{I}[q^8]^8$
               \\
               \hline

                \begin{minipage}{0.1\textwidth}
                \includegraphics[width=\linewidth, height=1.5cm]{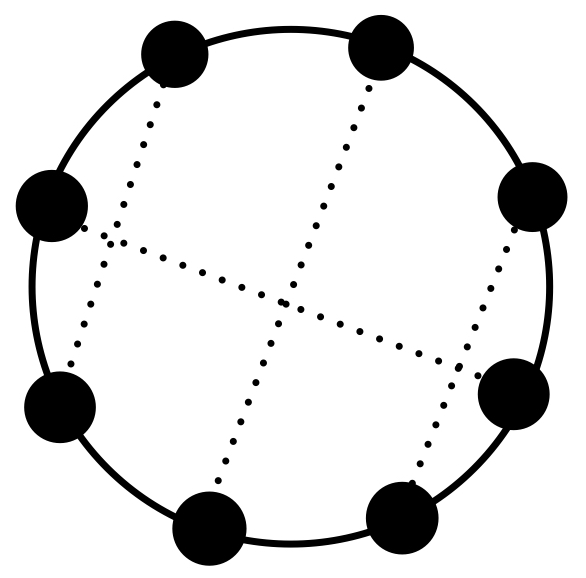}
                \end{minipage}&
                tr$\big(P_{\mu}P_{\nu}P_{\rho}P_{\nu}P_{\mu}P_{\sigma}P_{\rho}P_{\sigma}\big)$
                &$-i\frac{c_s}{2}\,2^8\mathcal{I}[q^8]^8$
               \\
               \hline

                \begin{minipage}{0.1\textwidth}
                \includegraphics[width=\linewidth, height=1.5cm]{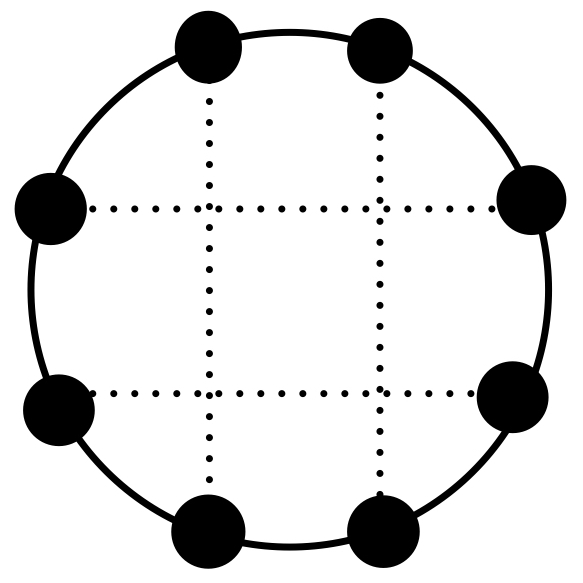}
                \end{minipage}&
                tr$\big(P_{\mu}P_{\nu}P_{\rho}P_{\mu}P_{\sigma}P_{\rho}P_{\nu}P_{\sigma}\big)$
                &$-i\frac{c_s}{4}\,2^8\mathcal{I}[q^8]^8$
               \\
               \hline

                \begin{minipage}{0.1\textwidth}
                \includegraphics[width=\linewidth, height=1.5cm]{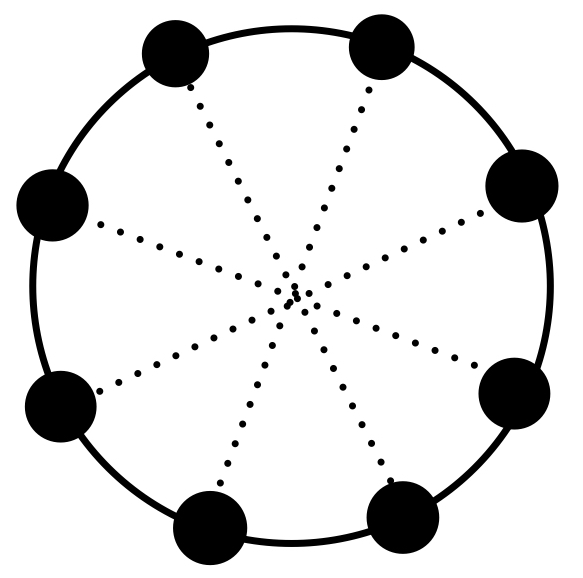}
                \end{minipage}&
                tr$\big(P_{\mu}P_{\nu}P_{\rho}P_{\sigma}P_{\mu}P_{\nu}P_{\rho}P_{\sigma}\big)$
                &$-i\frac{c_s}{8}\,2^8\mathcal{I}[q^8]^8$
               \\
               \hline
				
			\end{tabular}}
			\caption{\small All possible diagrams at the level of $O(P^8)$ containing eight $P_{\mu}$'s that are contracted among themselves. In the second column, we present their corresponding operator structures with open covariant derivatives. Their values are given in the third column.}
			\label{table:P8-diags}
		\end{table} 

\clearpage

\newpage

\subsubsection*{\Large$\mathbf{O(P^6\,U^2)}$}

\begin{table}[!h]
			\centering \scriptsize
			\renewcommand{\arraystretch}{2.6}
			\adjustbox{width = 0.8\textwidth}{\begin{tabular}{|c|c|c|c|}
				\hline
				\textsf{\quad Diagram}$\quad$&
				\textsf{\quad $O(P^6\,U^2)$ structure}$\quad$&
				\textsf{\quad Value}$\quad$\\
				\hline
                    \hline
                    
                \begin{minipage}{0.09\textwidth}
                  \includegraphics[width=\linewidth, height=1.4cm]{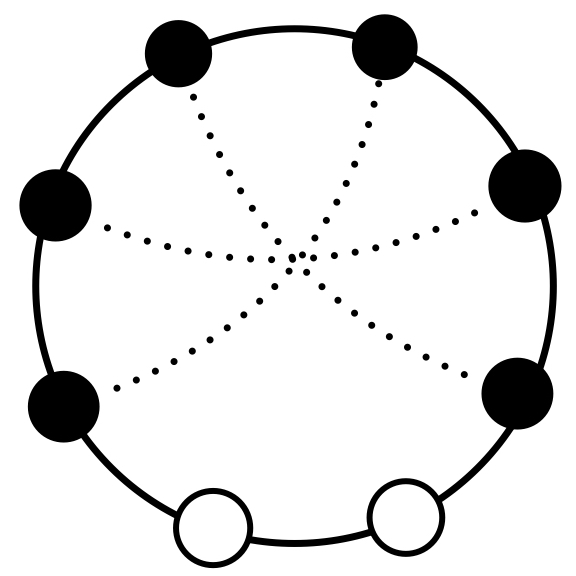}
                \end{minipage}&
                tr$\big(P_{\mu}P_{\nu}P_{\rho}UUP_{\mu}P_{\nu}P_{\rho}\big)$
                &$-ic_s\,2^6\,\mathcal{I}[q^6]^8$
               \\

             \hline

                 \begin{minipage}{0.09\textwidth}
                  \includegraphics[width=\linewidth, height=1.4cm]{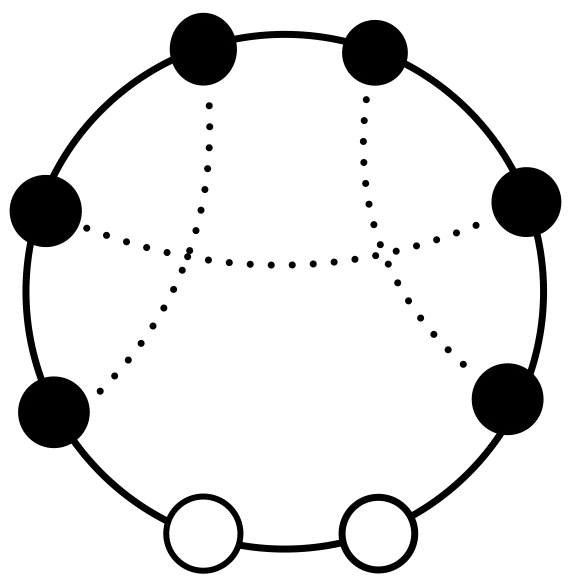}
                \end{minipage}&
                tr$\big(P_{\mu}P_{\nu}P_{\mu}UUP_{\rho}P_{\nu}P_{\rho}\big)$
                &$-ic_s\,2^6\,\mathcal{I}[q^6]^8$
               \\
               \hline

                 \begin{minipage}{0.09\textwidth}
                  \includegraphics[width=\linewidth, height=1.4cm]{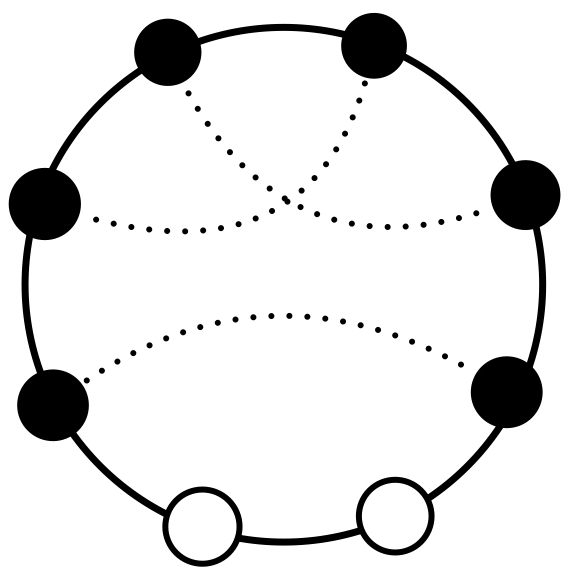}
                \end{minipage}&
                tr$\big(P_{\mu}P_{\nu}P_{\rho}UUP_{\rho}P_{\mu}P_{\nu}\big)$
                &$-ic_s\,2^6\,\mathcal{I}[q^6]^8$
               \\
               \hline

                   \begin{minipage}{0.09\textwidth}
                  \includegraphics[width=\linewidth, height=1.4cm]{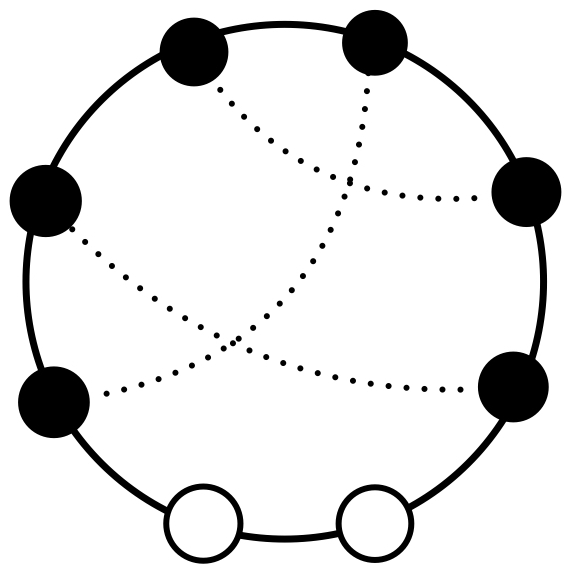}
                \end{minipage}&
                tr$\big(P_{\mu}P_{\nu}P_{\rho}UUP_{\mu}P_{\rho}P_{\nu}\big)$
                &$-ic_s\,2^6\,\mathcal{I}[q^6]^8$
               \\
               \hline

               \begin{minipage}{0.09\textwidth}
                  \includegraphics[width=\linewidth, height=1.4cm]{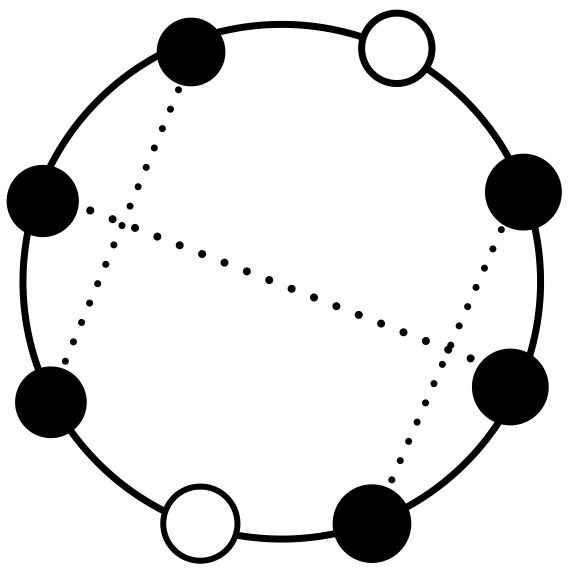}
                \end{minipage}&
                tr$\big(P_{\mu}P_{\nu}P_{\mu}UP_{\rho}P_{\nu}P_{\rho}U\big)$
                &$-i\frac{c_s}{2}\,2^6\,\mathcal{I}[q^6]^8$
               \\
               \hline

                \begin{minipage}{0.09\textwidth}
                  \includegraphics[width=\linewidth, height=1.4cm]{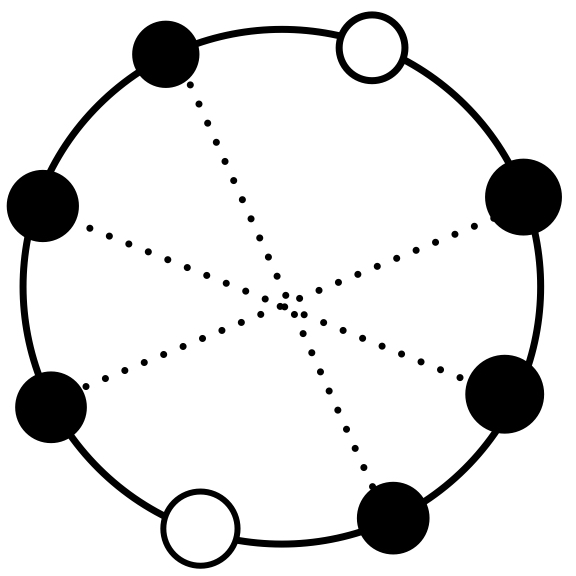}
                \end{minipage}&
                tr$\big(P_{\mu}P_{\nu}P_{\rho}UP_{\mu}P_{\nu}P_{\rho}U\big)$
                &$-i\frac{c_s}{2}\,2^6\,\mathcal{I}[q^6]^8$
               \\
               \hline

               \begin{minipage}{0.09\textwidth}
                  \includegraphics[width=\linewidth, height=1.4cm]{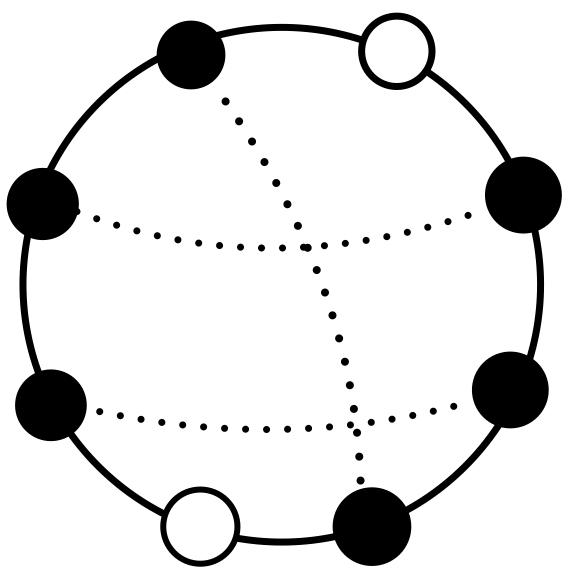}
                \end{minipage}&
                 tr$\big(P_{\mu}P_{\nu}P_{\rho}UP_{\nu}P_{\mu}P_{\rho}U\big)$
                &$-i\frac{c_s}{2}\,2^6\,\mathcal{I}[q^6]^8$
               \\
               \hline

               \begin{minipage}{0.09\textwidth}
                  \includegraphics[width=\linewidth, height=1.4cm]{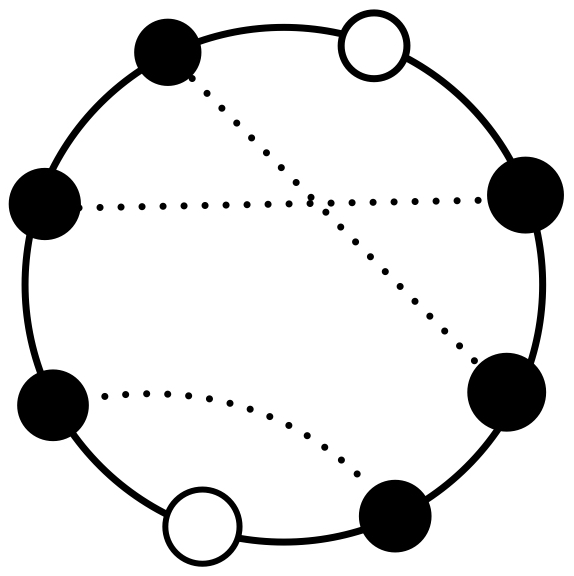}
                \end{minipage}&
                tr$\big(P_{\mu}P_{\nu}P_{\rho}UP_{\rho}P_{\mu}P_{\nu}U\big)$
                &$-ic_s\,2^6\,\mathcal{I}[q^6]^8$
               \\
               \hline

			\end{tabular}}
			\caption{\small All possible diagrams at the level of $O(P^6U^2)$ containing six $P_{\mu}$'s that are contracted among themselves and two $U$'s. In the second column, we present their corresponding operator structures with open covariant derivatives. Their values are given in the third column.}
			\label{tab:P6U2-diags-1}
		\end{table} 

\begin{table}[!h]
			\centering \scriptsize
			\renewcommand{\arraystretch}{2.6}
			\adjustbox{width = 0.8\textwidth}{\begin{tabular}{|c|c|c|c|}
				\hline
				\textsf{\quad Diagram}$\quad$&
				\textsf{\quad $O(P^6U^2)$ structure}$\quad$&
				\textsf{\quad Value}$\quad$\\
				\hline
                    \hline 

                        \begin{minipage}{0.09\textwidth}
                  \includegraphics[width=\linewidth, height=1.4cm]{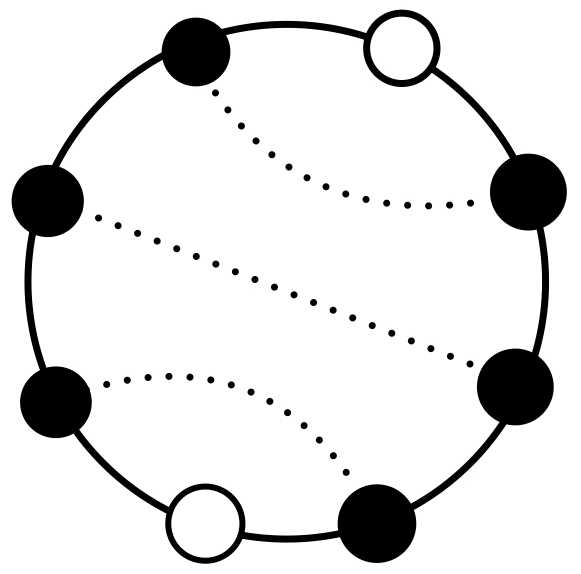}
                \end{minipage}&
                tr$\big(P_{\mu}P_{\nu}P_{\rho}UP_{\rho}P_{\nu}P_{\mu}U\big)$
                &$-i\frac{c_s}{2}\,2^6\,\mathcal{I}[q^6]^8$
               \\
               \hline

               \begin{minipage}{0.09\textwidth}
                  \includegraphics[width=\linewidth, height=1.4cm]{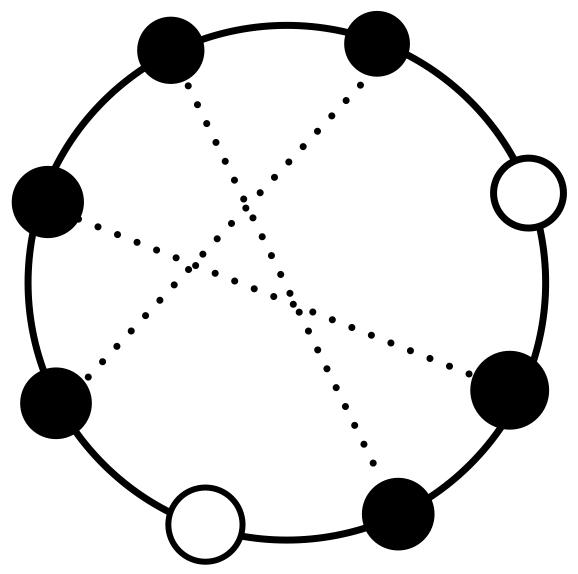}
                \end{minipage}&
                tr$\big(P_{\mu}P_{\nu}UP_{\rho}P_{\nu}P_{\mu}P_{\rho}U\big)$
                &$-i\frac{c_s}{2}\,2^6\,\mathcal{I}[q^6]^8$
               \\
               \hline

               \begin{minipage}{0.09\textwidth}
                  \includegraphics[width=\linewidth, height=1.4cm]{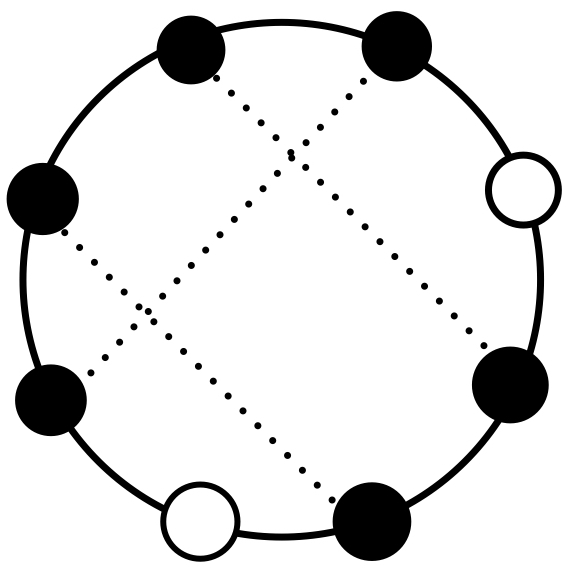}
                \end{minipage}&
                tr$\big(P_{\mu}P_{\nu}UP_{\rho}P_{\mu}P_{\nu}P_{\rho}U\big)$
                &$-i\frac{c_s}{2}\,2^6\,\mathcal{I}[q^6]^8$
               \\
               \hline

               \begin{minipage}{0.09\textwidth}
                  \includegraphics[width=\linewidth, height=1.4cm]{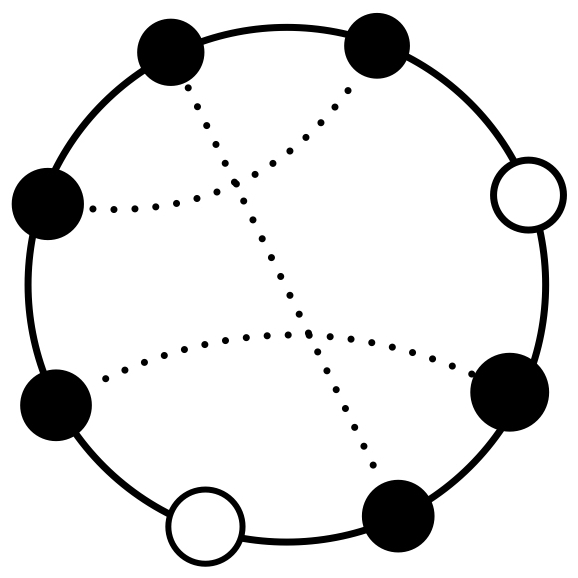}
                \end{minipage}&
                tr$\big(P_{\mu}P_{\nu}UP_{\mu}P_{\rho}P_{\nu}P_{\rho}U\big)$
                &$-ic_s\,2^6\,\mathcal{I}[q^6]^8$
               \\
               \hline

               \begin{minipage}{0.09\textwidth}
                  \includegraphics[width=\linewidth, height=1.4cm]{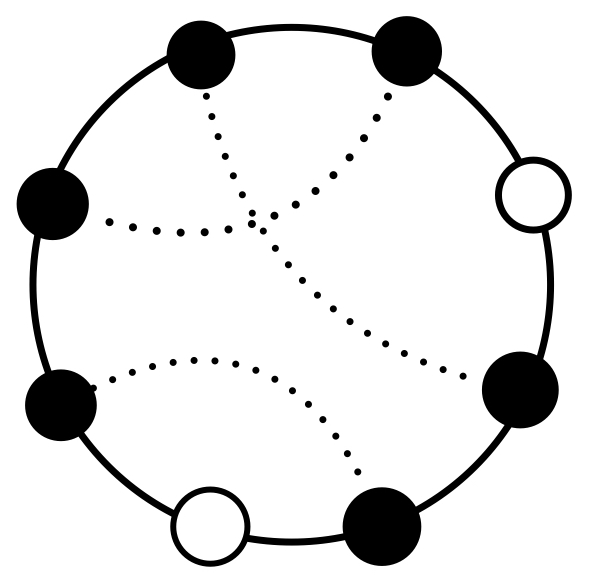}
                \end{minipage}&
                tr$\big(P_{\mu}P_{\nu}UP_{\nu}P_{\rho}P_{\mu}P_{\rho}U\big)$
                &$-ic_s\,2^6\,\mathcal{I}[q^6]^8$
               \\
               \hline

               \begin{minipage}{0.09\textwidth}
                  \includegraphics[width=\linewidth, height=1.4cm]{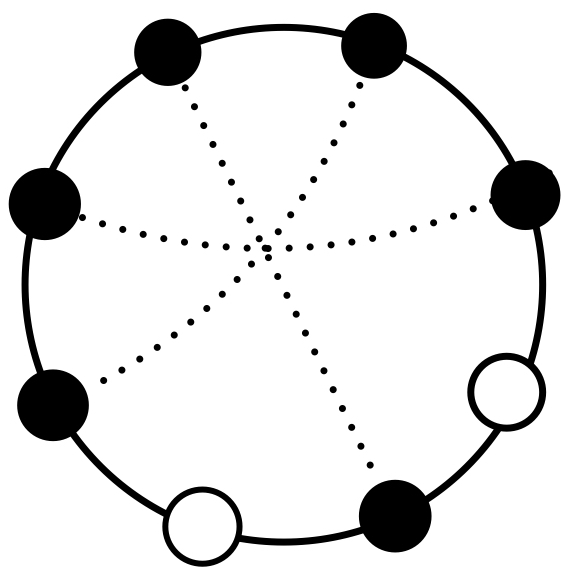}
                \end{minipage}&
                tr$\big(P_{\mu}P_{\nu}UP_{\rho}P_{\mu}P_{\nu}P_{\rho}U\big)$
                &$-ic_s\,2^6\,\mathcal{I}[q^6]^8$
               \\
               \hline

                    \begin{minipage}{0.09\textwidth}
                  \includegraphics[width=\linewidth, height=1.4cm]{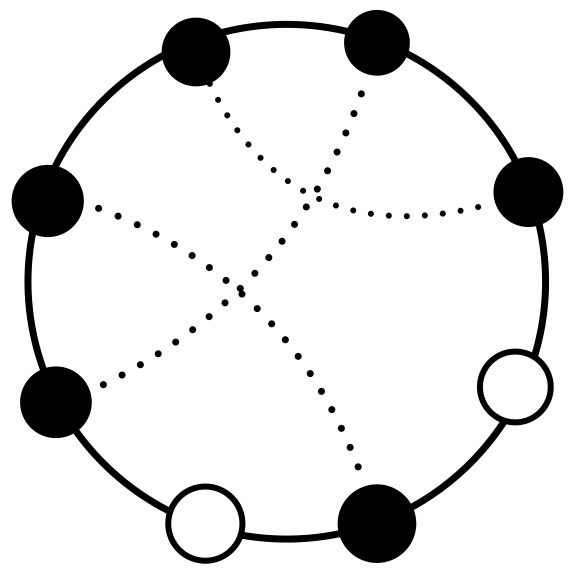}
                \end{minipage}&
                tr$\big(P_{\mu}P_{\nu}P_{\mu}UP_{\rho}UP_{\nu}P_{\rho}\big)$
                &$-ic_s\,2^6\,\mathcal{I}[q^6]^8$
               \\
               \hline

               \begin{minipage}{0.09\textwidth}
                  \includegraphics[width=\linewidth, height=1.4cm]{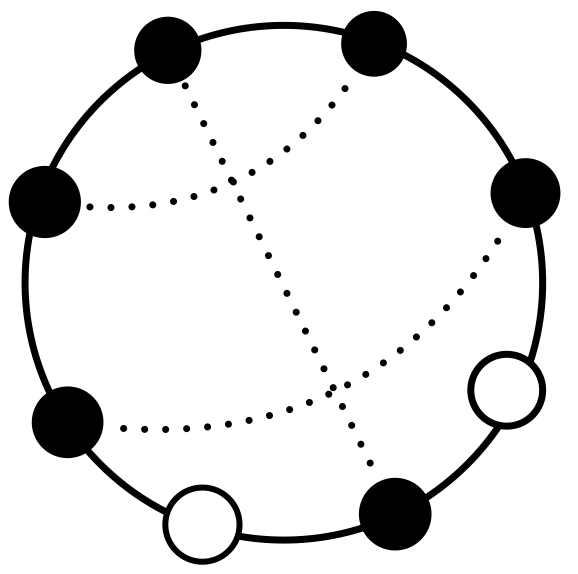}
                \end{minipage}&
                tr$\big(P_{\mu}P_{\nu}P_{\rho}UP_{\mu}UP_{\rho}P_{\nu}\big)$
                &$-ic_s\,2^6\,\mathcal{I}[q^6]^8$
               \\
               \hline

               \begin{minipage}{0.09\textwidth}
                  \includegraphics[width=\linewidth, height=1.4cm]{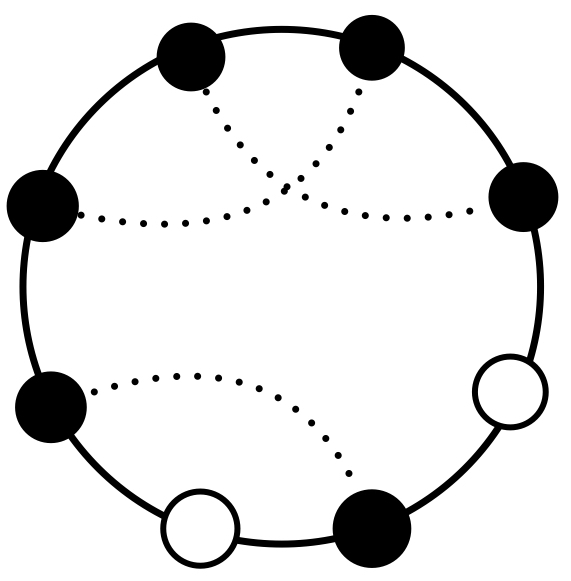}
                \end{minipage}&
                tr$\big(P_{\mu}P_{\nu}UP_{\rho}UP_{\rho}P_{\mu}P_{\nu}\big)$
                &$-ic_s\,2^6\,\mathcal{I}[q^6]^8$
               \\
               \hline
				
			\end{tabular}}
			\caption{\small Table~\ref{tab:P6U2-diags-1} continued.
                    }
			\label{tab:P6U2-diags-2}
		\end{table} 

\clearpage
\subsubsection*{\Large$\mathbf{O(P^6\,U)}$}

\begin{table}[!h]
			\centering \scriptsize
			\renewcommand{\arraystretch}{2.6}
			\adjustbox{width = 0.75\textwidth}{\begin{tabular}{|c|c|c|c|}
				\hline
				\textsf{\quad Diagram}$\quad$&
				\textsf{\quad $O(P^6U)$ structure}$\quad$&
				\textsf{\quad Value}$\quad$\\
				\hline
                    \hline

                \begin{minipage}{0.09\textwidth}
                  \includegraphics[width=\linewidth, height=1.4cm]{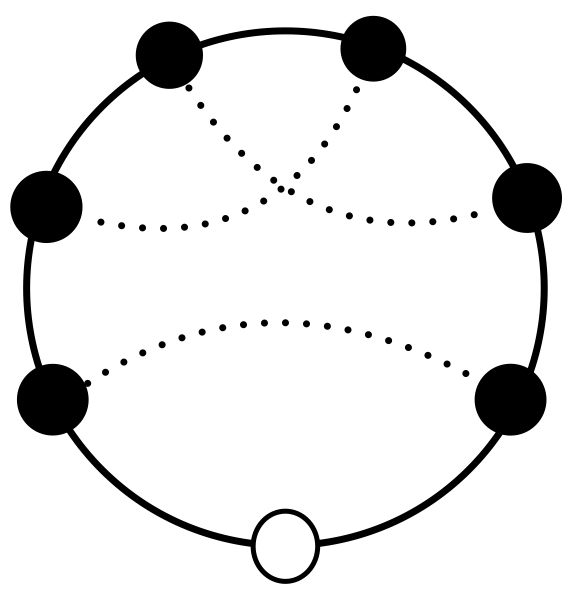}
                \end{minipage}&
                tr$\big(P_{\mu}P_{\nu}P_{\rho}P_{\nu}P_{\rho}P_{\mu}U\big)$
                &$-ic_s\,2^6\mathcal{I}[q^6]^7$
               \\

             \hline

                 \begin{minipage}{0.09\textwidth}
                  \includegraphics[width=\linewidth, height=1.4cm]{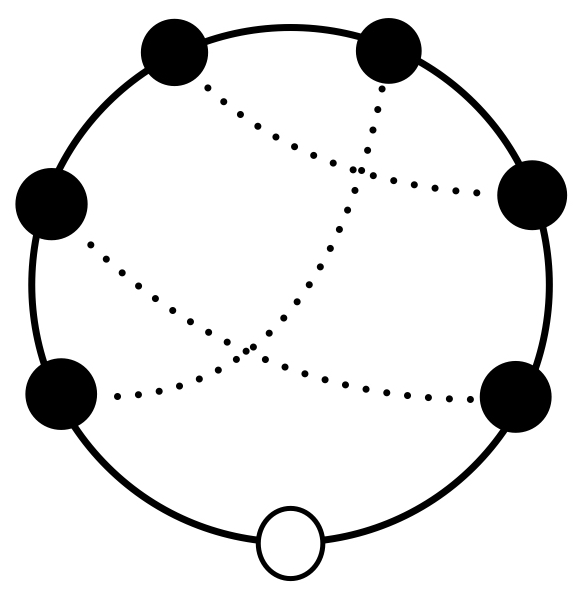}
                \end{minipage}&
                tr$\big(P_{\mu}P_{\nu}P_{\rho}P_{\mu}P_{\rho}P_{\nu}U\big)$
                &$-ic_s\,2^6\mathcal{I}[q^6]^7$
               \\
               \hline

                 \begin{minipage}{0.09\textwidth}
                  \includegraphics[width=\linewidth, height=1.4cm]{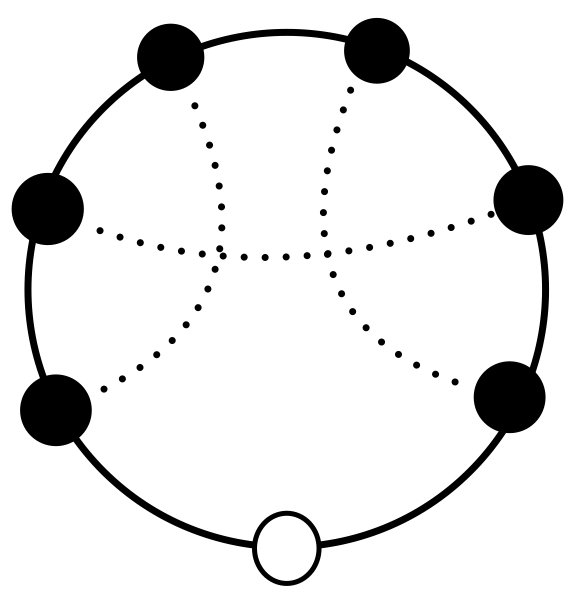}
                \end{minipage}&
                tr$\big(P_{\mu}P_{\nu}P_{\mu}P_{\rho}P_{\nu}P_{\rho}U\big)$
                &$-ic_s\,2^6\mathcal{I}[q^6]^7$
               \\
               \hline

                   \begin{minipage}{0.09\textwidth}
                  \includegraphics[width=\linewidth, height=1.4cm]{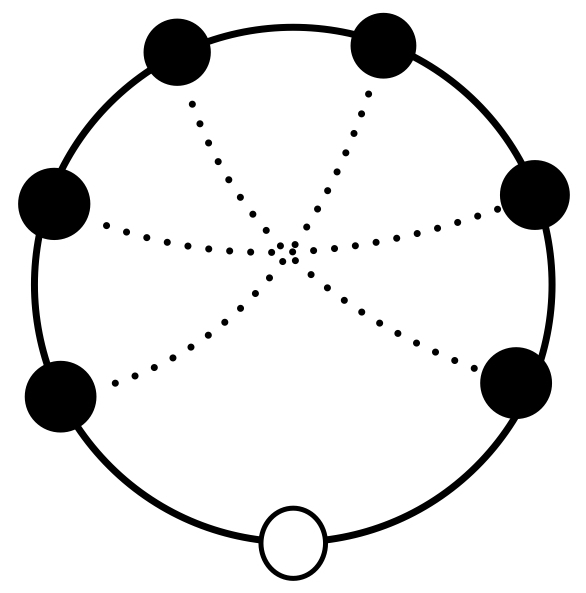}
                \end{minipage}&
                tr$\big(P_{\mu}P_{\nu}P_{\rho}P_{\mu}P_{\nu}P_{\rho}U\big)$
                &$-ic_s\,2^6\mathcal{I}[q^6]^7$
               \\
               \hline
				
			\end{tabular}}
			\caption{\small All possible diagrams at the level of $O(P^6U)$. In the second column, we present their corresponding operator structures with open covariant derivatives. Their values are given in the third column.}
			\label{table:P6U-diags}
		\end{table} 

\subsection{Master integrals for heavy loops}
\label{app:master-integral}

Each of the covariant diagrams mentioned in Sec.~\ref{sec:verification} corresponds to a loop integral with $n$ heavy propagators and $2n_c$ contractions which can be generalised in the following form
\begin{eqnarray}
    \int \frac{d^d q}{(2\pi)^d}\,\frac{q^{\mu_1}\cdots q^{\mu_{2n_c}}}{(q^2-M^2)^{n}}\,\equiv \,g^{\mu_1\cdots \mu_{2n_c}} \,\mathcal{I}[q^{2n_c}]^{n},
\end{eqnarray}
We have used \textit{Package-X}~\cite{Patel:2015tea,Patel:2016fam} to compute the loop integrals. In Table~\ref{tab:master-integrals}, we have listed the results for the loop integrals discussed in Sec.~\ref{sec:verification}. 
\begin{table}[!h]
			\centering \scriptsize
			\renewcommand{\arraystretch}{2.6}
			\adjustbox{width = 0.65\textwidth}{\begin{tabular}{|c|c|c|c|}
				\hline
				\textsf{\quad Integral}$\quad$&
				\textsf{\quad Value}$\quad$&
                    \textsf{\quad Mentioned in}$\quad$\\
				\hline
                    \hline

                $\mathcal{I}[q^2]^8$
                & $-\frac{i}{16\pi^2}\frac{1}{420\,\text{M}^{10}}$
                & Table~\ref{table:P2U6-diags}
               \\

             \hline

               $\mathcal{I}[q^4]^7$
                & $-\frac{i}{16\pi^2}\frac{1}{1440\,\text{M}^{6}}$
                & Table~\ref{table:P4U3-diags} 
               \\
               \hline

                $\mathcal{I}[q^6]^7$
                & $\frac{i}{16\pi^2}\frac{1}{5760\,\text{M}^{4}}$
                & Table~\ref{table:P6U-diags}
               \\
               \hline

               $\mathcal{I}[q^6]^8$
                & $-\frac{i}{16\pi^2}\frac{1}{20160\,\text{M}^{6}}$
                & Table~\ref{tab:P6U2-diags-1} and Table~\ref{tab:P6U2-diags-2}
               \\
               \hline

                $\mathcal{I}[q^8]^8$
                & $\frac{i}{16\pi^2}\frac{1}{80640\,\text{M}^{4}}$
                & Table~\ref{table:P8-diags}
               \\
               \hline

			\end{tabular}}
			\caption{\small The values corresponding to the master integrals associated with the covariant diagrams.}
			\label{tab:master-integrals}
		\end{table}



\newpage
\bibliographystyle{JHEP}
\bibliography{ref.bib}

\end{document}